\documentclass{article}

\usepackage{emulateapj}
\usepackage{onecolfloat}
\usepackage{graphicx}

\def\etal{{\it et\thinspace al.\ }}

\def\slantfrac#1#2{\hbox{$\,^#1\!/_#2$}}

\def\bv{\hbox{$B\!-\!V$}}

\def\ri{\hbox{$R\!-\!I$}}

\def\br{\hbox{$B\!-\!R$}}
\def\ABmag{AB\,mag}
\def\sqarcsec{arcsec$^2$}
\def\ion#1#2{\rm{#1}{\sc{#2}}\relax}
\def\nodata{ ~$\cdots$~ }
\def\rms{{\it rms}}
\def\MJysr{MJy sr$^{-1}$}
\def\escsa{ergs s$^{-1}$ cm$^{-2}$ sr$^{-1}$ \AA$^{-1}$}
\def\escsh{ergs s$^{-1}$ cm$^{-2}$ sr$^{-1}$ Hz$^{-1}$}
\def\escs{ergs s$^{-1}$ cm$^{-2}$ sr$^{-1}$}
\def\esca{ergs s$^{-1}$ cm$^{-2}$ \AA$^{-1}$}
\def\esch{ergs s$^{-1}$ cm$^{-2}$ Hz$^{-1}$}

\def\tto#1{$\times10^{#1}$}
\def\MJysr{ MJy sr$^{-1}$}
\def\micron{\hbox{$\mu$m}}

\def\arcdeg{\hbox{$^\circ$}}
\def\arcmin{\hbox{$^\prime$}}

\def\fs{\hbox{$.\!\!^{\rm s}$}}
\def\fdg{\hbox{$.\!\!^\circ$}}
\def\farcm{\hbox{.\kern -0.7ex\raisebox{.9ex}{\scriptsize$\prime$}}}
\def\farcs{\hbox{\kern 0.13ex.\kern -0.95ex%
  \raisebox{.9ex}{\scriptsize$\prime\prime$}\kern -0.1ex}}

\def\spose#1{\hbox to 0pt{#1\hss}}
\def\lesssim{\mathrel{\hbox{\rlap{\hbox{\lower4pt\hbox{$\sim$}}}\hbox{$<$}}}}
\def\gtrsim{\mathrel{\hbox{\rlap{\hbox{\lower4pt\hbox{$\sim$}}}\hbox{$>$}}}}
\def\lta{\mathrel{\spose{\lower 3pt\hbox{$\mathchar"218$}}
     \raise 2.0pt\hbox{$\mathchar"13C$}}}
\def\gta{\mathrel{\spose{\lower 3pt\hbox{$\mathchar"218$}}
     \raise 2.0pt\hbox{$\mathchar"13E$}}}

\lefthead{Bernstein, Freedman, \& Madore}
\righthead{Detections of the Optical EBL: Results}

\submitted{Submitted: 13 June 2000 ; 
Revised:  3 May 2001}

\begin{document}

\twocolumn[

\title{ The First Detections of the Extragalactic Background
	Light at 3000, 5500, and 8000\AA\ (I): 
	Results}

\author{Rebecca A. Bernstein\altaffilmark{1,2,3}}
\author{Wendy L. Freedman\altaffilmark{2}}
\author{Barry F. Madore\altaffilmark{2,4}}
 
\affil{\footnotesize 1) Division of Math, Physics, and Astronomy,
		California Institute of Technology,
                Pasadena, CA 91125}
\affil{\footnotesize 2) Carnegie Observatories,
                813 Santa Barbara St, 
                Pasadena, CA 91101}
\affil{\footnotesize 3) rab@ociw.edu, Hubble Fellow}
\affil{\footnotesize 4) NASA/IPAC Extragalactic Database, 
                California Institute of Technology, 
                Pasadena, CA 91125}
 
\begin{abstract}

We present the first detection of the mean flux of the optical
extragalactic background light (EBL) at 3000, 5500, and 8000\AA.
Diffuse foreground flux at these wavelengths comes from terrestrial
airglow, dust--scattered sunlight (zodiacal light), and
dust--scattered Galactic starlight (diffuse Galactic light).  We have
avoided the brightest of these, terrestrial airglow, by measuring the
absolute surface brightness of the night sky from above the Earth's
atmosphere using the Wide Field Planetary Camera2 (WFPC2) and Faint
Object Spectrograph (FOS), both aboard the Hubble Space Telescope
(HST).  On the ground, we have used the duPont 2.5\,m Telescope at Las
Campanas Observatory (LCO) to obtain contemporaneous spectrophotometry
of ``blank'' sky in the HST field of view to measure and then subtract
foreground zodiacal light from the HST observations.  We have
minimized the diffuse Galactic light in advance by selecting the HST
target field along a line of sight with low Galactic dust column
density, and then estimated the low--level Galactic foreground using a
simple scattering model and the observed correlation between thermal,
100\micron\ emission and optical scattered flux from the same dust.
In this paper, we describe the coordinated LCO/HST program and the HST
observations and data reduction, and present the resulting
measurements of the EBL.

Galaxies brighter than $V=23$\,\ABmag\ are not well sampled in an
image the size of the WFPC2 field of view.  We have therefore measured
the EBL from unresolved and resolved galaxies fainter than
$V=23$\,\ABmag\ by masking out brighter galaxies. We write as EBL23 to
emphasize this bright magnitude cut--off.  From absolute surface
photometry using WFPC2 and ground--based spectroscopy, we find mean
values for the EBL23 of
4.0 ($\pm2.5$),
2.7 ($\pm1.4$), and
2.2 ($\pm1.0$)
in units of \tto{-9}\escsa\ in the F300W, F555W, and F814W bandpasses,
respectively.  The errors quoted are $1\sigma$ combined statistical
and systematic uncertainties.  The total flux measured in resolved
galaxies with $V>23$\,\ABmag\ by standard photometric methods is
roughly 15\% of the EBL23 flux in each band. We have also developed
a new method of source photometry, uniquely suited to these data, with
which we can measure the ensemble flux from detectable sources much
more accurately than is possible with standard methods for faint
galaxy photometry.  Using this method, we have quantified systematic
biases affecting standard galaxy photometry, which prevent light from
being recovered in isophotes within a few percent of the sky
level. These biases have a significant effect on faint galaxy counts.
The flux from resolved sources as measured by our ensemble photometry
method is
3.2  ($\pm$ 0.22),
0.89 ($\pm$ 0.01), and 
0.76 ($\pm$ 0.01)
in units of \tto{-9} \escsa\ in the F300W, F555W, and F814W
bandpasses, respectively, with $1\sigma$ combined errors.  These
values, the total flux from resolved sources, represent {\it absolute
minima} for the EBL23 in each band, and are roughly 30\% of the
mean flux we measure for the total EBL23. 

\end{abstract}


 ]

\setcounter{footnote}{0}

\section{\uppercase{Introduction}}\label{intro}
 
\begin{figure}[t]
\begin{center}
\includegraphics[width=3.5in]{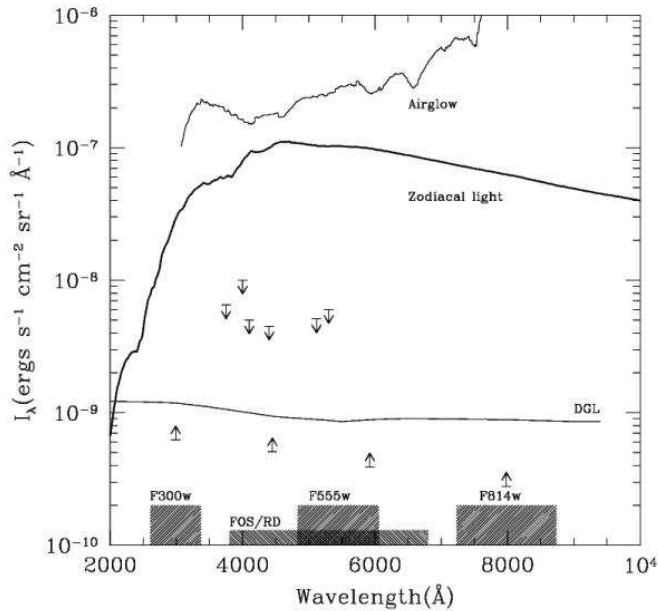}
\caption{\footnotesize Relative surface brightnesses of foreground
sources, upper limits on the EBL23 (see \S\ref{intro}), and lower
limits based on the integrated flux from resolved galaxies
($V_{555}>23$ AB mag) in the HDF (Williams \etal 1996). The spectral
shape and mean flux of zodiacal light and of diffuse galactic light
(DGL) are shown at the levels we detect in this work. The airglow
spectrum is taken from Broadfoot \& Kendall (1968) and is scaled to
the flux level we observe at 3800--5100\AA\ (see \S\ref{zl}). The
effective bandpasses for our HST observations are indicated at the
bottom of the plot.}
\label{fig:foregrndflux} 
\end{center}
\end{figure}

The Extragalactic Background Light (EBL) is the integrated, mean
surface brightness of both resolved and unresolved extragalactic
sources.  In wavelength range of these observations (2500--9500\AA),
the EBL is dominated by flux from stellar nucleosynthesis at redshifts
$z\lta 9$, and also includes flux from gravitational potential
energy such as accreting black holes (Fabian \& Iwasawa 1999), brown
dwarfs, gravitationally collapsing systems, and possibly decaying
particles.  As such, the EBL is the fossil record of the star
formation history of the universe and a fundamental measure of the
luminous energy content of the universe. As we show in Figure
\ref{fig:foregrndflux}, upper limits from previous attempts to measure
the EBL and lower limits from integrated galaxy counts constrain the
EBL to an expected level of roughly $1\times10^{-9}$ \escsa\ at
5500\AA, or 28.2 AB mag arcsec$^{-2}$.\footnote{All surface
brightnesses are specified in \escsa\ unless specifically noted
otherwise.  AB mag is defined as AB mag$=-2.5\log F_\nu - 48.6$, as
usual, with F$_\nu$ given in \esch.}  Because the combined flux from
foreground airglow, zodiacal light, and diffuse galactic light (also
plotted in Figure \ref{fig:foregrndflux}) is at least 100 times
brighter than the EBL, a detection of the EBL requires measurement
accuracies of better than 1\%.

Previous attempts to measure the optical EBL have employed a variety of
different approaches.  Mattila (1976) attempted to isolate the EBL by
differencing integrations on and off the line of sight to ``dark''
clouds at high Galactic latitude, under the assumption that the clouds
acted as a ``blank screens,'' spatially isolating all foreground
contributions from the background.  This pioneering work produced
upper limits and identified both the rapid temporal variability of
terrestrial airglow and extinction and the spatial variability of
diffuse Galactic light, which proved to be the primary obstacles to
these early efforts to measure the EBL.  Toller (1983) later attempted to
avoid both atmospheric and zodiacal foregrounds by using data taken
with the {\it Pioneer 10} spacecraft at a distance of 3 AU from the
Sun, beyond the zodiacal dust cloud.  Poor spatial resolution
($2^{\circ}$), however, prevented the accurate subtraction of discrete
Galactic stars, let alone the diffuse Galactic component from these
data.  Dube, Wilkes, \& Wilkinson (1977, 1979) made the first effort
to measure and subtract foreground contributions explicitly based on
geometrical modeling of airglow and Galactic foregrounds but including
spectroscopic measurement of the zodiacal light (ZL) flux by a
technique similar to that which we have adopted. Rapid variability
caused uncertainty in their airglow subtraction which dominated the
errors in their results.

In this work, we take advantage of the significant gains in technology
and in understanding of the foreground sources which have been
achieved since the last attempts to measure the optical EBL (see
Mattila 1990 for a review).  The most significant technological
advance is panoramic, linear CCD detectors.  Those on-board HST
allowed us to completely avoid bright, time--variable airglow and
provide sub--arcsecond spatial resolution. High spatial resolution
allowed us to resolve stars to $V\sim 27.5$ mag and thereby eliminate
the possibility of significant contamination from unidentified
Galactic stars in the field. Ground--based spectrophotometry with CCDs
also made possible much more accurate measurement of the foreground
zodiacal light than could be achieved with narrow--band filters and
photometers, as were used by Dube, Wickes, \& Wilkinson (1977,
1979). Finally, IRAS has provided maps of the thermal emission from
dust at high Galactic latitudes.  We have use the IRAS maps to select
a line of sight for these observations which has a low column density
of Galactic dust in order to minimize the DGL contribution caused by
dust--scattered starlight, and also to estimate the low--level DGL
which cannot be avoided.

Our measurement of the EBL utilizes three independent data sets.  Two
of these are from HST: (1) images from the Wide Field Planetary
Camera 2 (WFPC2) using the F300W, F555W, and F814W filters, each
roughly 1000\AA\ wide with central wavelengths of 3000, 5500, and
8000\AA, respectively; and (2) low--resolution spectra (300\AA\ per
resolution element) from the Faint Object Spectrograph (FOS) covering
3900--7000\AA.  The FOS data were taken in parallel observing mode
with the WFPC2 observations.  While flux calibration of WFPC2 images
and FOS spectra achieve roughly the same accuracy for point source
observations, the increase in spatial resolution, $10^4$ times larger
field of view, lower instrumental background, and absolute surface
brightness calibration achievable with WFPC2 make it better suited
than FOS to an absolute surface brightness measurement of the EBL.
Nonetheless, the FOS observations do provide a second, independent
measurement of the total background flux of the night sky, also free
of terrestrial airglow and extinction, but with greater spectral
resolution than the WFPC2 images.  The third data set consists of long--slit
spectrophotometry of a region of ``blank'' sky within the WFPC2 field
of view. These data were obtained at the 2.5m duPont telescope at Las
Campanas Observatory (LCO) using the Boller \& Chivens spectrograph
simultaneously with one visit of the HST observations (6 of 18
orbits).

The flow chart in Figure \ref{fig:flowchart} summarizes the reduction
procedures for each data set used in this measurement, the results
obtained from each data set individually, and the coordination of
those results to produce a measurement of the EBL.  In this paper, we
begin by describing the foreground sources in \S\ref{foreg} and the
details of HST scheduling in \S\ref{sched}.  The observations and data
reduction of both HST data sets, WFPC2 and FOS, are discussed in
detail in this paper.  WFPC2 observations and data reduction are
discussed in \S\ref{wfpc2}.  In \S\ref{wfpc2.backg}, we present the
first results from the WFPC2 data, which are measurements of the total
sky flux (foregrounds plus background) in each bandpass.  The FOS
observations, data reduction, and results are discussed in \S\ref{fos}
and \S\ref{fos.resul}. The modeling of diffuse Galactic light is
discussed in \S\ref{dgl}.  The LCO data and measurement of ZL are
discussed in Bernstein, Freedman \& Madore (2001b, henceforth Paper
II). A summary of that work is given in \S\ref{zl}.  In
\S\ref{ebl.minim}, we present a measurement of the minimum flux of the
EBL from resolved sources, which can be made using the WFPC2 data
alone.  The implications of that result are also discussed in
\S\ref{ebl.minim}.  Finally, in \S\ref{ebl.detec} we combine the
results of the individual data sets and modeling (horizontal
connections shown in the flow chart) to obtain a measurement of the
EBL. The implications of these results are discussed in Bernstein,
Freedman \& Madore (2001c, henceforth Paper III).

\section{\uppercase{Foregrounds}}\label{foreg}

The contribution from foreground sources to the flux of the night sky
is at least two orders of magnitude brighter than the expected
extragalactic signal. That is, the combined flux from foreground
sources is roughly 100\tto{-9} \escsa\ (or 23.2
AB\,mag\,arcsec$^{-2}$) (see Figure \ref{fig:foregrndflux}).  A
detection of the EBL therefore requires the measurement of both the
total background and the individual foreground sources with an
accuracy of roughly 1\% of the total sky surface brightness, or
$\sim1$\tto{-9} \escsa\ at 5500\AA\ (28.2 AB mag arcsec$^{-2}$). Each
of the foreground sources, and our approach to minimizing and
measuring them, are described briefly in the following sections.

\subsection{Terrestrial Airglow and Extinction}\label{foreg.terre}

\begin{figure*}[hp]
\begin{center}
\includegraphics[height=8in,angle=0]{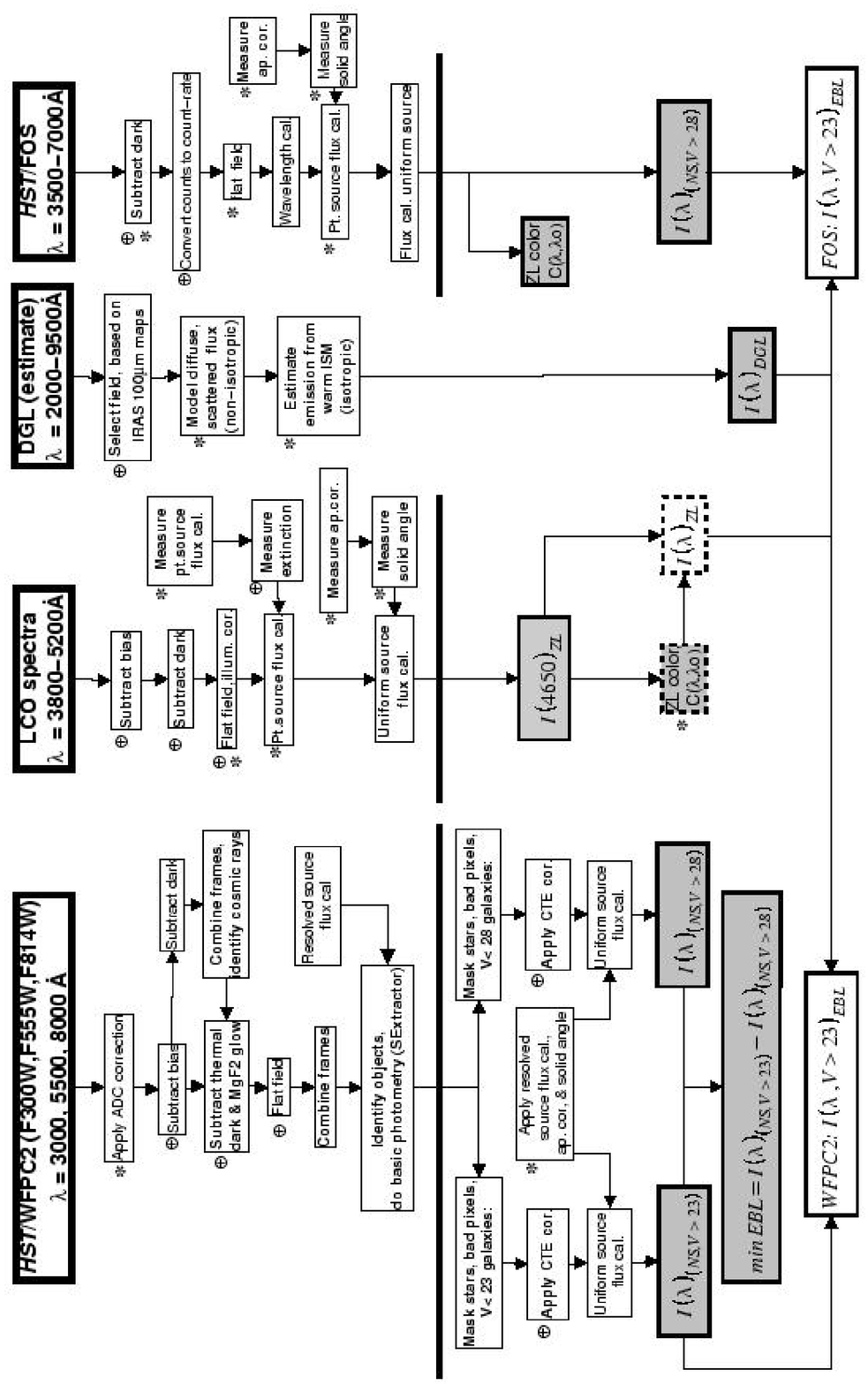}
\caption{ \footnotesize
 Flow chart of data reduction, analysis, and results from all
data sets used to measure the EBL.  Where appropriate, subscripts to
$I(\lambda)$ indicate the bright magnitude cut--off applied.  The
thick, horizontal bars divide the pre-reduction and analysis steps.
The symbols $\ast$ and $\oplus$ marking steps in data reduction
indicate that a systematic or statistical uncertainty is accrued at
that step.  Normal type--face indicates STScI pipeline data reduction
procedures; bold type--face indicates original procedures which we
developed for this work.  Dashed boxes mark estimates of the ZL color;
shaded boxes indicate results derived from one data set alone;
thick--lined boxes indicate final measurements of the EBL from
combined WFPC2/LCO and FOS/LCO data sets.}
\label{fig:flowchart}
\end{center}
\end{figure*}

Atmospheric emission is the brightest component of the night sky as
seen from the Earth's surface. The mean flux and line strengths of the
molecular, atomic, and continuum emission which produce ``airglow''
can vary by several percent on time-scales of minutes due to changes
in atmospheric column densities throughout the night, stimulation by
small meteorites, and  photo-chemical excitation and
de-excitation within several hours of twilight (see Chamberlain 1966
for background, Dube \etal 1979, and Paper II).  The mean airglow flux
can be predicted based on the line-of-sight path-length through the
atmosphere, but the accuracy of this method is limited to a few
percent. Airglow subtraction has dominated the errors in several
previous attempts to measure the EBL (see Dube~et~al.~1979 and
the discussion in Mattila 1990).  However, the only significant airglow
emission seen in the upper atmosphere from HST occurs at wavelengths
shorter than 2000\AA\ and only on the daytime side of the orbit (see
Lyons~et~al.~1993 and reference therein).  We have therefore entirely
avoided optical airglow by using HST as the primary instrument
in measuring the EBL.

\subsection{Zodiacal Light (ZL)}\label{foreg.zl}

Zodiacal light is sunlight scattered off of dust grains in the solar
system, and it can be as bright as 1500\tto{-9}{\escsa} in the
ecliptic plane. The ZL is faintest at viewing angles (heliocentric
longitude) 130--170 degrees away from the Sun, where scattering angles
are large, and at ecliptic latitudes greater than 30 degrees,
where the interplanetary dust (IPD) column densities are lowest. To
within 10\%, the mean flux of the ZL is predictable as a function of
scattering geometry and ecliptic latitude; however, this is not
accurate enough for our purposes.  Instead, we have measured the ZL
flux directly using detailed a priori knowledge of its intrinsic spectrum.
Observations of the ZL (Leinert \etal 1981, Murthy \etal 1990,
Matsuura \etal 1995) show that it has a nearly solar spectrum from
the UV to the near--IR (1500\AA--10\micron): scattering strength is
only weakly dependent on wavelength, so that the absorption lines and
narrow--band features of the Sun are very accurately reproduced in the
ZL. The broad--band color of the ZL is usually described as the ratio
of the zodiacal to the solar spectrum as a function of wavelength,
\begin{equation}
C(\lambda,\lambda_0) = \frac{I_{ZL}(\lambda)/ I_\odot(\lambda)}
{I_{ZL}(\lambda_0)/I_\odot(\lambda_0)},
\label{eq:intro:C}
\end{equation}
in which the reference wavelength, $\lambda_0$, is typically around
5500\AA.  Observations of the ZL to date find $C(\lambda,\lambda_0)$
changing by only 5\% per 1000\AA\ at the ecliptic orientation and
scattering angles of interest to us. 

The solar--type spectrum of the ZL is readily explained by Mie
scattering theory, which describes the scattering of light off solid
particles larger than the wavelength of the incident light. It
predicts the mild wavelength dependence for scattering by particles
with the composition and size distribution of the interplanetary dust
(IPD) cloud (see R\"oser \& Staude 1978, Berriman \etal 1994, and
Leinert \etal 1998 for a recent summary).  Because the equivalent
widths of the solar Fraunhofer lines are known with high accuracy, we
can uniquely determine the continuum level (mean flux) of the ZL at a
given wavelength using the apparent equivalent width of the Fraunhofer
lines in the ZL spectrum.  However, high signal--to--noise spectra at
$\sim1$\AA\ resolution are required to measure Fraunhofer lines in the
spectrum of the ZL.  At the time of these observations, such data
could only be obtained from the ground.  
We therefore measured the absolute flux of the
ZL using spectrophotometry between 3900--5100\AA\ obtained at Las
Campanas observatory simultaneously with the November 1995 HST
observations.  To identify the ZL flux over the full wavelength range
of the HST observations, we combine these observations with an
estimate of the ZL color obtained from the HST and LCO data together,
as described in \S\ref{zl}.  The measurement of the ZL is presented in
full in Paper II.  In \S\ref{ebl.detec}, we discuss the use of those
results in the EBL detection we present here.

\subsection{Discrete Stars and Diffuse Galactic Light (DGL)}\label{foreg.dgl}

\begin{figure}[t]
\begin{center}
\includegraphics[width=3.0in,angle=0]{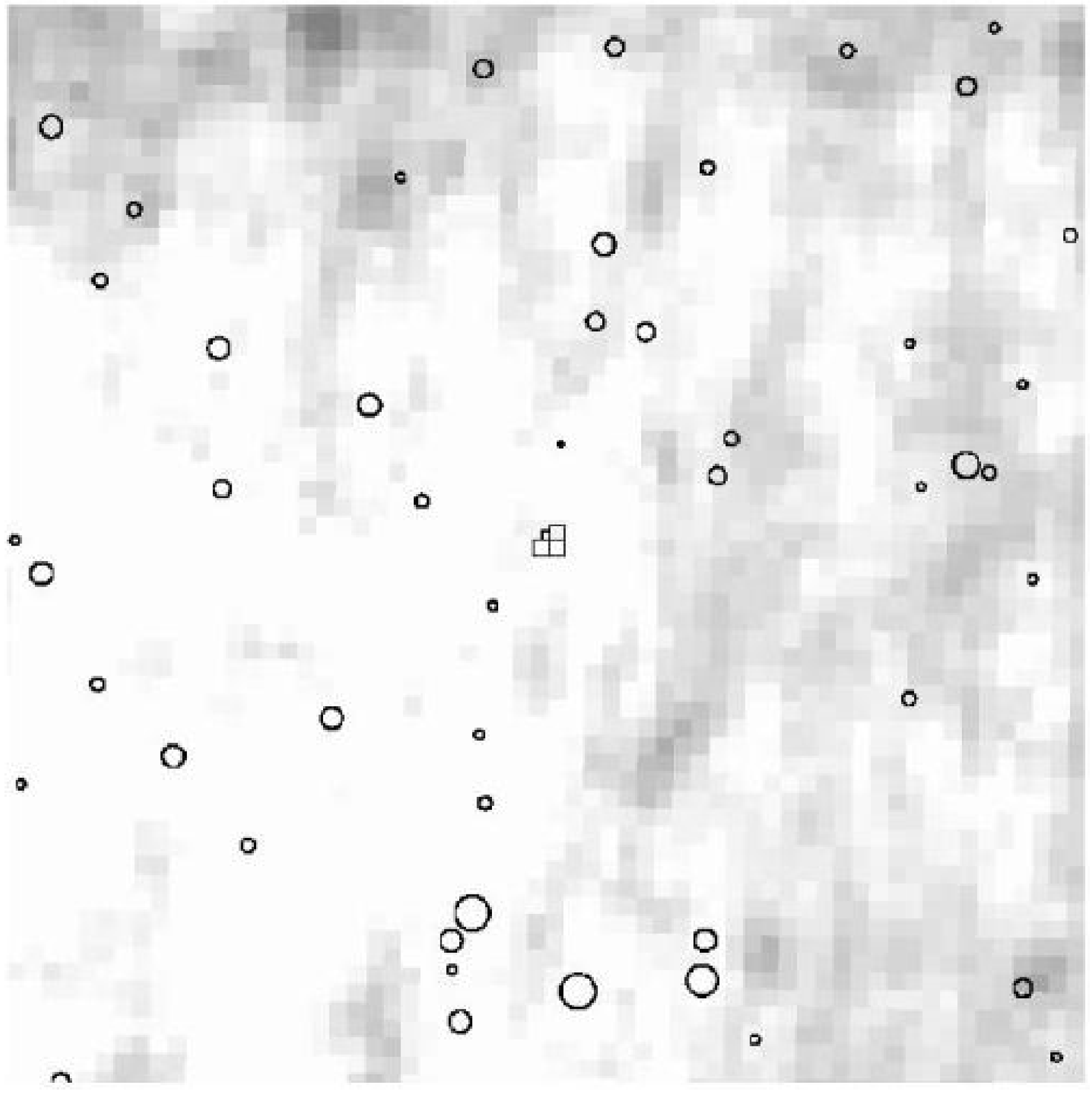}
\caption{\footnotesize
{\it IRAS} 100\,\micron\ map (Beichmann \etal 1986) covering
1.5\arcdeg$\times$1.5\arcdeg centered on the WFPC2 EBL field.  The
grey--scale is linear: white indicates 0.6\MJysr at 100\,\micron,
corresponding to N(H{\sc i})$\sim$ 0.6\tto{20} and E(\bv)$\sim 0.01$
mag; black indicate 1.6\MJysr, corresponding to N(H{\sc i})$\sim$
1.6\tto{20}, E(\bv)$\sim$ 0.03 mag (see Berriman \etal
1994). HST/WFPC2 and FOS footprints are overlaid. Stars with
$7<V<12$\,mag are marked with circles whose radii are linearly
proportional the magnitude of the stars.}
\label{fig:iras.siapers}
\end{center}
\end{figure}

The Galactic contribution to the diffuse night sky comes from discrete
stars, starlight scattered off interstellar dust,  and line emission
from the warm interstellar medium.  Light from resolved stars near the
optical axis of the telescope can also scatter into the field of view.

In some previous attempts to measure the optical EBL, even light from
discrete, resolved stars was difficult to subtract due to poor
detector resolution (e.g. Dube \etal 1979, Toller \etal 1983).  In
HST/WFPC2 images, however, Galactic stars in the field can be easily
identified and subtracted, with negligible residual contribution to
the errors in our final results.

Guhathakurta \& Tyson (1989, hereafter GT89), Murthy \etal (1990), and
others have demonstrated that thermal IR emission from Galactic dust
and the scattered optical flux are well correlated, as both are a
function of the column density of dust and the strength of the
interstellar radiation field which both heats the dust and is
scattered by it.  We therefore used the IRAS 100$\mu$m maps to select
a field with low 100$\mu$m emission to ensure that the diffuse
galactic light (DGL) contribution would be minimal. The approximate
100\micron\ flux in the surrounding $1.5^{\circ}\times1.5^{\circ}$ region
of our observations is shown in Figure \ref{fig:iras.siapers}, in
which the position of our target field is shown by the HST WFPC2 and
FOS footprints overlayed the IRAS 100\micron\ map.  We have estimated
the low--level DGL (which contributes even along the most favorable
lines of sight) by using a simple scattering model which accurately
reproduces the observed scaling relations between 100\micron\ thermal
emission and the optical scattered light (see Equation
\ref{eq:dgl.scat.model}).  The model combines an estimate of the dust
column density based on 100\micron\ emission with empirical values for
the interstellar radiation field and dust scattering parameters.
Unlike the ZL, the flux from DGL is only roughly equal to the EBL in
surface brightness, so that uncertainty in this modeling does not
prohibit a detection of the EBL.

\section{\uppercase{Field Selection and HST Scheduling}} \label{sched}

\begin{figure}[t]
\begin{center}
\includegraphics[width=3.0in,angle=0]{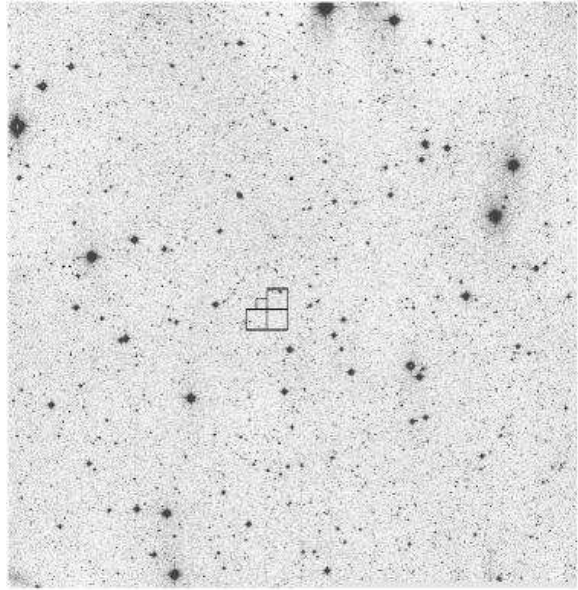}
\caption{\footnotesize
The HST/WFPC2 and FOS footprints overlaid on a mosaic
Gunn-$r$--band image ($0.6^\circ\times0.6^\circ$) taken with the
1m Swope Telescope at the Las Campanas Observatory.}
\label{fig:40in.siapers}
\end{center}
\end{figure}


\begin{deluxetable}{l l l}
\tablewidth{20pc}
\tablecaption{EBL Field Coordinates
\label{tab:coords}}
\tablehead{ \colhead{System} & \colhead{longitude} & \colhead{latitude}}
\startdata
Equatorial(J2000) & $3^{\rm h}00^{\rm m}20\fs{40}$
                & ${-20}\arcdeg{10}\arcmin{45}\farcs{3} $ \nl
Galactic &  206\fdg{6}  &        -59\fdg{8} \nl
Ecliptic &  35\fdg{5}   &        -35\fdg{5} \nl
\enddata
\end{deluxetable}

We chose the HST/WFPC2 field (see Table \ref{tab:coords} for
coordinates) at an ecliptic latitude $|\beta| > 30^\circ$ (to minimize
contributions from ZL) and near the Galactic pole in a region of low
Galactic 100\micron\ emission (to minimize diffuse Galactic light).
We also selected the field to avoid bright stars, in an effort to
minimize the scattered light from stars near the optical axis of the
telescope. The positions of stars relative to our field
can be seen in the $r$-band CCD image shown in HST Figure
\ref{fig:40in.siapers} and 
in the IRAS map in Figure \ref{fig:iras.siapers}.  No stars
brighter than 12 AB mag fall within 10 arcmin of the center of our
field, and no stars brighter than 7 AB mag fall within 2.5 degrees.
Based on measurements of the point spread function (PSF) of HST done
with WFPC1, the attenuation factor at 1 arcmin from the center of a
point source is $10^{-8}$ (STScI Technical Memos RSB--85--03,
RSB--85--02, ISR/OTA 06.1). Furthermore, the ``large--angle'' ($>3$
arcsec) scattering is roughly one order of magnitude lower for WFPC2
than for WFPC1.  Based on these results, the total contribution from
stars closer than 5\,arcmin to the field is at most $10^{-9}$ of their
total flux, which amounts to one 35\,mag star on--axis. The EBL
signal, by comparison, is approximately 27.5 mag arcsec$^{-2}$. The
flux from off--axis sources is therefore an insignificant contribution
to the background and a negligible source of error.

The exact scheduling of exposures was also critical to this program,
as stray solar and terrestrial light in HST observations are a strong
function of the orbital position of the satellite. Sunlight scattered
off the limb of the Earth can increase the background level by a
factor of 10 when the bright limb is near the optical axis of the
telescope.  All of our science observations were therefore scheduled
to execute exclusively in the shadow of the Earth.  Because our field
was at a viewing angle angle greater than $135^\circ$ from the Sun
during the months of our observations, the satellite was pointed in a
direction greater than $45^\circ$ away from the satellite's direction
of motion when our field was observed from within the Earth's
shadow. This guaranteed that no upper--atmosphere glow from the
satellite's flight path would affect our observations.  Scattered
moonlight was also explicitly avoided by exposing only with the Moon
greater than $65^{\circ}$ from the optical axis of the telescope,
which is the angular separation at which models predict that the
attenuation function for off--axis scattered light becomes flat (STScI
Technical Memo RSB--85--03).  Based on those models, confirmed by
on--orbit measurements (Burrows 1991, Hasan \& Burrows 1993), the
light from the moon at $65^\circ$ from the optical telescope assembly
(OTA) is less than $10^{-5}$ photons s$^{-1}$ arcsec$^2$ per 100\AA,
which is $1000$ times smaller than the lower limit expected for the
EBL, and is therefore also a negligible source of background during
our observations. 

Our 18 orbits were divided into three visits, staggered by one month
each between October and December, 1995. The ZL flux changes with the
Earth's orbital position about the Sun and the line of sight through
the interplanetary dust during the three visits.  We therefore expected
to see changes in the total sky flux between visits, at a level
predicted by the models of the interplanetary dust.  During all three
months, our target field has a ZL surface brightness which is within
20\% of the absolute minimum intensity of the ZL at any orientation.
By separating the orbits for this program into three visits, we were
also able to observe the field at two different roll angles -- V3
position angle $222^\circ$ for the November and December visits, V3
angle $132^\circ$ for the October visit --- thereby changing the
position of the off--axis stars with respect to the OTA. This allows
us to further rule out significant off-axis scattered light and other
possible photometric anomalies.

\section{\uppercase{HST/WFPC2 Data}}\label{wfpc2}

\subsection{Observations} \label{wfpc2.obser}

The 18 orbits scheduled for this program executed on 28 October, 29
November, and 16 December, 1995.  During each orbit, a single long
integration (1800 sec) was taken while the satellite was in the
Earth's shadow and one additional short integration (300--400 sec) was
taken to fill the time between target acquisition and entry into the
Earth's shadow.  Two orbits per visit were spent observing with each
of three WFPC2 filters (F300W, F555W, and F814W).  The short exposures
were combined and used to improve cosmic ray rejection.  In all, five
1700 sec darks and ten bias images were taken during the bright
portion of the six orbits comprising each visit.  The gain setting for
these data was 7\,$e^-$/DN (bay 4) and the read noise was roughly
5.2\,$e^-$ for each of the four CCDs.  The field of view of each WFPC2
image is roughly 4.4\,arcmin$^2$ ($725\times725$ well--exposed pixels
in each of the 3 WF CCDs, with 0.0996 arcsec$^2$ per pixel).

\subsection{Basic Data Reduction and Calibration}\label{wfpc2.basic}

The reduction of these data involved a mixture of the standard
pipeline and our own methods.  The pipeline procedures used included
the corrections for analog--to--digital conversion errors, overscan
subtraction, bias image subtraction, flat--fielding, and point-source
calibration. The accuracy of these steps is largely provided in STScI
documentation. Whenever possible, we have independently verified the
errors we quote. All steps in the reduction are indicated in the flow
chart in Figure \ref{fig:flowchart} with associated errors explicitly
included in our error budget.  The reduction is discussed briefly
below; greater detail can be found in Bernstein (1998).  The following
section should be read by those who are interested in the reduction
procedures and detailed error discussions.  As the goal of these
observations is an {\it absolute} measurement of the total flux of the night
sky, zero point calibration is as critical to the accuracy of the
results as flux calibration.  Table \ref{tab:wfpc2.errors} summarizes
the errors discussed below and indicates the dominant sources of
error.

\begin{deluxetable}{ l c c c c c c c c c c c }
\tablewidth{40pc}
\scriptsize
\tablecaption{Sample of $V_{\rm 555}$ Source Catalog
\label{tab:source.cat.sample}}
\tablehead{
\colhead{ID\#} & 
\colhead{x} & \colhead{y} & 
\colhead{RA(h:m:s)} &  \colhead{Dec(h:m:s)} & 
\colhead{$m_{\rm tot}$} & \colhead{$\sigma(m_{\rm tot})$} &
\colhead{$m_{\rm iso}$} & \colhead{$\sigma(m_{\rm iso})$} &
\colhead{A$_{\rm iso}$} & \colhead{$a$} & \colhead{elong.}  
}
\startdata
EBL555-2-1   &   129 &  55   &     3:00:21.66 & -20:10:37.4 & 25.51 & 0.04 &
25.45 &  0.05  &  11 & 0.53 & 1.08 \nl
EBL555-2-2   &   483 &  55   &     3:00:24.14 & -20:10:35.4 & 25.94 & 0.10 &
26.06 &  0.09  &  20 & 1.72 & 2.04 \nl
EBL555-2-3   &   623 &  58   &     3:00:25.13 & -20:10:35.1 & 26.32 & 0.13 &
26.37 &  0.10  &  16 & 1.50 & 2.03 \nl
EBL555-2-4   &    74 &  60   &     3:00:21.28 & -20:10:38.2 & 27.79 & 0.26 &
28.27 &  0.31  &   3 & 0.68 & 1.57 \nl
EBL555-2-5   &   600 &  61   &     3:00:24.97 & -20:10:35.4 & 26.62 & 0.13 &
26.71 &  0.12  &  11 & 1.04 & 1.43 \nl
EBL555-2-6   &   279 &  63   &     3:00:22.72 & -20:10:37.3 & 27.12 & 0.17 &
27.09 &  0.15  &   9 & 0.83 & 1.15 \nl
EBL555-2-7   &   298 &  57   &     3:00:22.85 & -20:10:36.6 & 25.46 & 0.20 &
26.16 &  0.11  &  29 & 4.81 & 7.60 \nl
EBL555-2-8   &   306 &  65   &     3:00:22.91 & -20:10:37.3 & 27.47 & 0.24 &
27.73 &  0.23  &   6 & 1.08 & 3.65 \nl
EBL555-2-9   &   155 &  68   &     3:00:21.85 & -20:10:38.5 & 26.60 & 0.12 &
26.70 &  0.12  &  10 & 0.95 & 1.29 \nl
EBL555-2-10  &   302 &  69   &     3:00:22.88 & -20:10:37.7 & 27.22 & 0.34 &
27.70 &  0.23  &   7 & 1.15 & 2.62 \nl
EBL555-2-11  &   650 &  68   &     3:00:25.32 & -20:10:35.9 & 25.09 & 0.05 &
25.08 &  0.05  &  33 & 1.30 & 1.09 \nl
EBL555-2-12  &   756 &  70   &     3:00:26.06 & -20:10:35.7 & 25.60 & 0.12 &
25.93 &  0.10  &  34 & 2.33 & 1.53 \nl
EBL555-2-13  &   380 &  75   &     3:00:23.43 & -20:10:37.9 & 26.75 & 0.12 &
26.70 &  0.11  &   9 & 0.79 & 1.25 \nl
EBL555-2-14  &   393 &  74   &     3:00:23.53 & -20:10:37.8 & 27.42 & 0.21 &
27.40 &  0.17  &   7 & 0.82 & 1.39 \nl
EBL555-2-15  &   729 &  75   &     3:00:25.88 & -20:10:36.3 & 27.24 & 0.20 &
27.51 &  0.20  &   7 & 0.92 & 1.67 \nl
EBL555-2-16  &   680 &  77   &     3:00:25.54 & -20:10:36.7 & 26.87 & 0.17 &
27.29 &  0.17  &   9 & 1.15 & 1.85 \nl
EBL555-2-17  &   759 &  82   &     3:00:26.09 & -20:10:36.8 & 27.79 & 0.49 &
28.22 &  0.33  &   4 & 0.95 & 1.76 \nl
EBL555-2-18  &   534 &  85   &     3:00:24.52 & -20:10:38.1 & 25.83 & 0.12 &
26.22 &  0.11  &  28 & 1.79 & 1.31 \nl
EBL555-2-19  &    86 &  89   &     3:00:21.37 & -20:10:40.9 & 27.05 & 0.19 &
27.22 &  0.16  &   8 & 0.93 & 1.41 \nl
EBL555-2-20  &   178 &  95   &     3:00:22.02 & -20:10:41.0 & 26.72 & 0.17 &
26.97 &  0.15  &  10 & 1.15 & 1.35 \nl
EBL555-2-21  &    96 &  98   &     3:00:21.45 & -20:10:41.8 & 27.56 & 0.23 &
28.02 &  0.28  &   4 & 0.72 & 1.45 \nl
EBL555-2-22  &   670 & 100   &     3:00:25.48 & -20:10:39.0 & 27.19 & 0.23 &
27.56 &  0.22  &  10 & 0.94 & 1.21 \nl
EBL555-2-23  &   321 & 100   &     3:00:23.03 & -20:10:40.7 & 26.30 & 0.12 &
26.76 &  0.14  &  13 & 1.16 & 1.51 \nl
EBL555-2-24  &   276 & 105   &     3:00:22.71 & -20:10:41.4 & 27.72 & 0.28 &
27.81 &  0.21  &   6 & 0.75 & 1.71 \nl
EBL555-2-25  &   599 & 106   &     3:00:24.98 & -20:10:39.8 & 27.07 & 0.21 &
27.31 &  0.18  &  10 & 1.02 & 1.40 \nl
EBL555-2-26  &   213 & 107   &     3:00:22.27 & -20:10:42.0 & 25.91 & 0.13 &
26.32 &  0.11  &  24 & 2.19 & 1.93 \nl
EBL555-2-27  &   666 & 112   &     3:00:25.45 & -20:10:40.1 & 26.98 & 0.21 &
27.31 &  0.19  &   9 & 1.32 & 3.09 \nl
EBL555-2-28  &   311 & 113   &     3:00:22.96 & -20:10:42.1 & 27.48 & 0.23 &
27.80 &  0.21  &   6 & 0.94 & 2.00 \nl
EBL555-2-29  &   645 & 106   &     3:00:25.30 & -20:10:39.6 & 22.01 & 0.01 &
22.01 &  0.01  & 375 & 4.82 & 1.46 \nl
\enddata
\noindent\tablenotetext{Note:}{Coordinates are J2000. Total magnitudes
($m_{\rm tot}$), isophotal magnitudes ($m_{\rm iso}$), and the
associated {\it rms} errors ($\sigma$) are in AB magnitudes. The
isophotal area ($A_{\rm iso}$) and flux--weighted, second order moment
along the major axis ($a$) are in units of WF pixels, which are 0.0996
arcsec on a side. The isophotal, weighted elongation ($e$) is the
ratio of the second order moments along the major and minor axes, as
calculated by SExtractor.}
\end{deluxetable}

\subsubsection{Pipeline}\label{wfpc2.pipel}

The WFPC2 analog--to--digital conversion (ADC) has a bias towards DN
values which correspond to the setting of all low--order bits in the
digitization, with larger signal levels being more strongly affected.
The correction for this bias involves replacing the output value with
a value corresponding to the scaled average input signal for that
output.  DN values in the range of our data have an average additive
correction of 0.86\,DN, with the correction smoothly varying for
levels in this range.  The same correction is used for each of the
four CCDs in WFPC2.  The data used to determine this correction were
taken pre-- and post--dynamic testing, with only 0.02\,DN
variation. The correction is thought to be stable under normal usage
and insensitive to temperature fluctuations (STScI Technical Memo
RSB--85--01).  At the signal levels of our data, the error after the
ADC correction applied is 0.02\,DN.

The bias level was removed with a two--step process of overscan
subtraction and bias--image subtraction. We employed the usual method
established for WFPC2 data of subtracting overscan from odd and even
columns using the overscan from each individual exposure.  For
bias--image subtraction, we used the ``superbias'' frames produced for
the reduction of the Hubble Deep Field (HDF, see Williams \etal
1996).  This bias frame was produced from the average of 200 frames,
and has lower read noise than we could achieve using the 30 bias
frames taken during our orbits. Instead, bias frames from our
orbits were test--reduced (overscan and bias subtraction using the HDF
superbias) to verify that the superbias produced corrected levels
consistent with zero, with a scatter of $0.002$\,DN.  As this error is
inseparable from our estimate of the error in the dark subtraction, it
is included with the dark subtraction in our error budget.

We used the pipeline flat--fielding images provided by the WFPC2 GTO team
in May 1996.  The pixel--to--pixel flux error in those images is reported
in the WFPC2 Data Handbook (V3.0) to be roughly 0.3\% (\rms) for the WF
chips, and 0.5\% for the PC1. The errors over spatial scales greater
than 10\,arcsec are less than 0.5\% for all four chips.  As we
excluded data within 75 pixels of the edge of each chip from our
analysis, issues of geometrical distortion and vignetting are avoided.
The photometric calibration of WFPC2 is tied to the mean level of the
flat--fielding images, so that no systematic error is introduced due
to flat--fielding.

\subsubsection{Subtraction of Dark Backgrounds}\label{wfpc2.dark}

\begin{figure}[t]
\begin{center}
\includegraphics[width=3.0in,angle=0]{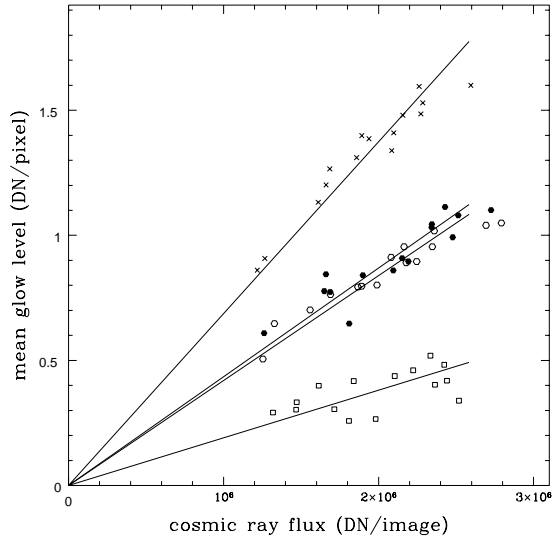}
\caption{\footnotesize 
Correlation between the cumulative flux in cosmic ray events
and the mean dark level (thermal dark plus MgF$_2$ glow) for each
WFPC2 chip in 15 exposures: $\times$'s for PC1 fluxes, open boxes for WF2,
circles for WF3, and filled circles for WF4. Statistical errors in both
quantities plotted are negligible; scatter in the correlation probably
reflects statistical variations in the energy level of the cosmic rays
hitting the CCD and the MgF$_2$ plates.}
\label{fig:wfpc2.glowvscr}
\end{center}
\end{figure}

We have developed a new method for dark subtraction because the
pipeline subtraction ignores the fact that the signal which
accumulates on the WF chips while the shutter is closed varies between
0.5--1.1~DN per 1800~sec exposures for the WF3 and WF4 chips (more for
WF1, less for WF2). This variability is not caused by thermal dark
current, but rather by variations in the dark ``glow'' contributed by
the MgF$_2$ field--flattening lenses, positioned immediately above
each of the WFPC2 chips.  When struck by even low--energy (200--2000
eV) ions and electrons (which are abundant in the upper atmosphere in
the form of N$_2^+$ and cosmic rays), MgF$_2$ will produce broad--band
fluorescence on a time--scale $<<1$~sec (Qi \etal 1991).  We have
found that the cumulative flux in cosmic ray hits recorded in a dark
(shutter--closed) exposure correlates strongly with the mean dark
level in that frame.  This correlation is shown Figure
\ref{fig:wfpc2.glowvscr} in which we plot the cosmic ray flux and dark
glow for each WFPC2 CCD in each of 15 dark exposures.  The scatter around
the linear fit for each chip is $\sim0.06$\,DN.  This plot also shows
clearly that the dark background in a 1700\,sec exposure can vary by
as much as $\pm0.35$~DN.  As the total background sky signal through
the F300W is only 0.3~DN/1700\,sec, accurate subtraction of the dark
glow is essential for this measurement.

In collaboration with H. Ferguson (STScI), we developed a method which
uses the correlation between the MgF$_2$ glow and the flux in cosmic
rays to isolate the thermal dark component from the dark glow.  We
obtained a final dark solution using 80 individual dark images taken
between October and December 1995, and have tested the accuracy of this
method by test-reducing the 15 dark exposures taken between the science
exposures of our program.  These 15 darks were not used in determining
the dark solution itself.  After ADC correction, overscan subtraction,
bias subtraction, and dark subtraction by the prescription described
above, the average mean level of the 15 darks is consistent with
zero, with a scatter of $<0.05$\,DN per 1700\,sec exposure. The
statistical error in the pipeline dark-subtraction, by comparison, is
roughly 0.15--0.25 DN for the WF3.

\subsubsection{Charge Transfer Efficiency (CTE)}\label{wfpc2.cte}

Accurate photometry depends on stable charge transfer efficiency
(CTE), which is the efficiency with which electrons are transferred
between pixels during read out. Variable CTE as a function of total
charge level, time, or position over the chip results in non--linear
sensitivity.  Unfortunately, the WFPC2 chips are known to have a ``a
small parallel CTE problem'' (Holtzman \etal 1995b, henceforth H95b;
Whitmore, Heyer, \& Casertano 1999): the percentage of electrons which
are read--out of a pixel depends on the row number of the pixel in
question, the number of electrons in that pixel when read--out begins,
and the mean charge level in the pixels through which the charge
travels during read--out.

Based on in--flight point source photometry and on laboratory tests
using a CCD from the same silicon wafer as WFPC2 CCDs, the CTE
variability is known to be caused by electron traps which were in the
silicon itself before the pixel mask was etched into the wafer
(J. Trauger, private communication).  Although electrons can be
trapped in the silicon at the pixel where they are detected, a greater
surface area of silicon is encountered during read--out, providing
greater opportunity for trapping if the traps are not already filled
before readout begins.  When a bright point source is imaged against a
faint background, many new traps are encountered during
read--out. For example, 4\% of a 10,000 $e^-$ point source are lost
when the charge is transferred over 800 rows if the background level
is $\sim$10 $e^-$/pixel.  If the traps are already filled by a high
background level over the whole chip, fewer electrons will be lost
from sources during read--out.  Surface photometry is less affected by
this CTE problem because all pixels are filled to the same charge
level and, hence, new traps are not encountered during read-out.

The signal level read out for our images is roughly 80e-/pixel for the
F555W and F814W images and 2e-/pixel for the F300W images.  We have
identified the CTE losses in these data by two methods.  First, data
from WFPC2 and laboratory tests conducted by J. Trauger of point
sources show that $\sim$0.35\,$e^-$/pixel are lost from a point source
which is imaged against a background level of roughly 80\,$e^-$/pixel,
while $\sim$1.1\,$e^-$/pixel are lost when the background level is
zero (J. Trauger, private communication).
The difference between the number of electrons lost at the two
different background levels indicates that the number of
single--electron traps available in a uniform background of
80\,$e^-$/pixel is $0.75$\,$e^-$/pixel ($\sim$1.1-0.35).  Second, to
confirm that this trapping reflects the CTE losses for uniform
sources, we have conducted another set of tests with the help of
J.Trauger to measure the non-linear response of the spare WFPC2 CCD to
uniform, low--level backgrounds.  In this test, we simply expose the
CCD to a light source which is stable to better than 0.1\% for varying
lengths of time and look for non-linearities in the detection rate.
The results of these uniform-source test are in excellent agreement
with results from point--source tests.  We therefore conclude that
0.75 ($\pm 0.25$)\,$e^-$/pixel are lost when a mean level of $\sim
80$\,$e^-$/pixel are read-out, and 0.1\ ($\pm 0.05$) $e^-$/pixel are
lost from a mean level of $\sim 2$\,$e^-$/pixel.  The maximum
uncertainty in these corrections is a negligible contribution to our
final results (see Table \ref{tab:wfpc2.errors}).  For more detail on
the result of the laboratory tests, see Bernstein (1998).

\begin{figure} [t]
\begin{center}
\includegraphics[width=3.0in,angle=0]{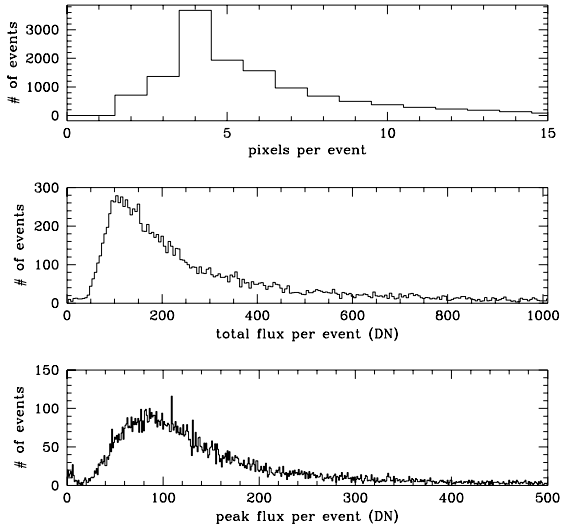}
\caption{\footnotesize 
Histogram of cosmic ray events as a function of
the number of affected pixels (top panel), total flux (middle panel),
and peak flux (bottom panel).  The number of events falls off in each
plot before an event would become difficult to detect in these data.}
\label{fig:wfpc2.crhists}
\end{center}
\end{figure}

\subsubsection{Cosmic Ray Detection}\label{wfpc2.crays}

If low--energy cosmic ray events were left unidentified in WFPC2 data,
we would unintentionally include the signal from those events in the
inferred background level.  Figure \ref{fig:wfpc2.crhists} shows the
total number of cosmic ray events as a function of total flux, peak
flux, and affected pixels per event.  In each histogram, the number of
cosmic ray events falls off sharply well before cosmic rays would
become difficult to detect: both peak and total fluxes for cosmic ray
events are significantly higher than the mean level of the background
($\lta 10$ DN), and the total number of pixels affected by a single
cosmic ray event rarely drops below three.  We therefore conclude that
the cosmic ray removal process detects virtually all cosmic rays which
hit the chips.

\subsubsection{Flux Calibration}\label{wfpc2.fluxc}

The final accuracy of a surface brightness measurement depends on
four independent facets of the calibration: point source calibration;
aperture correction, compensating for flux not recovered in the point
source calibration; the calibration of the fiducial standard star system;
and the solid angle subtended by each pixel. The solid angle of
the pixels is, of course, very well known and is a negligible source
of error for the WFPC2 data.  However, the other three calibration
steps require quantities of data and observing time only available to
the WFPC2 Instrument Team. Details of the WFPC2 calibration are
thoroughly discussed in Holtzman \etal (1995a, henceforth H95a), H95b
and subsequent STScI Instrument Science Reports. Below, we summarize
their results. 

Sensitivity to the same fiducial standard is stable to $\sigma <1$\%
for the F555W filter, and $\sigma <1.5$\% for the F814W and F300W between
February 1995 and March 1997 (STScI TIR/WFPC2 97--01).  Observations
of the same standard may be variably affected by CTE due to variations
in the peak flux of the observation which will arise from variations
in the position of the PSF within a single pixel. The real response of
WFPC2 is, therefore, likely to be more stable than this result.  The
synthetic photometry package (SYNPHOT), including system throughput
curves and synthetic zero--points, is based on the HST secondary
standard system and is estimated by H95b to be accurate for point
sources to roughly 1\% for a source with typical stellar
colors. Point--source calibration should therefore be good to roughly
1\%.

Aperture corrections based on the encircled energy curves are given in
H95a (see Figures 5a \& 5b, and Tables 2a \& 2b in H95a).  For the
optical filters, the aperture correction is nearly 10\% in moving from
a 0.5\,arcsec aperture to an infinite aperture, with uncertainties of
1\% for the redder bandpasses, and 1-2\% for the F300W.

Finally, the secondary standard star system which is used for all the
HST instruments is ``conservatively estimated'' to be within 1\% of
the Hayes 1985 optical calibration of Vega (Bohlin 1995; Colina \&
Bohlin 1994; Hayes 1985), with internal agreement of the same order.
The agreement between the HST secondary system and the Hamuy \etal
(1992) secondary system to which our ground-based observations are
tied is estimated to be 1--2\%, as discussed in Paper II.

\subsection{Combined Images}\label{wfpc2.comb.im}

\begin{figure} [t]
\begin{center}
\includegraphics[width=3.0in,angle=0]{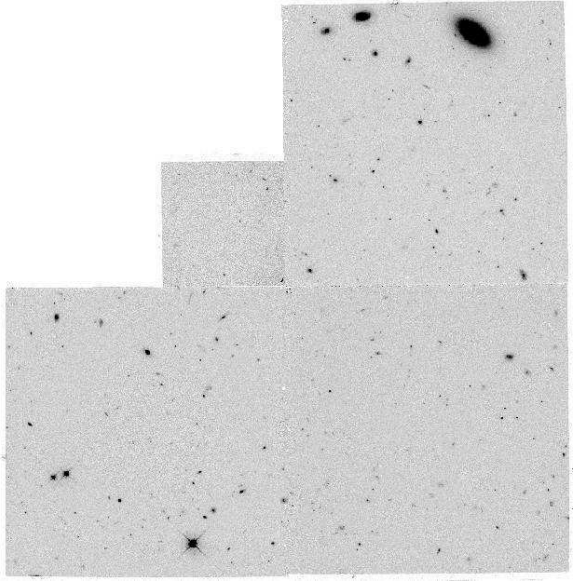}
\caption{\footnotesize
 The combined F555W images from November and December
visits. The total exposure time is $4\times1800$ sec.}
\label{fig:mosaic}
\end{center}
\end{figure}

We produced combined images in which individual resolved sources could
be identified and photometered for two purposes: (1) to allow bright
galaxies and stars to be identified and excluded from the EBL
measurements, (2) to allow comparison of the galaxy sample and EBL
results to other data sets.  The measurements of the mean sky flux
discussed in \S\ref{wfpc2.backg} and \S\ref{ebl.minim} are obtained
from individual WFPC2 exposures.  The deepest combined image for each
filter is obtained from the four exposures taken during the November
and December visits, during which the telescope roll angles were the
same. The small differences in the alignment of the WF CCDs prevent
data taken during the October visit from being usefully included in
combined images for the purposes of accurate photometry.  The combined
images and resolved source catalogs are available on the NASA/IPAC
Extragalactic Database (NED).\footnote{Web address:
http://nedwww.ipac.caltech.edu/level5/} A sample of the catalogs is
shown in Table \ref{tab:source.cat.sample} and the combined F555W
image is shown in Figure \ref{fig:mosaic}  The F300W, F555W, and
F814W catalogs contain 140, 687, and 644 objects, respectively.  The
photometry of these sources is discussed in the next subsection.

\begin{deluxetable}{llr}
\tablewidth{30pc}
\tablecaption{WFPC2 Background Flux: Errors Per Image
\label{tab:wfpc2.errors}}
\tablehead{
\colhead{}              & \colhead{F555W, F814W}        & \colhead{F300W}}
\startdata
\multicolumn{1}{c}{\underline{Statistical  (per image)}}
        &  & \nl
Off--axis scattered light (\S\ref{sched})
                                & $ <0.001$\,DN         & $ <0.001$\,DN \nl
Dark and bias subtraction (\S\S\ref{wfpc2.dark}, \ref{wfpc2.pipel})
                                &  $\pm0.05$\,DN        & $\pm0.05$\,DN \nl 
Flat--fielding (\S\ref{wfpc2.pipel})
                                & $\cdots$              & $\cdots$      \nl
CTE (\S\ref{wfpc2.cte})
                                & $\pm 0.03$\,DN        & $\pm 0.007$\,DN \nl
\cline{2-3}
Cumulative Statistical Error\tablenotemark{a}
                                & 0.06\,DN              & 0.05\,DN  \nl
\multicolumn{1}{c}{\underline{Systematic }}  &  & \nl
A--to--D conversion\tablenotemark{b} (\S\ref{wfpc2.pipel})
                &                $\pm 0.02$\,DN (0.2\%) &$\pm0.02$\,DN (6\%)\nl
Point source flux cal.(\S\ref{wfpc2.fluxc})
                                & 1\%                   & 1.5\% \nl
Aperture correction (\S\ref{wfpc2.fluxc})
                                & 1\%                   & 1\%   \nl
Solid angle (pixel scale)(\S\ref{wfpc2.fluxc})
                                & 0.1\%                 & 0.1\% \nl
\cline{2-3}
Combined Systematic Error\tablenotemark{c} 
                                & 1.2\%                 & 5.6\%\nl
\enddata
\tablenotetext{a}{Individual random errors have been added in
quadrature to obtain a cumulative, one-sigma random error in DN.  To
obtain fractional errors, compare to the mean flux per exposure
($\sim$0.3 DN per pixel for the F300W, $\sim$12.5 DN per pixel for
F555W and F814W).}
\tablenotetext{b}{The A--to--D conversion is an additive correction,
        the associated error is therefore the ratio of the A--to--D
        error and the mean level in the frame.  All other systematic
        uncertainties listed are multiplicative.}
\tablenotetext{c}{Systematic errors have been combined assuming a flat 
probability distribution for each contributing source of error.
The resulting systematic error is roughly Gaussian distributed,
and the quoted value is the 68\% confidence interval.  For a detailed
discussion see \S\ref{ebl.detec}.}
\end{deluxetable}

\subsection{Resolved Objects} \label{wfpc2.objec}

\begin{figure*}[t]
\begin{center}
\includegraphics[width=2.0in,angle=0]{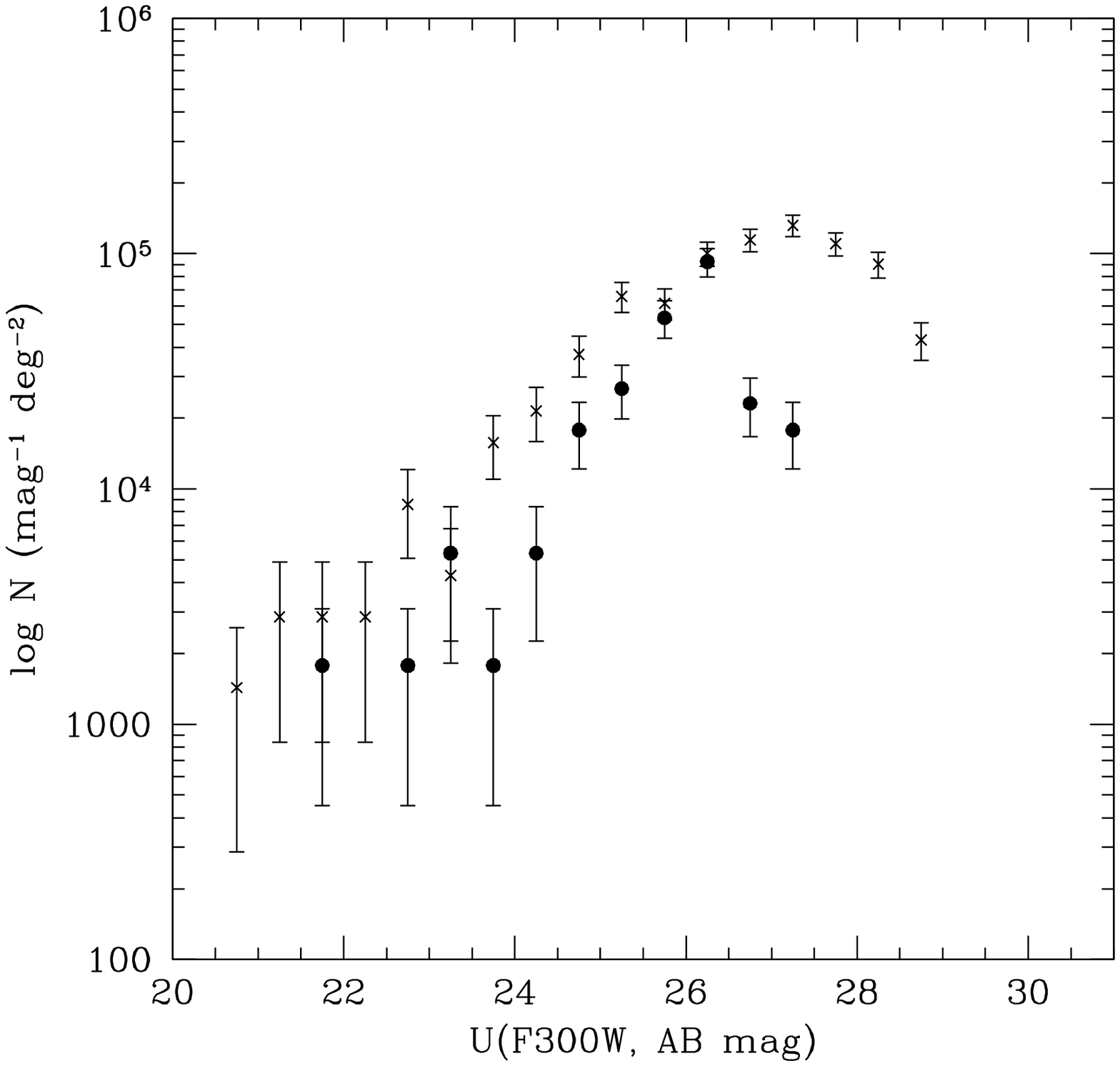}
\includegraphics[width=2in,angle=0]{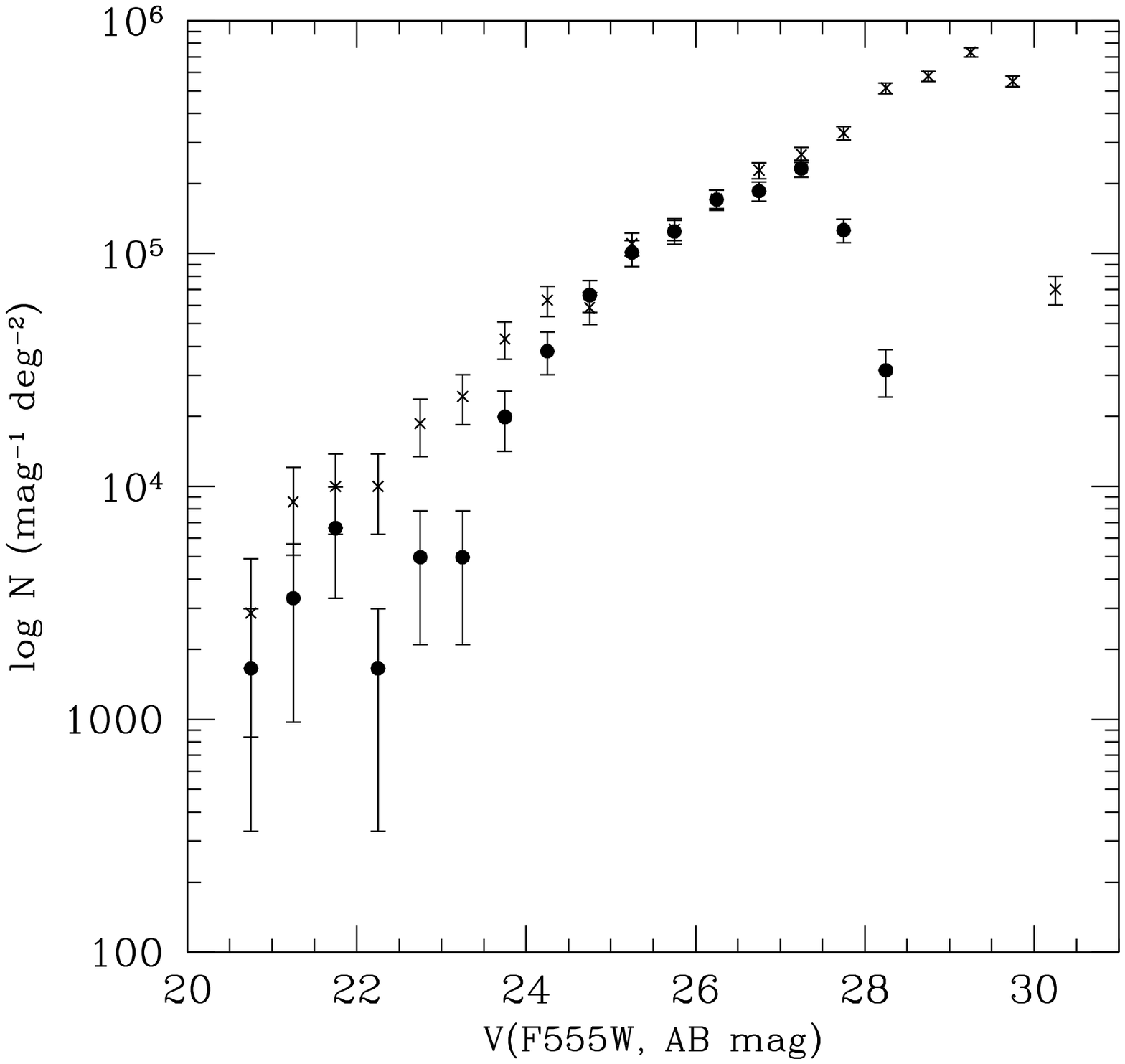}
\includegraphics[width=2.0in,angle=0]{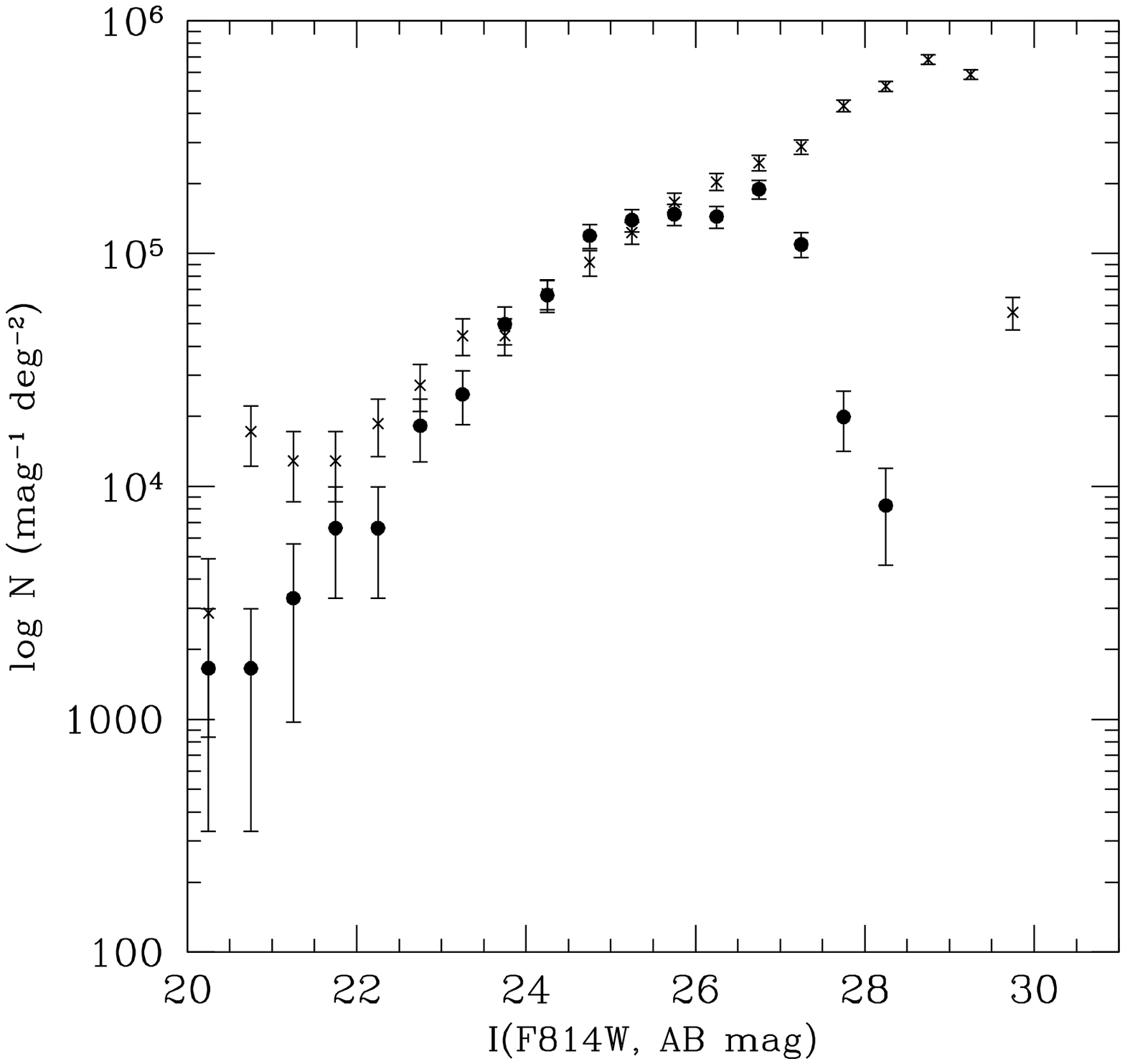}
\caption{\footnotesize
Differential galaxy counts with $\sqrt{N}$
error bars from WFPC2 EBL images (filled circles), and the HDF images at
F300W, F606W, and F814W ($\times$'s). HDF counts are taken directly
from the published catalogs (Williams \etal 1996).  The difference in
completeness limits between the two data sets reflects the difference
in total exposure times, which are roughly $\times 21$, $\times 16$,
and $\times 17$ at $U$, $V$, and $I$, respectively. }
\label{fig:wfpc2.glxycnts}
\end{center}
\end{figure*}

We used the image analysis package SExtractor (Bertin \& Arnouts 1996)
for object identification and photometry. The requirement for
detecting an object in our data is a core surface brightness,
$\mu_{\rm core}$, of at least $2\sigma_{\rm sky}$ within the core 4
pixels. The isophotal detection threshold, $\mu_{\rm iso}$ was defined
as $1.5\sigma_{\rm sky}$ over 6 pixels, after smoothing with a
$3\times3$ boxcar filter.  This corresponds to isophotal detection
thresholds of 24.7, 25.8, and 25.3 ST mag arcsec$^{-2}$, at F300W,
F555W, and F814W, respectively.  The shape of the smoothing kernel and
the detection threshold have little affect on detection of objects
with $V_{555}<27.5$\,AB\,mag, and photometry is not strongly affected
for objects with $V_{555}<27$\,AB\,mag.  Total magnitudes are defined
as the flux within an area at least three times larger than the
ellipse defined by the first moment radius (see Kron 1980) and
isophotal elongation. For well--detected sources, the detection and
photometry parameters we employ are similar to those used by the HDF
team in producing the HDF catalogs (see the photometry discussion in
Williams \etal 1996).  Isophotal detection limits are roughly 1.5 mag
brighter than the HDF, in keeping with the difference in exposure
time, which is a factor of $15-21$ in the various
bandpasses. Differential number counts for our field are plotted in
Figure \ref{fig:wfpc2.glxycnts}. Our detection limits are illustrated
by the turn--over in the counts in each band.  The $\sqrt{N}$--error
bars show that $V \sim 23$\,AB\,mag is reasonable bright--magnitude
cut-off for observations in a field of this size.  The galaxy counts
from the HDF are also plotted in Figure \ref{fig:wfpc2.glxycnts} to
demonstrate that the field we have observed contains a typical
field--galaxy population in both color and number density.  The HDF is
a convenient benchmark for this comparison simply because the data are
publicly available, widely studied, and of a similarly ``blank''
field.  The HDF counts show twice as many galaxies at magnitudes
brighter than 23\,AB\,mag in each of the three band, while the
difference in number density fainter than 23\,AB\,mag is modest. Such
differences at bright magnitudes are well within the typical
field--to--field fluctuations for galaxy counts and illustrate the
need for a bright magnitude cut--off.

Four stars are detected in the WF chips, three in WF2 with
$V_{555}=19.0$, 20.5 and 20.8\,AB\,mag, respectively, and one in
WF4 with $V_{555}=$22.0\,AB\,mag.  Star--galaxy separation poses
no difficulty; stars brighter than $V_{555}\sim23.0$\,AB\,mag are
masked out, regardless, and the flux from stars beyond that limit is
$<10$\% of that from detected galaxies at the same apparent magnitudes
(Infante 1997).

\subsection{On--Axis Scattered Light}\label{wfpc2.scatt}

\begin{figure}[t]
\begin{center}
\includegraphics[width=3.0in,angle=0]{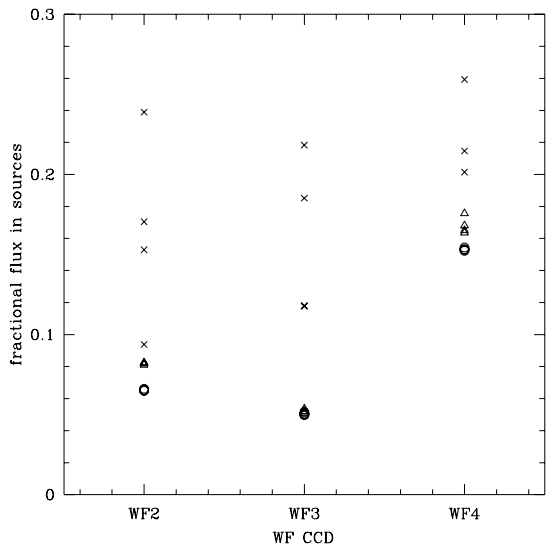}
\caption{\footnotesize The flux from bright ($V_{555}<23$ AB mag)
sources relative to the total flux in each WFPC2, EBL image is plotted
for four exposures through three filters (twelve images total).  The
total flux is just the sum of all counts in the calibrated images. The
flux from bright sources is a background--subtracted flux within four
times the isophotal radius ($4r_{\rm iso}$) for galaxies (see
\S\ref{ensemb.phot}) and within a 5 arcsec radius for stars.  The
F555W, F814W, and F300W flux ratios are marked by open circles
(overlapping), triangles, and crosses, respectively.}
\label{fig:wfpc2.flux.in.sources}
\end{center}
\end{figure}

Scattered light from the ``bright'' ($19<V<23$ AB mag) objects imaged
in the field is one possible source of error in a measurement of the
background from fainter objects.  To quantify this effect, we begin by
considering how much light these sources produce relative to the total
flux in an image. In Figure \ref{fig:wfpc2.flux.in.sources}, we plot
the ratio of the flux from bright sources to the total flux detected
in each of the three WF CCDs during each of four exposures. The total
flux is just the sum of all counts in the calibrated images. The flux
from bright sources is a background--subtracted flux within four times
the isophotal radius ($4r_{\rm iso}$) for galaxies (see
\S\ref{ensemb.phot}) and within a 5 arcsec radius for stars.  For each
bandpass, the ratio of flux in bright sources to the total flux varies
from chip to chip.  In the WF2 or WF3 chips, bright objects contribute
only $\sim 5$\% of the total integrated flux.  As several of the
brightest galaxies in the field are imaged on the WF4 chip, almost
$\sim 15$\% of the flux in F555W and F814W in that field comes from
objects with $V<23$ ST mag.

In order exclude these $V<23$ AB mag objects from the measured
background, we masked regions around each resolved galaxy which extend
to four times the isophotal detection radius of the galaxy, and masked
regions with radii of 5 arcsec for stars. Based on the encircled
energy curves in H95a and growth curves in our own images (see
\S\ref{ebl.minim} and Appendix B), we estimate that less than 2\% of
the light from these objects is imaged beyond the extent of the masked
regions.  As an upper limit on their scattered light, 2\% of the flux
from these objects constitutes only 0.1\tto{-9}\,\escsa, less than 3\%
of the total flux at 5500\AA\ in our estimate of the EBL. As discussed
previously in \S\ref{sched}, the scattered light from off--axis
sources is also negligible.

At the wavelength of the F300W filter, the flux from bright objects on
the WF4 chip is 15--20\% of the total flux, and variation in the
percentage flux contributed by objects in the F300W images is greater
than in the other two bands.  These characteristics are explained by
two facts: (1) errors in dark glow subtraction cause noticeable
variation in the background level between exposures in this band; and
(2) the ZL falls off rapidly below 4000\AA, so that the ZL
contribution to this band is a factor of three smaller than in the
F555W or F814W and variations in source flux from chip to chip are
fractionally larger in this band than in the other two. In contrast,
ZL contributes roughly 95\% of the diffuse background at
5500--9000\AA\ (see \S\ref{zl} and Paper II).  Nonetheless, the
contribution to the background from detected objects in the F300W
images, again assuming 2\% scattered light, is still a minor
uncertainty in our results, being 0.3\tto{-9}\escsa, or 7\% of the
EBL we detect at 3000\AA.

\section{\uppercase{ HST/WFPC2 Total Measured Background}}\label{wfpc2.backg}

\begin{figure}[t]
\begin{center}
\includegraphics[height=3in,angle=90]{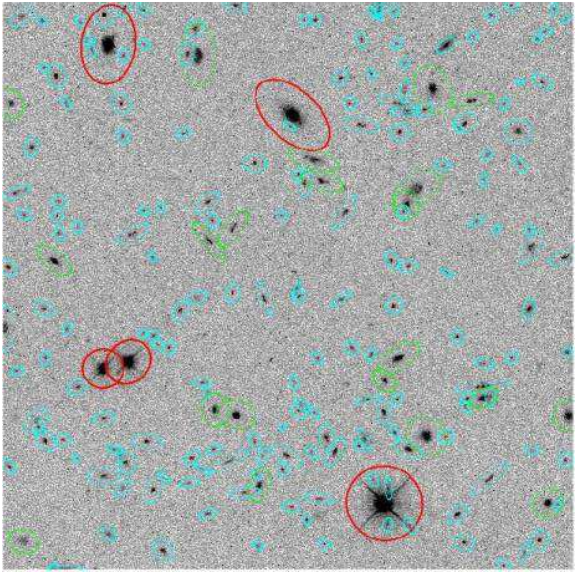}
\caption { \footnotesize Combined F555W WF2 image. North is up and
East is left.  The ellipses indicate the region extending to $4\times$
the isophotal detection radius, as discussed in \S\ref{ebl.minim}.
Red ellipses mark galaxies with $V_{555}\lta 23$ AB mag; green
ellipses, $23 < V_{555} \lta 25$ AB mag; and blue ellipses, $25 <
V_{555} \lta 28.5$ AB mag. All sources in the catalog are shown. The
completeness limit is $V_{555}\sim27.5$.}
\label{fig:wfpc2.ellips}
\end{center} 
\end{figure}

\begin{figure}[t]
\begin{center}
\includegraphics[height=3in,angle=180]{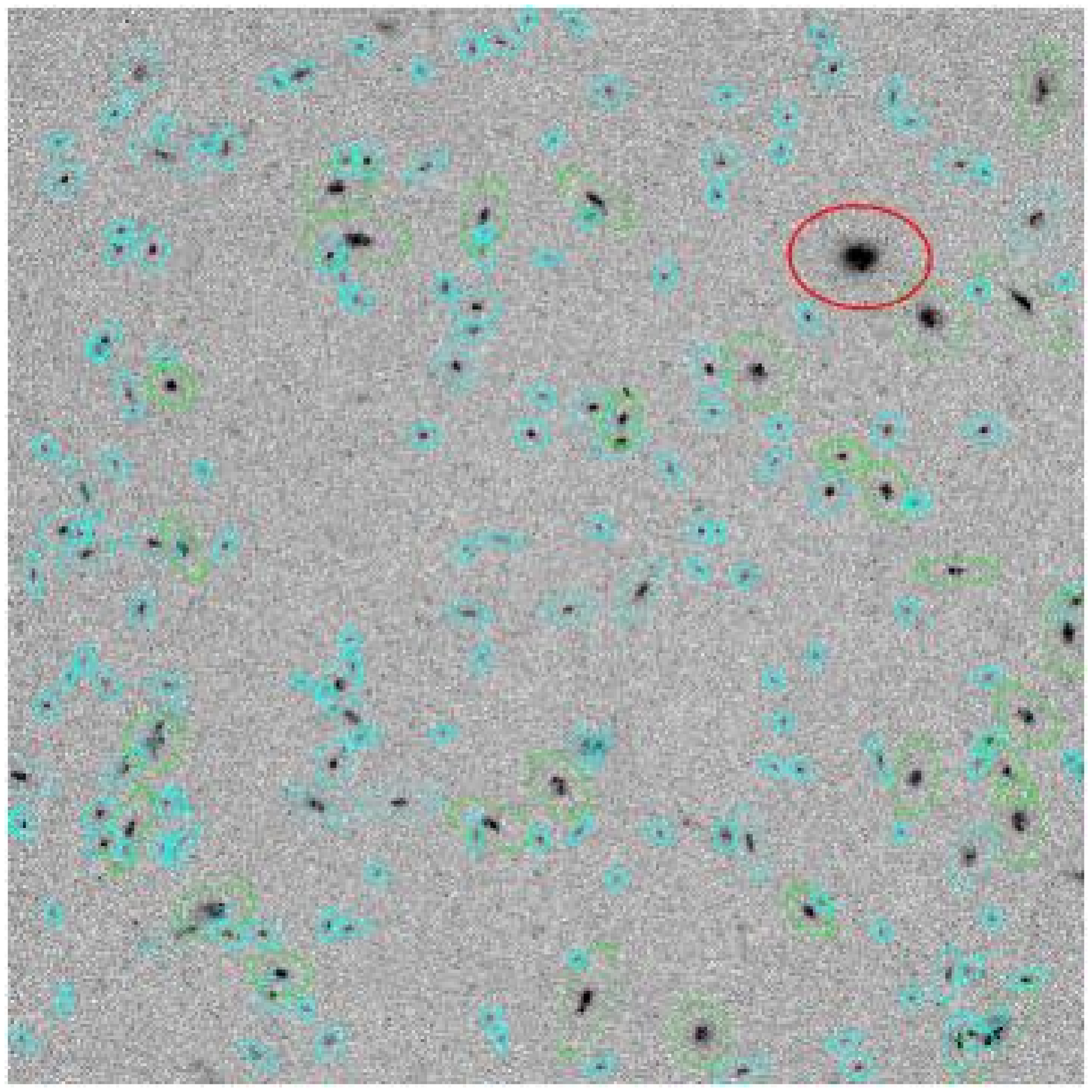}
\caption { \footnotesize Combined F555W WF3 image. North
is up and East is left.  Ellipses are as in Figure 10.}
\label{fig:wfpc2.ellips.wf3} 
\end{center} 
\end{figure}

\begin{figure}[t]
\begin{center}
\includegraphics[height=3in,angle=-90]{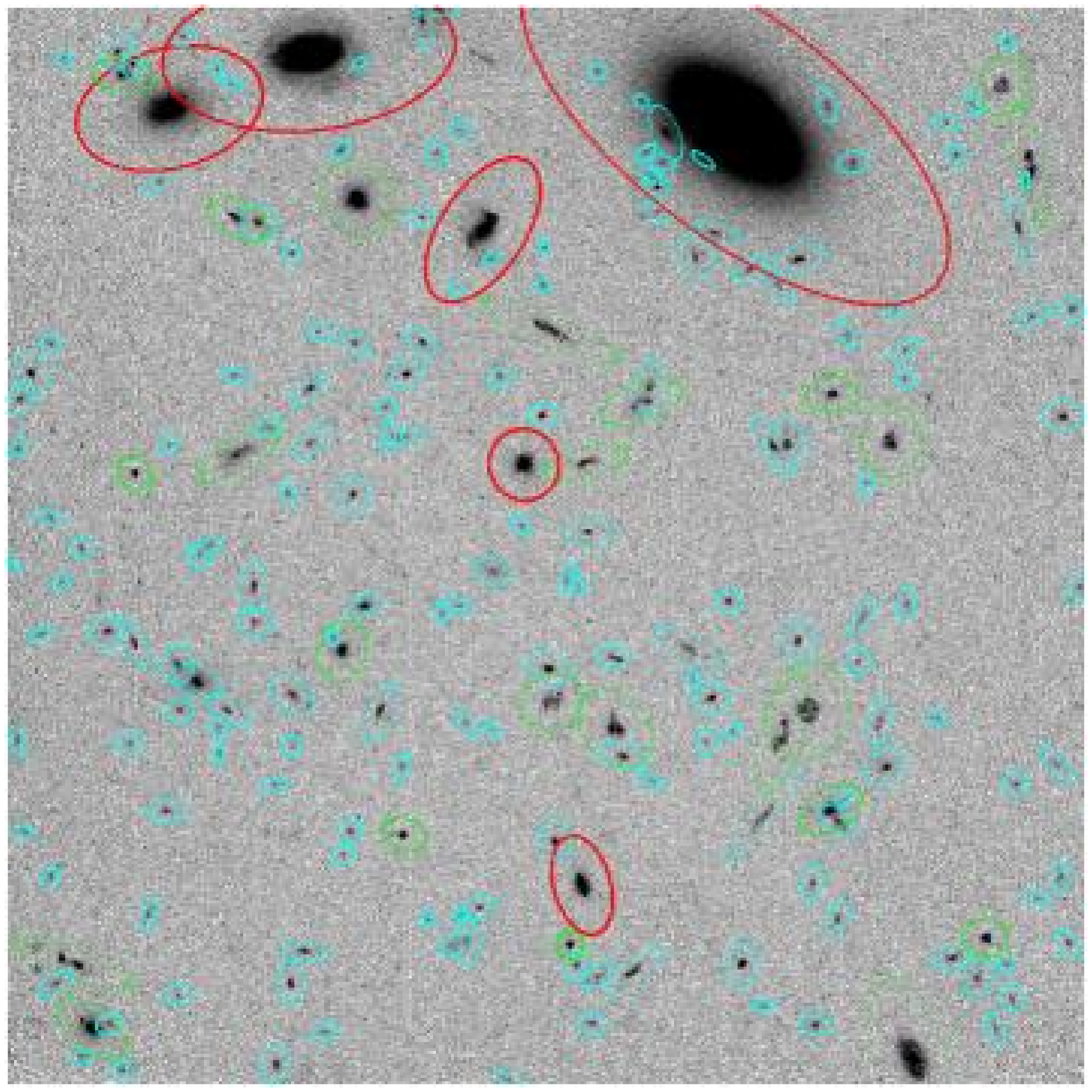}
\caption { \footnotesize Combined F555W WF4 image. North
is up and East is left. Ellipses are as in Figure 10.
}
\label{fig:wfpc2.ellips.wf4} 
\end{center} 
\end{figure}

After calibration, we measure the total background signal by simply
taking the average of the detected flux per pixel, excluding pixels
which fall into any of the following three categories: those flagged
as bad in the data quality file for any reason (see WFPC2 Data
Handbook V3.0), those within two pixels of a cosmic ray event, or
those within the masked region associated with a star or galaxy
brighter than $V_{555}=23$\,AB\,mag, as described in the previous
section.  We show the masked regions in Figures
\ref{fig:wfpc2.ellips}--\ref{fig:wfpc2.ellips.wf4}, for each of the WF
chips. Masks were defined based only on the F555W images; the same
object masks were applied to all three bandpasses.  The detected EBL
at all wavelengths is therefore defined by contributions from the same
sources.

\begin{figure}[t]
\begin{center}
\includegraphics[width=3.0in,angle=0]{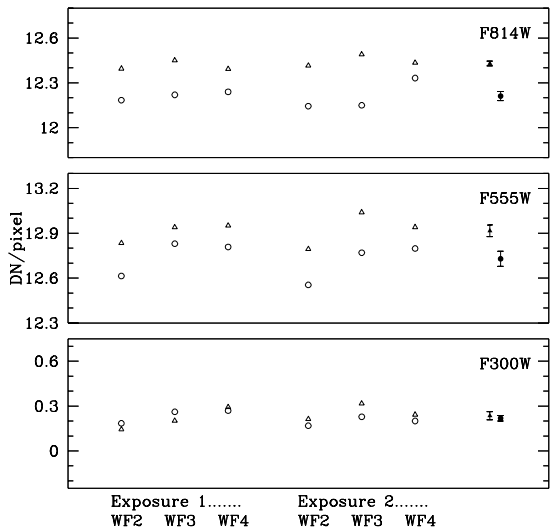}
\caption{\footnotesize 
Total background detected (DN/pixel in 1800 sec) in each of
two exposures per filter obtained in the November ($\triangle$) and
December ($\circ$) 1995 visits.  Bright galaxies ($V_{555}<23$ mag),
stars, and bad pixels are excluded from this average.  Filled symbols
indicate the mean for each visit with error bars showing empirical
$1\sigma$ scatter around the mean, in good agreement with tabulated
statistical errors (see Table \ref{tab:wfpc2.errors}).  For this
comparison, DN/pixel for each chip has been normalized to match the
gain of the WF3 chip (see H95b).  Decrease in the mean from November
to December is in good agreement with the predicted change in the ZL
flux due to viewing geometry.}
\label{fig:wfpc2.bckgrnd.dn}
\end{center}
\end{figure}

\begin{figure}[t]
\begin{center}
\includegraphics[width=3.0in,angle=0]{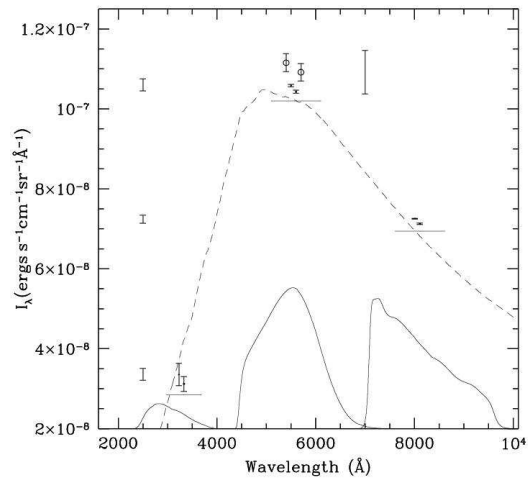}
\caption{\footnotesize The total sky flux detected through each of the
WFPC2 filters in the November and December 1995 data sets, excluding
bright galaxies ($V_{555}<23$\,AB\,mag).  November 1995 results are
indicated at the central wavelength of each filter; the December 1995
data are offset by +100\AA\ for clarity. The dashed line shows the ZL
spectrum at the appropriate flux; horizontal bars show ZL flux
convolved with WFPC2 bandpass profiles. The total background detected
with the FOS spectra are marked by $\circ$'s and offset left
(November) and right (December) of the central F555W wavelength.
Error bars on all points indicate one--sigma statistical errors (see
Tables \ref{tab:wfpc2.errors} and \ref{tab:fos.errors}).  The floating
error bars, arbitrarily plotted at 2500\AA, show systematic
uncertainties in the WFPC2 results at the indicated fluxes. The
floating error bar at 7000\AA\ shows the systematic uncertainty in the
FOS results.  Solid curves show the effective F300W, F555W, and F814W
bandpass profiles.}
\label{fig:wfpc2.bckgrnd.ergs}
\end{center}
\end{figure}

 

\begin{deluxetable}{l l  c r r}
\tablewidth{30pc}
\tablecaption{Total Sky Flux (Foregrounds and EBL) Measured from WFPC2
\label{tab:wfpc2.results}}
\tablehead{
\colhead{Filter} & 
\colhead{$\lambda_0 ({\rm\small FWHM})$ \AA}& 
\colhead{Total Background} &
\colhead{($\pm$ {\rms})\tablenotemark{a}}& 
\colhead{[$\pm$ sys]\tablenotemark{b}}
}
\startdata
November 1995&  \hfil & \hfil &\hfil &\hfil \nl 
\cline{1-1}
F300W   & 3000\,(740)
                        & 3.35\tto{-8} &$(\pm$ 9\%) &$[\pm$ 5.6\%]\nl
F555W   & 5500\,(1230)
                        & 1.06\tto{-7} &$(\pm$ 0.3\%) &$[\pm$ 1.4\%]\nl
F814W   & 8000\,(1490)
                        & 7.24\tto{-8} &$(\pm$ 0.2\%) &$[\pm$ 1.4\%]\nl
\hfil   & \hfil  & \hfil        & \hfil & \hfil \nl
December 1995&   \hfil  &\hfil &\hfil &\hfil \nl \cline{1-1}
F300W   & 3000\,(740)
                        & 3.12\tto{-8} & ($\pm$ 7\%) & [$\pm$ 5.6\%] \nl
F555W   & 5500\,(1230)
                        & 1.04\tto{-7} & ($\pm$ 0.4\%) & [$\pm$ 1.4\%]\nl
F814W   & 8000\,(1490)
                        & 7.11\tto{-8} & ($\pm$ 0.2\%) & [$\pm$ 1.4\%]\nl
\enddata
\tablenotetext{a}{Statistical errors indicate one--sigma scatter
        in the six measurements per visit (three
        images per exposure, two exposures). Compare to estimated
        errors in Table \ref{tab:wfpc2.errors}.}
\tablenotetext{b}{Systematic uncertainty as tabulated in Table 
        \ref{tab:wfpc2.errors}.}
\end{deluxetable}

Each WF image from each 1800\,sec exposure produces a measurement of
the mean background. We have averaged the six measurements from each
visit to obtain a single background measurement from each visit. The
results for the November and December 1995 visits are compared in
Figure \ref{fig:wfpc2.bckgrnd.dn} to illustrate the change in
background due to the geometric path length through the IPD, and thus
the difference in ZL contribution during the two sets of observations.
The same modulation in background flux was identified in the FOS data
between November and December 1995, as discussed in \S\ref{fos.resul}.
The predicted change from October to November is of order 10\%.
However, as there are no FOS or ground--based measurements in October,
the interpretation of the modulation as a change in ZL is not
conclusive for that epoch.  We do not include the PC data in the final
average because the dark subtraction is least accurate for that chip
due to the smaller pixels and higher dark current (see
\S\ref{wfpc2.dark}), and because the field of view of the PC is a
negligible increase over that of the other three chips. However, the
PC chip gives the same result as the WF chips to within 2\%, which is
consistent with the variation in object distribution over the 4 chips
and dark subtraction errors.

The error bars plotted in Figure \ref{fig:wfpc2.bckgrnd.dn} show the
statistical variation between the six measurements (three WF chips,
two exposures per filter) from each visit.  This scatter is well
matched to the statistical errors that we predict based on our
assessment of the errors accrued at each stage in the data reduction
(see Table \ref{tab:wfpc2.errors}).  Systematic errors are dominated
by the flux calibration from DN to physical units and are not shown in
the comparisons in Figure \ref{fig:wfpc2.bckgrnd.dn} as they are, of
course, identical for all points.  In Figure
\ref{fig:wfpc2.bckgrnd.ergs} we plot the total background flux
measured in each bandpass in the November and December visits in
physical flux units (\escsa) with statistical and random $1\sigma$
errors plotted separately. 
The measurements are summarized in Table \ref{tab:wfpc2.results}.
The zodiacal light spectrum is shown at
the flux level we measure in our ground--based spectrophotometry (see
\S\ref{zl}) to demonstrate that background observed from HST is
clearly dominated by zodiacal light.

\section{\uppercase{ HST/FOS Observations and Data Reduction }}\label{fos}

The FOS field of view during our parallel observations is determined
by the WFPC2 coordinates and the roll angle of the telescope, which we
carefully specified.  Ground based images taken with the Swope 1 meter
telescope at Las Campanas Observatory were used to assure that the FOS
field of view was free of detectable sources brighter than $V=25$ AB
mag.

Based on calculations using an early version of the FOS exposure--time
calculator, the FOS/RED was used in parallel observing mode and
configured with the 1.0-PAIR aperture and the G570H disperser during
our October visit, producing spectra with $\sim4.5$\AA\ resolution
from 4500 to 6800\AA.  No exposures were taken during transits of the
South Atlantic Anomoly. These data were intended to provide a
measurement of the ZL by the method outlined in \S\ref{foreg.zl}, but
the count rate from the sky in this configuration was comparable to
the instrument dark rate ($\sim0.01$ DN/sec per diode).  Accordingly,
the FOS observations during the November and December 1995 visits used
the FOS/RED, A--1 aperture ($3.66\times3.71$ \sqarcsec), and G650L
disperser, which provided spectra with $\sim12$ diodes ($\sim300$\AA)
per resolution element for an aperture--filling source (25\AA\ per
resolution element for a point source) from 3800--7000\AA\ and a
signal--to--noise ratio of almost 20 for the night sky.  While
narrow--band imaging as such is not useful for measuring the ZL flux
by its spectral features, these data do provide better spectral
resolution than WFPC2 images, and a second, independent measurement of
the total background.  We limit our discussion to the data from the
November and December visits.  One 1300 sec FOS exposure was obtained
per orbit during these visits, executing within the time-span of the
1800 sec WFPC2 observations to avoid conflicts in writing the data to
on--board recorders.

Many of the pipeline calibration procedures (described thoroughly in
the HST Data Handbook) contribute negligible errors for our
purposes and were adopted directly. In order of their application, the
pipeline procedures used include the conversion from counts to
count--rate ({\sc cnt\_corr}), flat fielding ({\sc flt\_corr}),
wavelength calibration ({\sc wav\_corr}), and flux calibration ({\sc
ais\_corr}, {\sc apr\_corr}, and {\sc tim\_corr}).  The A--1 aperture
defines the throughput standard for other apertures, so that the
relative correction ({\sc apr\_corr}) for the A--1 aperture is unity.
The G650L has no observed time variation, so that the correction
factor applied by the {\sc tim\_corr} procedure is also unity. We were
able to improve on the dark subtraction procedure, and accordingly
used our own methods for that step, as described below.
Unfortunately, much of the information needed to obtain accurate
surface photometry with the FOS, namely the solid angle of the
apertures and PSF of the instrument, has never been provided by
STScI.  We have therefore supplemented the pipeline calibration with
an explicit measurement of the PSF of the instrument. The final flux
calibration has a systematic uncertainty of 4.5\%. Errors at each step
in the calibration are summarized in Table \ref{tab:fos.errors}, with
the total systematic error tabulated as described in \S
\ref{wfpc2.backg}.  Below, we summarize the calibration of the FOS
data and briefly discuss the results.  Greater detail can be found in
Bernstein (1998).  Once again, the following section is intended for
readers who are interested in the details of the data analysis.

\begin{deluxetable}{lcc}
\tablewidth{20pc}
\tablecaption{FOS Background Flux: Errors per Resolution Element
\label{tab:fos.errors}}
\tablehead{
\colhead{} & 
\colhead{Random} & 
\colhead{Systematic} }
\startdata
Poisson noise    (\S\ref{fos})
                                & 2\%           & $\cdots$ \nl
Dark subtraction (\S\ref{fos.dark})
                                & 0.5\%         & 0.4\% \nl
Fiducial standards  (\S\ref{wfpc2.fluxc})
                                & $\cdots$      & 1\%   \nl
Point source flux cal.  (\S\ref{fos.fluxc})
                                & $\cdots$      & 1.5\% \nl
Aperture correction (\S\ref{fos.fluxc})
                                & $\cdots$      & 2\%   \nl
Solid angle     (\S\ref{fos.fluxc})
                                & $\cdots$      & 2\%   \nl
\cline{2-3}
Cumulative\tablenotemark{a}   &   $(\pm 2.1\%)$ &  [$\pm$ 2.8\%]\nl
\enddata
\tablenotetext{a}{\footnotesize Statistical errors have been combined in
quadrature to obtain a cumulative, one-sigma error. 
Systematic errors have been combined assuming a flat 
probability distribution for each contributing source of error.
The resulting systematic error is roughly Gaussian distributed,
and the quoted value is the 68\% confidence interval.  For a detailed
discussion see \S\ref{ebl.detec}.}
\end{deluxetable}

\subsection{Dark Subtraction} \label{fos.dark}

\begin{figure}[t]
\begin{center}
\includegraphics[width=3.0in,angle=0]{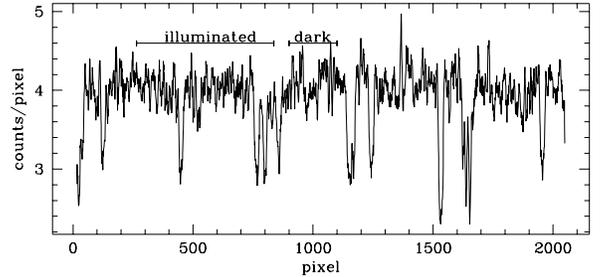}
\caption{ \footnotesize
Average of $25$ FOS dark exposures (1300\,sec),
demonstrating the stability of the pixel-to-pixel structure of the
dark signal.  Strong features appear where diodes have been turned off
due to poor performance. Sub--stepping along the diodes allows
complete spectral coverage over these dead diodes.}
\label{fig:fos.dark}
\end{center}
\end{figure}

The majority of the FOS instrumental background is Cerenkov radiation,
caused by cosmic rays striking the photocathode, rather than thermal
dark current.  Consequently, the instrumental background can vary by
factors of two between exposures, a fact ignored in the pipeline
calibration.  As with the WFPC2, the relative dark rate is a function
of instrument geometry and is quite stable over the detector. The
G650L low--resolution grating illuminates only 144 of the 512 diodes;
therefore, the unilluminated portion can be used to identify the dark
rate during a given exposure.  On the suggestion of L. Petro (private
communication), we produced a ``superdark'' using 25 dark frames taken
in the HDF parallel program (Program 6342, Freedman; Program 6339,
Petro) and scaled this to the level indicated by a subset of the
unexposed diodes (pixels 900--1100) in each of our spectra. The
exposed and unexposed regions of the diode array are labeled in Figure
\ref{fig:fos.dark}, where we plot the averaged dark spectrum. 
To determine the accuracy of this dark--subtraction method, we have
reduced darks which were taken during our own orbits expressly for
this purpose.  The test reduction of these darks shows no systematic
error in the dark subtraction by this method, and statistical errors
are dominated by shot noise ($\pm2.2$ DN/pixel).  The mean dark level
can be determined in an individual frame to roughly $\pm0.3$DN/pixel.
With a sky signal of roughly 30 DN/pixel in the low-dispersion
configuration, this dark--subtraction method introduces a 0.5\% random
error to each spectrum overall and 0.4\% systematic uncertainty to
each resolution element.

\subsection{Flux Calibration} \label{fos.fluxc}

The surface brightness of an aperture--filling source is
a function of the pipeline--calibrated spectrum,   $F(\lambda)$
in {\esca}, the detector solid angle, $\Omega$, and the aperture
correction, $T$(A--1)$\times D$, which 
includes a term for flux lost
at the A-1 science aperture ($T(A-1)$) and at the detector ($D$):
\begin{equation}
I(\lambda) =  \frac{F(\lambda)\; T({\rm A-1})\; D}{\Omega}.
\end{equation}
The pipeline calibration (calibration of the spectrum as appropriate
for a point source) depends on the stability of the instrument for
relative flux calibration and on the accuracy of the fiducial
standards for absolute flux calibration.  The accuracy of the
secondary standard star system does not dominate the FOS calibration
uncertainty; it was already discussed briefly in \S\ref{wfpc2.fluxc} and
is discussed further in Paper II.

The pipeline calibration converts DN sec$^{-1}$ per diode to ergs
sec$^{-1}$ \AA$^{-1}$, as appropriate for point sources, achieving
repeatability of 1--2\% for point sources only if they are centered in
the aperture to within $\pm0.2$ arcsec (Keyes 1995).  This sensitivity to
centering implies at least two serious complications for surface
photometry.  First, the transmission efficiency across the
photocathode can vary by as much as 20\% over surface areas
corresponding to 10 diodes.  Second, and more importantly, the PSF is
not well contained within either the aperture, the detector, or both.
An accurate aperture correction is therefore crucial to surface
photometry.

The aperture dilution factor, $T({\rm ap})\times D$, given in the FOS
Instrument Handbook is an estimate produced by modeling based on the
OTA with post--COSTAR configuration.  While no official Instrument
Science Report exists, unofficial estimates for the monochromatic
transmission of the A--1, post--COSTAR configuration at 6500\AA\ range
from 97\% (R. Bohlin, private communication) to 95\% (The Data
Handbook V3.0).  Because this factor is crucial to our result, we have
recalculated it using data taken for this purpose as part of the FOS
calibration program (Proposal 5262, Koratkar).  These data were taken
in ACQ/IMAGE mode, which uses no dispersing element, providing a two
dimensional image in the diode plane.  The stepping pattern used for
the observations was kindly provided to us by E. Smith at STScI. Our
reduction and analysis of these data are described in Bernstein
(1998).  We find that 98\% of the flux from a point source is
contained within the A--1 aperture at the focal plane, and 96.5\% of
that flux is then imaged onto the 1.29 arcsec spatial extent of the
diode array: $T$(A--1)=0.98 and $D=0.965$. Thus,  we find 
\begin{equation} 
T({\rm A-1})\times D +0.002 \approx 0.95,
\end{equation}
in which we have included a small (0.002) correction for conversion
from the ``white'' light of the ACQ/IMAGE data to our central
wavelength of 5500\AA.  The statistical error in this estimation is
much less than 1\%, but the systematic uncertainty may be as large
as 2\%.

The A--1 science aperture measures $3.63\times3.71$\,\sqarcsec, with
1\% errors in both dimensions. In the spatial direction the solid
angle is determined by the diodes, which are 1.289($\pm$1\%) arcsec in
the spatial direction.  This values is based on a laboratory
measurement made before launch (Instrument Science Report ISR
CAL/FOS--019), and has been corrected for the measured change in the
FOS plate scale before and after COSTAR was installed (ISR
CAL/FOS--123,141).  The effective solid angle through the A-1 aperture
is then $3.63\times1.29$\,\sqarcsec, or 4.68\,arcsec$^2$ with an
uncertainty of $\sim 2$\%.

\section{\uppercase{HST/FOS: Results}} \label{fos.resul}

\begin{figure}[t]
\begin{center}
\includegraphics[width=3.0in,angle=0]{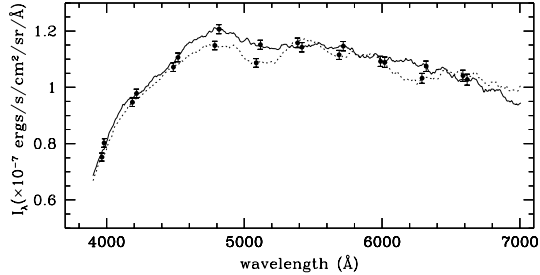}
\caption{\footnotesize
Surface brightness of the total background
detected in the FOS spectra.  The solid and dashed lines show the
average of all six spectra taken during the November and December 1995
visits, respectively.  The error bars indicate $1\sigma$ statistical
errors of 1.5-2\% per 300\AA\ resolution element.}
\label{fig:fos.spec}
\end{center}
\end{figure}

The combined averages of the six spectra taken during the November and
December visits, respectively, are shown in Figure \ref{fig:fos.spec}.
With roughly 80 counts per diode, and twelve diodes per resolution
element, the statistical error per resolution element for one spectrum
is roughly 4\%.  Statistical errors in the averaged spectra are
roughly 1.5--2\%, indicated by the error bars which are placed one per
resolution element (300\AA).  The dark subtraction contributes an
error of less than 0.5\%.  We do not show systematic uncertainties, as
they will affect both November and December data in the same way and
are irrelevant for this comparison.  We find that the integrated
background flux was brighter in November than in December by
$2\pm0.5$\%, in good agreement with the expected change based on the
increase in the path--length through the zodiacal plane between visits
and empirical estimates from ground-based observations
(Levasseur--Regourd \& Dumont 1980).

\begin{figure}[t]
\begin{center}
\includegraphics[width=3.0in,angle=0]{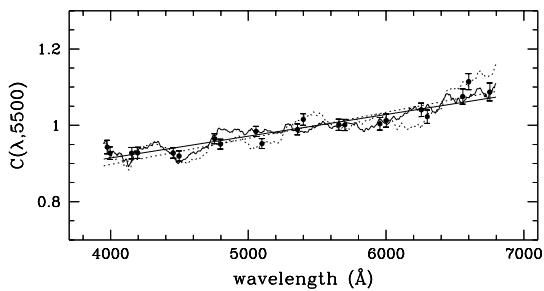}
\caption{\footnotesize Color of the detected background with respect
to solar.  Solid and dashed lines show the detected background in the
November and December 1995 data sets, respectively, divided by the
Neckel \& Labs (1984) solar spectrum at matched resolution. Straight
lines show the linear fits.  The error bars show the statistical error
per resolution element in the FOS spectra. The adopted solar spectrum
is simply the fiducial spectrum from which the ZL color is defined,
and so it contributes no error here. For further discussion, see Paper
II.}
\label{fig:fos.color}
\end{center}
\end{figure}

Employing the usual definition of the ZL color relative to the solar
spectrum (see \S\ref{foreg.zl}), we find $C(7000,4000) = 1.044$ for
the November visit, and 1.075 in the December data, as shown in Figure
\ref{fig:fos.color} with $1\sigma$ statistical uncertainty of 0.05\%.
The reddening of the ZL relative to the solar spectrum is a function
of the scattering angle and, thus, also of time of year. Again, the
trend we observe is in good agreement with the Helios Space Probe
observations and others (see Leinert \etal 1981, and Leinert \etal
1998 for a review).  We stress, however, that the FOS spectrum
includes both the EBL and the ZL, and therefore we cannot measure the
color of the ZL explicitly from these observations. It is only
possible to determine the ZL color separately from the EBL by explicit
measurement of ZL absorption features over a wide range in wavelength.
All published colors of the ZL to date rely on broad--band
observations.  Separation of the ZL from other the EBL and DGL is
discussed further in \S\ref{ebl.detec} and in detail in Paper II.

An error budget for the absolute flux calibration of these data is
shown in Table \ref{tab:fos.errors} as the error per resolution
element.  The random error in the mean flux over the entire spectrum
goes down as the square root of the number of resolution elements
($\sqrt{10}$).  As discussed in \S\ref{ebl.detec}, random and
systematic errors have been combined assuming Gaussian and flat
distributions, respectively.  For comparison with the WFPC2
observations, we have integrated the FOS spectrum through the F555W
bandpass (see Figure \ref{fig:wfpc2.bckgrnd.ergs}) and find the two
results in good agreement.  This comparison is subject to the
systematic uncertainties in both data sets.

\section{\uppercase{Diffuse Galactic Light (DGL)}}\label{dgl}

\begin{deluxetable}{ l c c r r r}
\tablewidth{30pc}
\footnotesize
\tablecaption{Observed Correlations: 
        $I_{\lambda}(0.16\mu{\rm m})/I_{\nu}(100\mu{\rm m})$
\label{tab:dgl.uv.ir}}
\tablehead{
\colhead{Reference} & \colhead{$l^\circ$ \tablenotemark{a}} & 
\colhead{$|b^\circ|$\tablenotemark{a}} & 
\colhead{$I_{\lambda}(0.16)/I_\nu(100)$\tablenotemark{b}} &
\colhead{$I_{\lambda}(0.16)/I_\nu(100)$\tablenotemark{c}} &
\colhead{$I_{\nu}(100)$\tablenotemark{d}}
}
\startdata
Witt~et~al.\ (1997)\tablenotemark{e}            &
$\langle 145\rangle$    & $ \langle 30\rangle$  & 72 & 0.86\tto{-9}
& $2-8$ \nl
Hurwitz~et~al.\ (1991)\tablenotemark{f}         & 
$135-220$               & $>40$                 & $80 (\pm 10)$
&$0.96\times10^{-9}$    & $1-5$ \nl
Sasseen \etal (1996)  & 
$\langle 270\rangle$    & $\langle 45\rangle$   & $< 233$
& $< 2.8$\tto{-9}       & $2-8$                 \nl
Jacobsen~et~al.\ (1987)                         & 
$\langle 70\rangle$     & $\langle 50\rangle$   & $65 (\pm 25)$ 
& 0.78\tto{-9}          & $1-2$                 \nl
Witt~et~al.\ (1997)                             &
$\langle 290\rangle$    & $\langle 45 \rangle$  & 258
& 3.10\tto{-9}          & $2-8$                 \nl
\enddata
\tablenotetext{a}{Bracketed values of Galactic longitude ($l$) and 
        latitude ($b$) indicate average coordinates for the data
        used in the calculation.}
\tablenotetext{b}{$I_\lambda$ is in 
        photons s$^{-1}$ cm$^{-2}$ sr$^{-1}$\AA$^{-1}$.
        $I_\nu$ is in units of MJy\,sr$^{-1}$. }
\tablenotetext{c}{$I_\lambda$ is in \escsa.}
\tablenotetext{d}{$I_\nu(100\mu{\rm m})$ is in MJy\,sr$^{-1}$ .}
\tablenotetext{e}{
        Witt~et~al.\ (1997) use a model which is 
        based on their observations and which includes the average
        scattering angle and phase function along the line of
        sight to predict 
        $I_{\lambda}(0.16\mu{\rm m})/I_\nu(100\mu{\rm m})$ at
        the Galactic position indicated.}
\tablenotetext{f}{We calculate 
        $I_{\lambda}(0.16\mu{\rm m})/I_\nu(100\mu{\rm m})$ for 
        Hurwitz~et~al.\ (1991) from the points plotted in their
        Figures 2a and 2b.}
\end{deluxetable}

\begin{deluxetable}{ l r r  r r r r }
\tablewidth{40pc}
\scriptsize
\tablecaption{Observed Correlations: $I_{\nu}(\lambda)/I_{\nu}(100\mu{\rm m})$ 
\label{tab:dgl.op.ir}}
\tablehead{
\colhead{Reference} & 
\colhead{$l(^\circ)$} & 
\colhead{$b(^\circ)$} & 
\colhead{$I_{\nu}(0.45)/I_{\nu}(100)$\tablenotemark{a}} & 
\colhead{$I_{\nu}(0.65)/I_{\nu}(100)$\tablenotemark{a}} & 
\colhead{$I_{\nu}(0.90)/I_{\nu}(100)$\tablenotemark{a}} & 
\colhead{$I_{\nu}(100)$\tablenotemark{a}}}
\startdata
GT89 - ir1\tablenotemark{b}                     & 
174     & $-$42 & 0.36\tto{-3}  & 1.1\tto{-3}   &$\cdots$  & $11.4$ \nl
Laureijs~et~al.\ (1987)\tablenotemark{c}        &  
211     & $-$37 & 0.48\tto{-3}  & $\cdots$      & $\cdots$      & 6.3 \nl
GT89 - ir2\tablenotemark{b}                     &
235     &  37   & 1.1\tto{-3}   & 2.2\tto{-3}   & $<1.6$\tto{-3} &$ 3.6$ \nl
Paley~et~al.\ 1991                              & 
104     & $-$32 & 4.8\tto{-3}   & 8.0\tto{-3}   & 11.0\tto{-3}  & $2.5$ \nl
GT89 - ir3\tablenotemark{b}                     &  
38      &  45   & 2.6\tto{-3}   & 4.4\tto{-3}   & 6.0\tto{-3}   & $5.9$ \nl
\hfil   & \hfil & \hfil & \hfil & \hfil & \hfil \nl
\enddata
\tablenotetext{a}{$I_\nu$ is in units of MJy\,sr$^{-1}$, throughout. 
        1 MJy\,sr$^{-1}$ = $10^{-17} c/{\lambda}^2$ \escsa.}
\tablenotetext{b}{Guhathakurta \& Tyson (1989) -- field designation.}
\tablenotetext{c}{From the data in Table 1 of  Laureijs \etal (1987)}
\end{deluxetable}

\begin{deluxetable}{ l c c c  r  r r}
\tablewidth{35pc}
\tablecaption{Model Correlations: $I(\lambda)/I(100\mu{\rm m})$ 
\label{tab:dgl.mymodel}}
\tablehead{
\colhead{$\lambda(\mu{\rm m})$}  & 
\colhead{$\tau_\lambda$/{N(\ion{H}{i})}\tablenotemark{a}}& 
\colhead{$\omega_\lambda$\tablenotemark{b}} & 
\colhead{$S(g,b)$\tablenotemark{c}} & 
\colhead{${I_\lambda}(\lambda)/I_\nu(100)$\tablenotemark{d}}&
\colhead{${I_\nu}(\lambda)/I_\nu(100)$\tablenotemark{e}}&
\colhead{${I_\lambda}(\lambda)/I_\nu(100)$\tablenotemark{f}}
}
\startdata
0.16 & 0.120 & 0.410 &  0.42 &181.96  &  1.86\tto{-4} & 2.18\tto{-9}\nl
0.25 & 0.130 & 0.550 &  0.50 & 94.54  &  3.69\tto{-4} & 1.77\tto{-9}\nl
0.30 & 0.105 & 0.581 &  0.54 &118.09  &  7.97\tto{-4} & 2.66\tto{-9}\nl
0.45 & 0.068 & 0.600 &  0.56 & 68.75  &  1.57\tto{-3} & 2.32\tto{-9}\nl
0.55 & 0.047 & 0.600 &  0.59 & 42.85  &  1.78\tto{-3} & 1.77\tto{-9}\nl
0.65 & 0.042 & 0.600 &  0.61 & 32.66  &  2.24\tto{-3} & 1.59\tto{-9}\nl
0.80 & 0.032 & 0.540 &  0.68 & 18.82  &  2.41\tto{-3} & 1.13\tto{-9}\nl
0.90 & 0.025 & 0.500 &  0.71 & 11.76  &  2.14\tto{-3} & 7.94\tto{-10}\nl
\enddata
\tablenotetext{a}{Optical depth as a function of  \ion{H}{i} column
        density in units of $10^{20}$\,cm$^{-2}$. Values are 
        from Savage \& Mathis (1979) and are in good
        agreement with the standard value of 
        A$_V$/N(\ion{H}{i})=0.06\,mag/($10^{20}$\,cm$^{-2}$) 
        (Bohlin, Savage \& Drake 1978).}
\tablenotetext{b}{Albedo values from models of Draine \& Lee (1984).} 
\tablenotetext{c}{Scattering phase function in terms of asymmetry
parameter, $g$ and Galactic latitude ($b=50^\circ$), as calculated by  Draine \& 
Lee (1984).}
\tablenotetext{d}{$I_\lambda(\lambda)$ is in 
        photons s$^{-1}$cm$^{-2}$sr$^{-1}$\AA$^{-1}$.
        $I_\nu(100\mu{\rm m})$ is in MJy\,sr$^{-1}$. }
\tablenotetext{e}{$I_\nu(\lambda)$ is in MJy\,sr$^{-1}$.
        $I_\nu(100\mu{\rm m})$ is in MJy\,sr$^{-1}$. }
\tablenotetext{f}{$I_\lambda(\lambda)$ is in \escsa.
        $I_\nu(100\mu{\rm m})$ is in MJy\,sr$^{-1}$. }
\end{deluxetable}

\subsection{Structured Component}\label{dgl.struc}

A diffuse, non--isotropic, optical background is produced in the Milky
Way by scattering of the optical interstellar radiation field (ISRF)
by Galactic dust. The same dust is heated by the UV ISRF, causing it
to produce thermal IR emission.  It is not surprising, therefore, that
the thermal Galactic emission seen in the IRAS 100\,\micron\ maps
correlates well with the surface brightness of the optical diffuse
Galactic light (DGL), as both are proportional to the column density
of the dust and the intensity of the ambient ISRF along the line of
sight.  In Tables \ref{tab:dgl.uv.ir} and \ref{tab:dgl.op.ir}, we give
a representative summary of the observed correlations between optical
and 100\micron\ fluxes for regions with low to moderate 100\,\micron\
intensities (N(\ion{H}{i})$<5$\tto{20} cm$^{-2}$, $I_{100} < 5$\MJysr)
and a range of Galactic orientations.

As evident from those results, there is only moderate agreement
concerning the exact scaling relations between the optical DGL and
thermal emission at any wavelength. Measurement errors in the IR,
optical, and UV intensities are $>10$\% in most cases and are one
cause for variations between results. However, asymmetry in the
scattering phase function of Galactic dust also contributes to the
variable scaling relations seen between different lines of
sight. Strong forward scattering causes lower optical surface
brightnesses at both high latitudes ($|b|> 50^{\circ}$) and at
longitudes away from the Galactic Center ($130^{\circ} < l <
230^{\circ}$) (see Draine \& Lee 1984, and references therein; Onaka \&
Kodaira 1991; Witt, Friedmann, \& Sasseen 1997).  Both trends are
evident from the data shown in Tables \ref{tab:dgl.uv.ir} and
\ref{tab:dgl.op.ir}.

While these results suggest a range of appropriate correlation
factors, they do not identify a single appropriate scaling law for our
purposes for two reasons.  First, variability in the measured
IR--optical and UV--optical correlations is evident within a single
cloud, as well as between clouds (see, for example, Figures 5 \& 6 in
GT89 and Table 1 in Laureijs, Mattila, \& Schnur 1987). This suggests
that the observed systems may be dense enough that self-shielding and
further complications come into play. These clouds have been selected
precisely because the optical and IR emission is bright enough to be
readily observed: while the IR flux levels and N(\ion{H}{i}) column
densities of the clouds listed in Tables \ref{tab:dgl.uv.ir} and
\ref{tab:dgl.op.ir} are low enough that the molecular gas fraction
does not affect the correlation between dust column density (or
extinction) and N(\ion{H}{i}), they are still roughly a factor of 10
higher than the values for our observed field, for which $I_{100}
\sim$0.4\MJysr (N(\ion{H}{i}) $\sim$0.47\tto{20} cm$^{-2}$, or
E(\bv)$\sim0.009$\,mag).  Second, while empirical relations between
the scattered and thermal DGL have have been published in the far--UV
and at optical $B$-- and $R$--bands, the expected surface brightness
from scattering at 3000\AA\ is not clear from these results. Neither
the optical depth of interstellar dust nor the ISRF is a monotonic
function of wavelength between 1600\AA\ and 4500\AA\ (see Savage \&
Mathis 1979 and Mathis, Mezger, \& Panagia 1983).  

To better understand the contribution of non--isotropic DGL over the
full range of our observations, we have used a basic scattering model
to predict the scattered light from dust.  We then compare the results
of this model to the observed DGL at UV and optical wavelengths.

Assuming the Galactic cirrus along the line of sight in question is
optically thin (extinction, $A_\lambda < 1.08$\,mag), the surface 
brightness of scattered light off of interstellar dust can be expressed as 
\begin{equation}
I_\lambda =  j_\lambda\, \omega_\lambda\, \tau_\lambda\, 
[1-1.1g\sqrt{\sin b}],
\label{eq:dgl.scat.model}
\end{equation}
in which $j_\lambda$ is the flux of the radiation field in {\escsa};
$\omega_\lambda$ is the effective albedo of the dust; $\tau_\lambda$
is the optical depth; and the term in brackets is the back--scattered
intensity in terms of Galactic latitude, $b$, and the average phase
function of the dust, $g$ (Jura 1979). For strong forward scattering,
$g\sim1$; for isotropic scattering, $g\sim0$.  We take the ISRF flux,
$j_\lambda$, from the Mathis~et~al.\ (1983) estimate for the Solar
Neighborhood (10\,kpc from the Galactic center).  As our observations
are $b=60^\circ$ from the Galactic plane and $l=$206\fdg6 from the
Galactic center, this estimate is probably slightly high.  We take the
dust albedo from the results of Draine \& Lee (1984), which are based
on an exponential distribution in grain sizes suggested by Mathis,
Rumpl \& Nordsieck (1977).

The optical depth of Galactic dust, $\tau$, is well known to correlate
strongly with hydrogen column density (see Savage \& Mathis 1979,
Boulanger \& P\'{e}rault 1988, and references therein).  It is not
surprising, then, that the thermal emission, $I_{100}$, also
correlates well with hydrogen column density.  While optical depth is
a physical manifestation only of the column density of dust, $I_{100}$
is also affected by the strength of the ISRF.  We therefore use the
observed $I_{100}$ and $I_{100}$/N(\ion{H}{i}) as calibrated by
Boulanger \& P\'{e}rault (1988) from the IRAS 100\,\micron\ maps to
obtain an effective optical depth for our observations as follows.
Optical depth can be written as a function of optical extinction and
dust column density as
\begin{equation}
\tau_\lambda = 0.921 \frac{R_\lambda} {N({\rm H}{\sc i})/E(B\!-\!V)}
	\,N({\rm H}{\sc i}),
\end{equation}
in which $R_\lambda = A_\lambda$/E(\bv) is the usual expression for
the normalized extinction.  Several groups find N(\ion{H}{i})/E(\bv)
between 48\tto{20} and 50\tto{20} cm$^{-2}$\,mag$^{-1}$ from
measurements of the \ion{H}{i} densities from 21\,cm line emission
strength and the reddening to globular clusters and star counts
(Bohlin, Savage \& Drake 1978, Burstein \& Heiles 1982, Knapp \& Kerr
1974).  To get an effective optical depth (weighted by the ISRF field
strength which is at issue for scattering), we use the relation found
by Boulanger \etal (1996) for the low--column density regime
(N(\ion{H}{i}) $<$5\tto{20} cm$^{-2}$): $I_{100}$/N(\ion{H}{i})$
\propto 0.85$\MJysr/($10^{20}$\,cm$^{-2})$.\footnote{ A slightly
different scaling, $I_{100}$/N(\ion{H}{i}) $\propto
0.53$\MJysr/($10^{20}$\,cm$^{-2})$, is seen in from the DIRBE results
(Boulanger \etal 1996).  The difference is attributed to a well
known calibration offset in the IRAS maps. Since we are using IRAS
fluxes, we use the IRAS correlation.}  The fluxes in our field are
roughly 0.4\MJysr, or 0.47\tto{20}\,cm$^{-2}$.  The predicted
scattered fluxes from this model are shown in Table
\ref{tab:dgl.mymodel}.  Scattering angle is not considered in this
model. Consequently, this estimate is conservative in the sense that it
should over-predict the DGL for our observations, as the line of sight
to our field is away from the Galactic center and the dust is forward
scattering.

This scattering model reproduces the observed flux ratios with
reasonable accuracy in the range 1600--4500\AA\ (see Tables
\ref{tab:dgl.op.ir} and \ref{tab:dgl.uv.ir} at $b>45$).  The phase
function changes by less than 10\% at latitudes $|b|>50^\circ$, so the
values shown in Table \ref{tab:dgl.mymodel}, for which $|b|=50^\circ$
was used, are generally representative for high latitude fields.
However, as noted by GT89, optical colors (\br) and (\ri) are redder
than a basic scattering model predicts.  GT89 find values of
$I_\nu(R)/I_\nu(B)=$ 3.2, 2, and 1.7 and $I_\nu(I)/I_\nu(R)=$ 2.3,
2.1, and $<1.5$ in three different fields. By comparison, the ratios
we predicted are $I_\nu(R)/I_\nu(B)$=1.4 and $I_\nu(I)/I_\nu(R)$ =
0.95.  A significant H$\alpha$ contribution as the explanation for the
red colors is ruled out by GT89.  Variable scattering asymmetry with
wavelength is another possible explanation, but strong wavelength
dependence in the range 4500--9000\AA\ has never been observed in the
lab or in space (Witt \etal 1997, Onaka \& Kodaira 1991, Laureijs
\etal  1987).  The most plausible explanation is suggested by
observations of reflection nebulae, which have high N(\ion{H}{i}) and
show red fluorescence from molecular hydrogen, hydrogenated amorphous
hydrocarbons, and polycyclic aromatic hydrocarbons.  The relevance of
such contributions to fields with 10 times lower N(\ion{H}{i}) and
$I_{100}$, as is the case for our data, is not clear, as the density
of molecular gas correlates only with high column densities,
N(\ion{H}{i}) $> 5$\tto{20}\,cm$^{-2}$.  The results of GT89, in fact,
do show that the degree of reddening is well correlated to the average
$I_{100}$ emission but not structure within the cloud.
Self--shielding, local optical depth and local ISRF may be responsible
for strong variations in the correlation between color and molecular
gas density both in and between fields (Stark 1992, 1995).  It seems
conservative, therefore, to adopt optical colors found for the fields
with the lowest IR flux in the GT89 sample, listed in Table
\ref{tab:dgl.op.ir}.  Note that the IR flux in the 2 lower flux fields
(denoted ``ir2'' and ``ir3'') is still more than a factor of 10 higher
than in our own.

In summary, we estimate the optical flux in our field using our
scattering model for $\lambda < 4500$\AA, and adjust the predicted
scattering model at redder wavelengths to match the average colors
observed by GT89: $I_\nu(R)/I_\nu(B)\sim$ 1.8 and $I_\nu(I)/I_\nu(R)$
$\sim$ 2.0.  We apply this correction in the sense of increasing the
long wavelength fluxes over that predicted by our models, so that the
DGL estimate we use is, if anything, {\it higher} than is appropriate,
although given the small total flux associated with the DGL even a
large fractional decrease in our estimate of the DGL would have a
negligible impact on our EBL results.  The resulting spectrum is flat in
$I_\lambda$, with a value of roughly 0.9--1.0\tto{-9} {\escsa} from
3000--9000\AA.  We note, also, that our scattering model was not
dependent on Galactic longitude, which, again, makes ours a
conservative overestimate of the DGL contribution to the total sky
background, and our measurement of the EBL, therefore, a conservative
underestimate in this regard.

\subsection{Isotropic Component}\label{dgl.isotr}

Line emission and continuum processes from warm ionized gas in the
Galaxy also contribute an isotropic component to the DGL.  For
$|b|>5^\circ$, Reynolds (1992) finds that H$\alpha$ emission strength
matches the prediction of a path--length through a slab model for the
galaxy, $I({\rm H}\alpha) \approx 2.9\times 10^{-7} \csc
|b|$\,{\escs}.  Fortunately, H$\alpha$ emission, specifically, is
irrelevant for us because the relative throughput of the F555W
bandpasses at H$\alpha$ ($\sim6562$\AA) is only $\sim$10\% of the peak
filter throughput. The strongest H$\alpha$ emission expected in our
field would contribute 0.01\tto{-9}{\escsa}, which corresponds to
0.01\% of the total background, and 1\% of the expected EBL.  The next
strongest line, [O{\sc iii}] at 5007\AA, is near the peak of the F555W
sensitivity, but it is fainter than H$\alpha$ by a factor of 20
(Reynolds 1985, Shields \etal 1981) and will contribute at most
0.05\% of the expected EBL.

More important than line emission is the two--photon, free--free, and
bound--free continua emitted by ionized gas with the density implied
by the detected H$\alpha$ emission.  The combined spectrum of
free--free, bound--free, and two--photon emission was calculated by
Aller (1987) as a function of electron and ion densities, and has been
expressed by Reynolds (1992) as a function of the observed H$\alpha$
emission; it is a function of the temperature of the warm ionized
medium.  For our purposes, a conservative estimates of the isotropic
continuum from gas with temperature $T\sim 10^{4}$ is given by Aller
(1987) and Reynolds (1992) as $I_\lambda(\lta
3700\AA)<0.3$\tto{-9}{\escsa} and $I_\lambda(\gta
3700\AA)<$0.01\tto{-9}{\escsa} (see Aller 1987 and Reynolds 1992 for
discussion). We include these contributions in our estimate of the
DGL.

\section{\uppercase{Measurement of Zodiacal Light}}\label{zl}

To measure the ZL contribution to the background flux that we measured
with HST, we obtained ground--based spectra at 3900-5100\AA\ within
the field of view of our WFPC2 images using the Boller \& Chivens
spectrograph on the duPont 2.5m Telescope at Las Campanas Observatory
in Chile on the nights of 26--29 November 1995, concurrently with the
HST observations of that field on 29 November 1995.  We have used
those spectra to measure the absolute flux of the ZL at 4650\AA\ with
a precision of 0.6\%, and a systematic uncertainty of 1.1\%, using the
method outlined in \S2. We also mesure the color of the zodiacal light
to be $C(5100,3900)=1.05\pm0.01$. That measurement is discussed in
detail in Paper II.

To identify the ZL flux contributing to the FOS and WFPC2 measurements
of the total sky flux, we need an absolute, flux--calibrated spectrum
of the ZL from roughly 2000--1$\mu$m (see WFPC2 bandpasses plotted in
Figure \ref{fig:wfpc2.bckgrnd.ergs}).  We obtain this spectrum
by scaling a solar spectrum to the surface brightness value we measure
for the ZL at 4650\AA\ and applying a small reddening correction
redward and blueward of 4650\AA\ to compensate for the changing
scattering efficiency of the interplanetary dust (see \S\ref{foreg.zl}
for discussion).  Although our LCO measurement of the ZL color is
quite accurate, it only covers a small fraction of the total
wavelength range covered by our WFPC2 observations.  We therefore use
our FOS observations to identify the appropriate reddening correction,
which cover the wavelength range 4000--7000\AA.  Published measurements
of the ZL color have absolute uncertainties as large as 10\% and are
in poor agreement with each other (see Leinert 1998).  Our FOS
observations, by contrast, are accurate to better than 2\% in relative
flux as a function of wavelength and are identical in line of sight
and epoch to the WFPC2 and LCO observations.  Like most measurements
of the ZL, however, the FOS observations clearly include the EBL flux.
This fact prevents the FOS and LCO results from unambiguously
determining the color of the ZL.  Fortunately, this is not an
insurmountable problem: the WFPC2 observations can be used as a
further constraint, as we describe in the following section.

\section{\uppercase{EBL from Resolved Sources in WFPC2 Images}}\label{ebl.minim}

\subsection{Ensemble Photometry}\label{ensemb.phot}

The total flux from resolved sources defines a lower limit to the EBL.
Typically, such minima are obtained by measuring the flux in
individual resolved sources using standard photometry packages, such
as FOCAS (Valdes 1982, Jarvis \& Tyson 1981) or SExtractor (Bertin \&
Arnouts 1996), and summing the flux in the resulting catalog of
objects (c.f. Tyson \etal 1989, Madau \etal 1996, 1998).  While it is
quite straightforward to measure an isophotal magnitude, it is very
difficult, if not impossible, to measure the total flux of an object
including that flux in regions fainter than the noise level of the
local sky.  Efforts at measuring galaxy counts often attempt some sort
of ``total magnitude'' correction, either by scaling the isophotal
aperture to include roughly twice the isophotal extent (see
Bertin~et~al.~1998 for discussion) or by applying a magnitude
correction to galaxies near the detection limit (c.f. Smail \etal
1995).

\begin{figure}[t]
\begin{center}
\includegraphics[width=3.0in,angle=0]{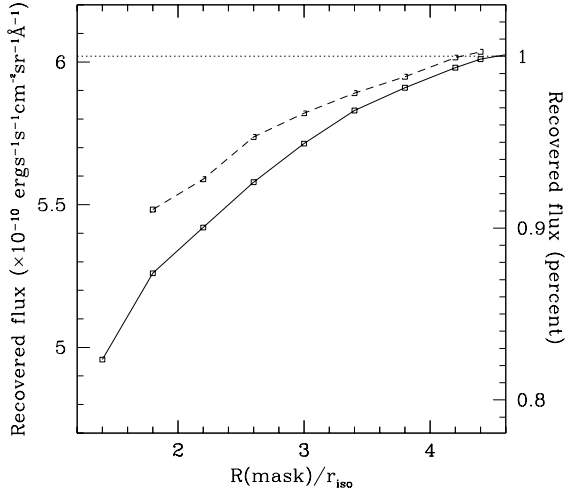}
\caption{\footnotesize
The ensemble flux recovered from apertures
around detected galaxies ($V>23$\,AB\,mag) as a function of the radius
of the apertures. Radius is given in terms of the limiting isophotal
radius ($r_{\rm iso}$) for each galaxy. The solid line shows the flux
recovered when $r_{\rm iso}$ is unconstrained; for the faintest
objects, $r_{\rm iso}$ may be less than 3 pixels (the 90\% encircled
energy radius of WFPC2). The dotted line shows the flux recovered when
a 3 pixel minimum is imposed for $r_{\rm iso}$. Note that the
curve has not yet converged at the limit of the mask size plotted here.}
\label{fig:growth.curve}
\end{center}
\end{figure}


Fortunately, individual source photometry is not necessary for
estimating a lower limit to the EBL in our data.  We have developed a
simplified method of aperture photometry with which we can measure the
flux from the ensemble galaxy population as a whole.  We first use
SExtractor to identify detectable sources and their isophotal radii
($r_{\rm iso}$) using detection and extraction parameters very similar
to those employed in Williams \etal (1996) (see \S\ref{wfpc2.objec}).
We can then measure the ``sky'' surface brightness, $\mu_{\rm sky}$,
for an image by masking all detected sources ($V_{555}\lta
27.5$\,AB\,mag) and computing the average surface brightness of the
remaining pixels.  In doing so, we implicitly assume that
extragalactic sources fainter than this limit contribute negligibly to
the EBL, so that the apparent background level in the image consists
only of foreground sources.\footnote{We point this out not because we
believe that the flux in galaxies beyond the detection limit is
negligible, but rather as a reminder that the sum in detected galaxies
is by definition a minimum value for the detected EBL.}  To estimate
the total surface brightness from extragalactic sources and
foregrounds, $\mu_{\rm obj+sky}$, we mask all stars and only galaxies
brighter than $V_{\rm cut}=23$AB mag and again compute the average
surface brightness of the remaining pixels.  We then isolate the flux
from resolved sources, the minimum value of EBL23, by differencing the
two surface brightness estimates: minEBL23 $\equiv \mu_{\rm obj}=
\mu_{\rm obj+sky} - \mu_{\rm sky}$.  In contrast to standard aperture
photometry, $\mu_{\rm obj+sky}$ will always include all of the light
from faint galaxies.  However $\mu_{\rm sky}$ can include galaxy flux if the
masks are too small.

\begin{figure}[t]
\begin{center}
\includegraphics[width=3.0in,angle=0]{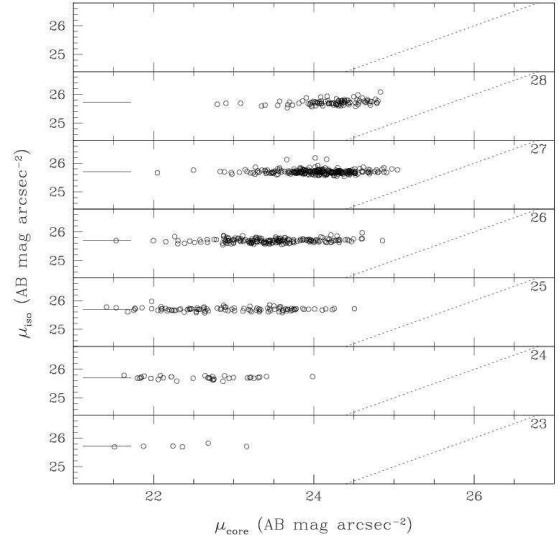}
\caption{\footnotesize 
Isophotal surface brightness, $\mu_{\rm iso}$,
versus core surface brightness (mean surface brightness in the
brightest 4 pixels), $\mu_{\rm core}$, for all detected galaxies in
the EBL images in unit magnitude bins $22<V_{555}<28$; the faintest
bin with data shows $27<V<28$. In each plot, the diagonal lines
(right) mark the limit $\mu_{\rm iso}=\mu_{\rm core}$; the horizontal
lines (left) show the mean $\mu_{\rm iso}$, defined by sky noise
($1\sigma_{\rm sky}$). As expected, this does not change with
magnitude.  The distribution in $\mu_{\rm core}$ narrows at fainter
apparent magnitudes, suggesting that
surface brightness biases and detection limits may be causing
incompleteness at the fainter magnitudes.  The detection limit occurs
at $V=27.5$ AB mag, at which the average galaxy has $\mu_{\rm iso} -
\mu_{\rm core}\sim 1.5$ mag arcsec$^{-2}$. The requirement for detection in our data is
$\mu_{\rm core}< 2\sigma_{\rm sky}$ in the central 4 pixels.}
\label{fig:detec.rabmumumag}
\end{center}
\end{figure}

\begin{figure}[t]
\begin{center}
\includegraphics[width=3.0in,angle=0]{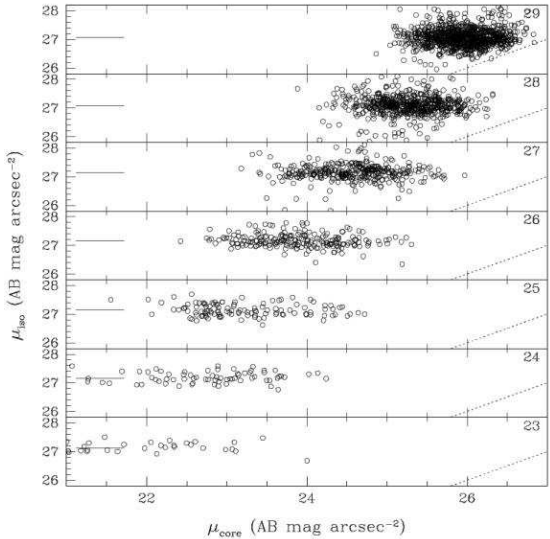}
\caption{\footnotesize 
The same quantities as plotted in Figure
\ref{fig:detec.rabmumumag} are shown for the galaxies in the HDF F606W
catalog (Williams \etal 1996).  The detection limit of the HDF catalog
for $V_{606}$ is quoted as 28.2 AB mag, which corresponds to a
$10\sigma$ detection in the drizzled images ($\sim2\sigma_{\rm sky}$
in 4 pixels in the original images).  At $V_{606}=28.2$AB mag, a
typical galaxy has $(\mu_{\rm iso} - \mu_{\rm core})\lta 1$ mag
arcsec$^{-2}$.  See Appendix B.2 for discussion.}
\label{fig:detec.hdfmumumag}
\end{center}
\end{figure}

Not surprisingly, minEBL23 from this method of ensemble photometry
is a strong function of the mask size.  The growth curve plotted in
Figure \ref{fig:growth.curve} shows that at least 20\% of the flux
from galaxies within 4.5 magnitudes of the detection limit
($23<V<27.5$\,AB\,mag) lies beyond $\sqrt{2}\times r_{\rm iso}$ for
individual galaxies. As discussed in Williams \etal (1996), it is
possible for the faintest galaxies detected to have isophotal radii
which are smaller than the WFPC2 90\% encircled energy radius for a
point source (0.3 arcsec, or 3 pixels). To explore the effect of this
resolution limit, we also show in Figure \ref{fig:growth.curve} the
growth curve which results if we impose 0.3 arcsec as a minimum for
$r_{\rm iso}$ and increase the mask size relative to that new starting
radius.  Naturally, the improvement in recovered flux is significant:
in the HDF catalog of Williams \etal, nearly all galaxies with 
$V_{606}>27.5$ ABmag are affected by this size requirement.  While the
instrumental PSF clearly contributes to the increasing recovered flux
with radius, the growth curve is not entirely due to PSF effects: 
the 100\% encircled energy radius would be reached by a mask of radius
$2.5r_{\rm iso}$ if the smallest galaxies were physically
contained within the original isophotal radii.

In addition to the fact that apertures extending to twice each
galaxy's isophotal area will not recover all of the galaxy light from
detected galaxies, the ``local sky'' estimates used in standard
photometry packages come from regions just beyond the apertures
for each source. These sky estimates will
undoubtedly include a significant fraction of this missed light,
doubly compounding the photometry errors.

While the growth curve in Figure \ref{fig:growth.curve} is beginning
to level off by $4 r_{\rm iso}$, it is clearly not flat. The covering
factor of galaxies to $4 r_{\rm iso}$ is roughly 20\% in these images,
and roughly 30\% of the galaxies defined by these apertures overlap.
In the HDF images, more than 1800 galaxies are found beyond our
detection limit, with a covering factor of roughly 80\% to $4 r_{\rm
iso}$.  These facts suggest that we have reached a confusion limit of
sorts, in the sense that the wings of detectable galaxies are
overlapping.  Moreover, these facts suggest that the wings of
detectable galaxies contribute flux to a significant fraction of the
pixels used to measure the foreground ``sky'' level, producing an
extragalactic ``pedestal'' level which blends with the diffuse flux
from ZL and DGL.  Even though the flux level of our growth curve in
Figure \ref{fig:growth.curve} is clearly converging, it cannot be used
to quantify this pedestal of overlapping galaxy wings.

To estimate the flux contained in the wings of galaxies beyond $4
r_{\rm iso}$ in our own images, we have constructed a Monte Carlo
simulation of the contribution to a random point on the sky from
randomly distributed, detectable galaxies.  The surface number density
of detectable galaxies was drawn from the published HDF catalog and
exponential light profiles were adopted for all galaxies, with scale
lengths and central surface brightness matching the galaxies in the
EBL and HDF images.  Efforts to characterize the light profiles of
faint galaxies have produced evidence for both exponential disks and
flatter, irregular profiles (for example, see Smail \etal\ 1995,
Driver \etal\ 1995, Brinchmann \etal\ 1998, Driver \etal\ 1998 and
references therein).  For the faintest galaxies, profiles are
unconstrained. We adopted exponential profiles here, in part because
they produce the most conservative (smallest) estimate of the light
beyond the $4r_{\rm iso}$ apertures.  Also, given that the physical
scales at those radii are large, we expect to be beyond any central
bulges. We also note that the additional flux identified by our models
from the faintest galaxies is less than 10\% of the flux in galaxy
counts by standard methods, and so is the adopted profiles for the
faintest galaxies are not a critical issue to the minimum EBL
estimates.  The simulations are described in Appendix B.  For the
F555W EBL images, we estimate that the additional flux from galaxies
beyond $4 r_{\rm iso}$ is roughly 1.1\tto{-10} \escsa.  The
simulations are affected only by the surface brightness limits, galaxy
profiles, and the galaxy surface densities adopted.  Similar levels
are therefore found for the F814W images, for which galaxy parameters
and detection limits are very similar to those at F555W. The F555W
aperture masks have been used to recover the flux from sources in all
three bands for reasons discussed below. In the very low
signal--to--noise ratio F300W images, only 20\% of the F555W sources
are detected, and the F555W aperture masks for those which are
detected extend to many more than four times the F300W isophotal
detection areas.  Due to the larger statistical uncertainties in the
flux recovered by ensemble photometry at F300W (see Table
\ref{tab:cum.errors}), simulations of the flux beyond the detection
apertures were not warranted.

\begin{deluxetable}{l l l r r r }
\tablewidth{30pc}
\tablecaption{Summary of Measurements
\label{tab:total.summary}}
\tablehead{
\colhead{Source} &\colhead{Bandpass}&\colhead{Data source}
& \colhead{Flux} & \colhead{Random} &\colhead{Systematic }}
\startdata
Total  
 & F300W 	& {\sc hst/wfpc2}
		& 33.5		&$(\pm$ 4.9\%)	&$[\pm$ 5.6\%]\nl
\ \ Background
 & F555W 	& {\sc hst/wfpc2}
		& 105.7 	&$(\pm$ 0.3\%) 	&$[\pm$ 1.4\%]\nl
 & F814W 	&  {\sc hst/wfpc2}
		& 72.4 		&$(\pm$ 0.2\%) 	&$[\pm$ 1.4\%]\nl
 & F555W\tablenotemark{a}
	 	&  {\sc hst/fos}
		& 111.5		&$(\pm$ 0.7\%) 	&$[\pm$ 2.8\%]\nl
Zodiacal
 & 4600--4700\AA& {\sc lco} 
		& 109.4		&$(\pm$ 0.6\%)	&[$\pm$ 1.1\%]	\nl
\ \ Light
 & F300W 	& {\sc lco}\tablenotemark{b} 
		& 28.5		&$(\pm$ 0.6\%)	&[-1.1\%,+1.2\%] \nl
 & F555W 	& {\sc lco}\tablenotemark{b} 
		& 102.2 	&$(\pm$ 0.6\%)	&[-1.1\%,+1.1\%] \nl
 & F814W 	& {\sc lco}\tablenotemark{b} 
		& 69.4		&$(\pm$ 0.6\%)	&[-1.3\%,+1.1\%] \nl
%
Diffuse	
 & F300W & {\sc dgl} model & 1.0	& \nodata  	& [+25\%,-50\%] \nl
\ \ Galactic
 & F555W & {\sc dgl} model & 0.8	& \nodata 	& [+25\%,-50\%] \nl
\ \ Light
 & F814W & {\sc dgl} model & 0.8	& \nodata 	& [+25\%,-50\%] \nl
\enddata
\tablecomments{All fluxes are in units of 1\tto{-9}\escsa.  For a
source with constant flux in $ F_{\lambda}$, filters F300W, F555W, and
F814W have effective wavelengths $\lambda_0(\Delta\lambda)$
=3000(700), 5500(1200), and 8100(1500)\AA. For a source with a solar
spectrum, effective wavelengths are $\lambda_0$ = 3200, 5500,
8000\AA.}
\tablenotetext{a}{Observed FOS spectrum, convolved with the
WFPC2/F555W bandpass to allow direct comparison with the WFPC2
results.}
\tablenotetext{b}{LCO measurement of zodiacal light, 
extrapolated to the WFPC2 bandpass by applying a correction for
changing zodiacal light color with wavelength relative to the solar
spectrum.  The zodiacal light flux through the WFPC2 bandpasses was
identified using 
SYNPHOT models, the uncertainty due to which is included in the
uncertainty for the filter calibration and is shared with the
systematic uncertainty for the total background flux.}
\end{deluxetable}


\begin{deluxetable}{lcccr}
\tablewidth{30pc}
\tablecaption{EBL Results and Uncertainties
\label{tab:cum.errors}}
\tablehead{
\colhead{Bandpass}& 
	\colhead{Random} &
	\colhead{Systematic}& 
	\colhead{Combined}&
	\colhead{EBL($\pm1\sigma$)}\\ 
\colhead{} &
	\colhead{$\sigma_R$ (68\%)} &
	\colhead{$\sigma_S$ (68\%)} &
	\colhead{$\sigma$ (68\%)} &
	\colhead{} }
\startdata
\multicolumn{5}{l}{\underline{Detected EBL23 (WFPC2 + LCO)}\tablenotemark{a}} \nl
F300W\hspace{0.5in}     & 2.1	& 1.5           & 2.5	& 4.0  $(\pm 2.5)$  \nl
        F555W           & 0.6	& 1.3           & 1.4	& 2.7  $(\pm 1.4)$  \nl
        F814W           & 0.4	& 0.9		& 0.0	& 2.2  $(\pm 1.0)$  \nl
\multicolumn{5}{l}{\underline{Minimum EBL (WFPC2)}\tablenotemark{a}} \nl 
        F300W           & 0.19	& 0.13		& 0.22	& 3.2  $(\pm 0.22)$ \nl
        F555W           & 0.003	& 0.009		& 0.01	& 0.89 $(\pm 0.01)$ \nl
        F814W           & 0.002	& 0.007		& 0.01	& 0.76 $(\pm 0.01)$ \nl
\multicolumn{5}{l}{\underline{Detected EBL23 (FOS + LCO)}\tablenotemark{a}} \nl
        F555W           & 0.7	& 2.7           & 2.8	& 8.5  $(\pm 5.6)$  \nl
\multicolumn{5}{l}{\underline{Flux from detected sources 
				in HDF ($m>23$ AB mag)}} \nl
        F300W           &	&		&	& 0.66	 	\nl
        F450W           &	&		&	& 0.51	 	\nl
        F606W           &	&		&	& 0.40	 	\nl
        F814W           &	&		&	& 0.27	 	\nl
\multicolumn{5}{l}{\underline{Published number counts}\tablenotemark{b}}\nl
        \multicolumn{3}{l}{F300W ($18<U_{300}<23$ AB mag)}
						&	& 0.27 $(\pm 0.05)$ \nl
        \multicolumn{3}{l}{F555W ($15<V_{555}<23$ AB mag)}
						&	& 0.49 $(\pm 0.10)$ \nl
        \multicolumn{3}{l}{F814W ($13<I_{814}<23$ AB mag)}
						&	& 0.65 $(\pm 0.13)$ \nl
 \enddata
\tablecomments{All fluxes and errors are given in units of $10^{-9}$
\escsa.}  
\tablenotetext{a}{The systematic and statistical errors have been
combined assuming a flat or Gaussian probability distribution,
respectively,  as discussed in \S\ref{ebl.detec}.  We equate
$1\sigma$  combined errors with the 68\% confidence interval, as the
combined errors are nearly Gaussian distributed. Individual sources of
error contributing to these totals are summarized in Tables 3 and 4 of
this paper  and in  Table 1 of Paper II.}
\tablenotetext{b}{Estimated errors correspond to 
uncertainties in the fits to published galaxy counts.  The values
given correspond to $0.081\times10^{-20}$, $0.46\times10^{-20}$, and
$1.5\times10^{-20}$ in units of \escsh\ and are consistent with those
used in Pozzetti et al.\ (1998).}
\end{deluxetable}

By using different values for $V_{\rm cut}$ (23, 24, 25 and 26 AB
mag), we can isolate the flux contributed from galaxies in successive
magnitude bins.  Comparing these measurements with the integrated flux
from standard photometry methods (see \S\ref{wfpc2.objec}), we find
that roughly 15\%, 25\%, 45\% and 65\% of the total flux in successive
1 magnitude bins between $23<V_{555}<27$ is contained at radii between
$\sqrt{2}r_{\rm iso}$ and $4r_{\rm iso}$.  In our data, galaxies with
magnitudes $23<V_{555}<27.5 $ have $1.5<\Delta\mu<3.5$ mag
arcsec$^{-2}$ as shown in Figure \ref{fig:detec.rabmumumag}.  These
measurements are in broad agreement with the models in the literature
for the fraction of flux which can be recovered as a function of
$\Delta\mu$ based on extrapolation of simple exponential profiles
(c.f. Disney \& Phillipps 1983, Davis 1990, Dalcanton 1998).

The total flux in resolved galaxies as measured by ensemble photometry
has two obvious implications for galaxy counts.  First, the total flux
in galaxy counts based on standard photometry will significantly
underestimate the EBL.  This point is discussed further in the
following sections. Second, because standard photometry will miss
fractionally more light from the faintest galaxies, the galaxy counts
which result will have an artificially shallow slope at the faint end.
Our results allow us to derive ``aperture corrections'' for faint
galaxy photometry as a function of central and isophotal surface
brightness.  This is done in Paper III. We can use these corrections
to recalculate the surface number density of galaxies as a function of
magnitude.  As we discuss in detail in Paper III, the corrected number
counts do not flatten out at the faintest limits of the HDF.

\subsection{Discussion: Minimum EBL23}\label{minebldiscussion}

Using the ensemble photometry method described above, we measure the
flux from detectable sources relative to the mean sky level in the
F300W, F555W, and F814W images of our own data to be 3.19\tto{-9},
6.02\tto{-10}, and 5.16\tto{-10}, respectively.\footnote{For
comparison, the flux measured by standard photometry for sources with
AB magnitudes between 23 and 27.5 in the HDF are 6.7, 5.1, 3.8, and
2.5\tto{-10}\escsa\ in the F300W, F450W, F606W, and F814W,
respectively.}  For the F300W and F814W images, we use the masks
derived from the F555W images in order to guarantee that the same
sources are contributing to the minimum EBL23 at all wavelengths.  The
1800 galaxies detected in the HDF with $V_{606}$ magnitudes in the
range 27.5--30\,AB\,mag (see Figure \ref{fig:wfpc2.glxycnts}) are
clearly not included in our estimate of the minimum EBL23 derived by
ensemble photometry.

The contribution from galaxies in the HDF catalog with
$V_{555}>27.5$\,AB\,mag is 0.57\tto{-10}\escsa\ at $V_{555}$ as
measured by standard photometric methods.  In keeping with the
discussion in the previous section, we estimate that only 35\% of the
light is recovered from galaxies with $27.7<V_{606}<30$\,AB\,mag in
the HDF, so that the true flux from these sources is roughly
1.8\tto{-10}\escsa.  Adding this to the flux from detected sources
($23 < V_{555} <27.5 $) in our EBL field, we find the total flux from
detected sources ($23<V_{555}<30$) to $4r_{\rm iso}$ apertures to be
7.8\tto{-10}\escsa.  Adding to this the estimated extragalactic
contribution to the background sky level beyond $4r_{\rm iso}$ (Figure
\ref{fig:sky.hist.ebl}), we identify the minimum flux from detected
galaxies with $V_{555}>23$\,AB\,mag to be 8.9\tto{-10}\escsa.
We emphasize that this estimate of the minimum EBL23 is indeed a
minimum estimate from which sources will be excluded due to the
surface brightness biases and detection limits of our own images as
well as the HDF (see Figures \ref{fig:detec.rabmumumag} and
\ref{fig:detec.hdfmumumag}).  Incompleteness due to surface brightness
detection limits is discussed further in Paper III.

Following the same method for F300W and F814W data leads to the
minimum EBL23 values summarized in Table \ref{tab:cum.errors}.  The
total combined statistical and systematic error for this minimum EBL23
measurement is roughly $\pm 1$\tto{-11} {\escsa}, $\times 100$ smaller
than the error for the EBL23 detections because the conversion to
physical units can occur after foregrounds are subtracted.  Comparing
the total flux from detected galaxies measured using ensemble
photometry versus standard methods, the photometry errors affecting
standard methods are clearly worse at UV wavelengths where
signal--to--noise ratio is lower, calibration is less accurate, and
the instrumental PSF is broader than in the other filters.

\section{\uppercase{EBL Detections}}\label{ebl.detec}

\begin{figure}[t]
\begin{center}
\includegraphics[width=3.0in,angle=0]{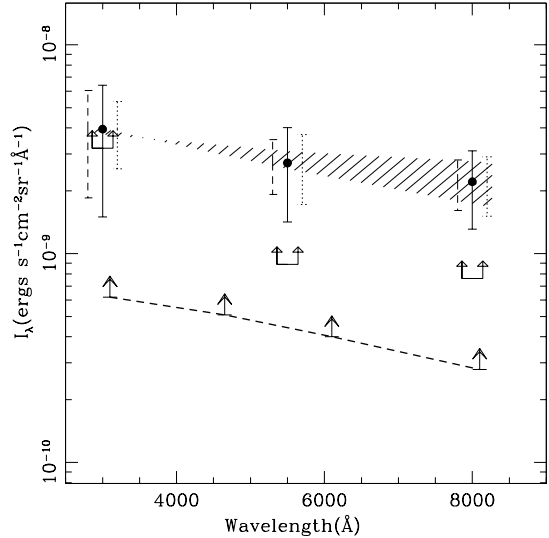}
\caption{\footnotesize Filled circles show the EBL23 obtained using
the HST/WFPC2 EBL images, the ZL flux measured at (4650\AA), and the
nominal ZL color as described in \S\ref{ebl.detec}.  The solid error
bars (centered on the filled circles) show the combined $1\sigma$
errors while the dotted (offset by +100\AA) and long--dashed (offset
by -100\AA) error bars show $1\sigma$ systematic and random errors,
respectively.  The hatch--marked region shows the $1\sigma$
uncertainty in the detected EBL due to uncertainty in ZL color.  The
lower limit arrows connected by a dashed line indicates the total flux
from individually photometered galaxies with magnitudes $23<V_{555}
<30$\,AB\,mag in the HDF catalog. The u-shaped lower limit arrows show
minEBL23, the flux from detectable galaxies in the EBL fields with $23
<V_{555}\leq28$\,AB\,mag as determined by ensemble photometry as
described in \S\ref{ebl.minim}.}
\label{fig:ebl.min.cnts}
\end{center}
\end{figure}

\begin{figure}[t]
\begin{center}
\includegraphics[width=3.0in]{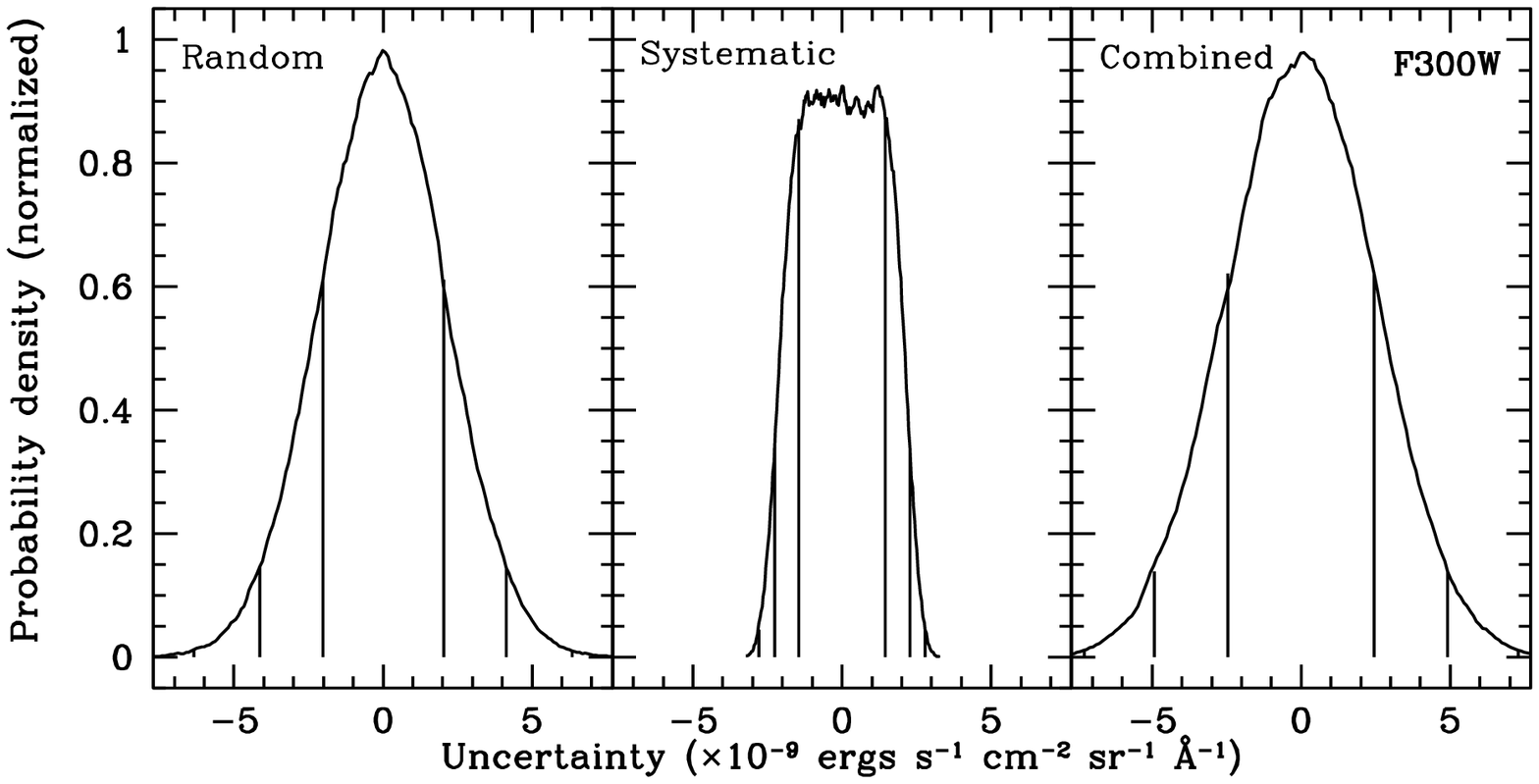}
\includegraphics[width=3.0in]{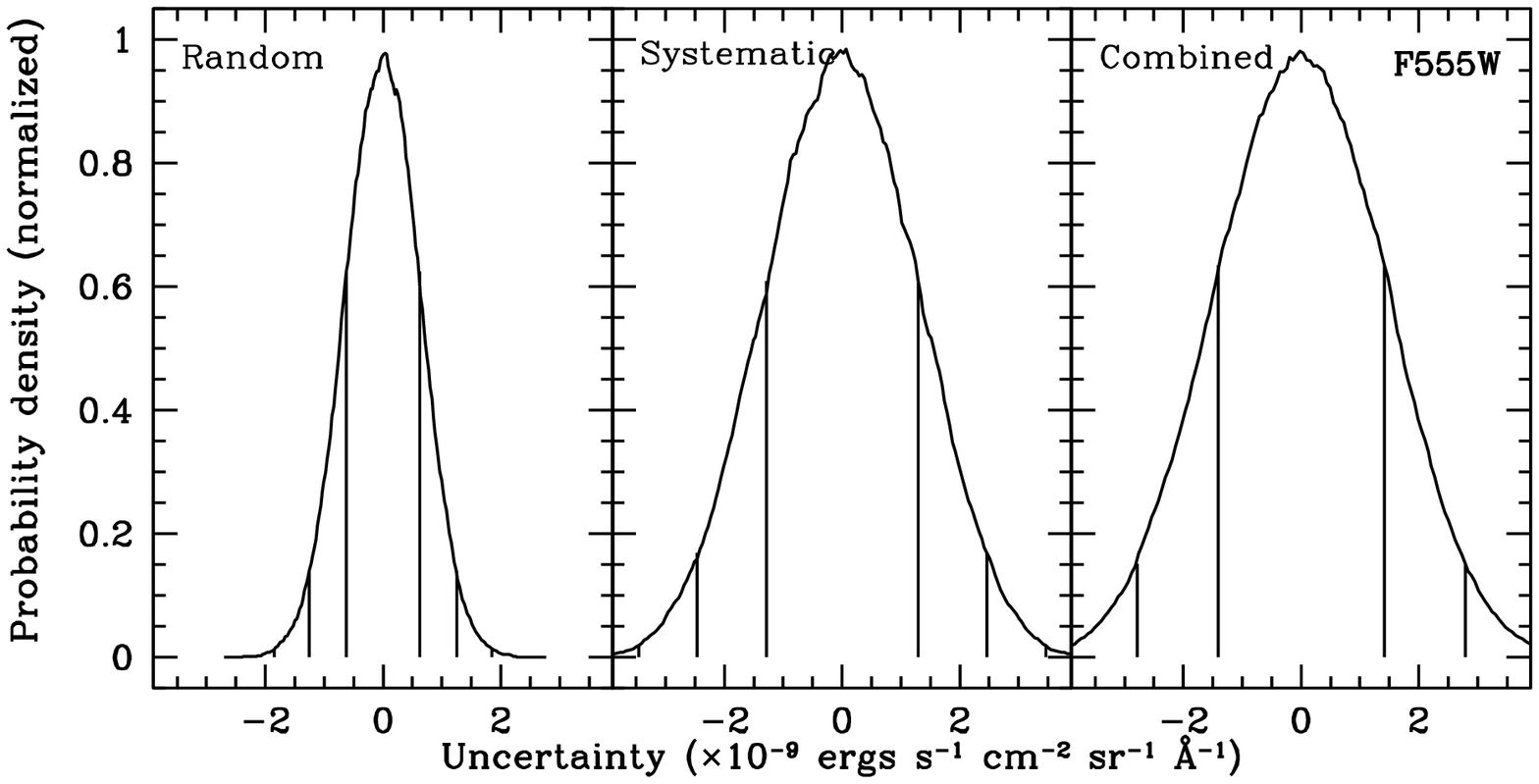}
\includegraphics[width=3.0in]{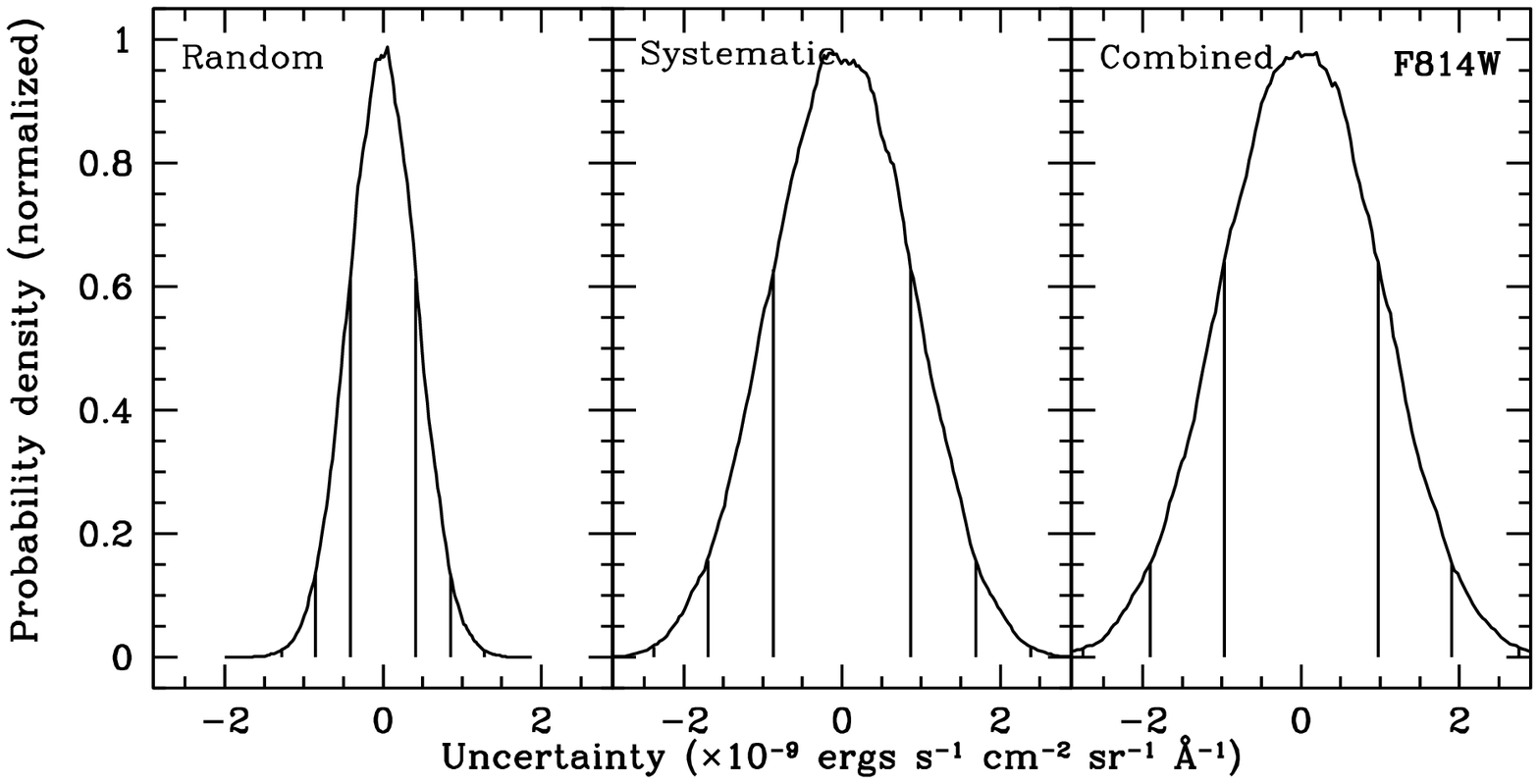}
\caption{\footnotesize
 Combined probability distributions of the
errors (systematic, random, and combined) for the EBL flux detected in
the F300W (panel a), F555W (panel b), and F814W (panel c) WFPC2 bandpasses.
A flat (Gaussian) probability distribution was assumed for each
contributing systematic (random) error.  The vertical lines show
68.3\%, 95.4\%, and 99.7\% confidence intervals, which we use to
define 1, 2 and 3$\sigma$ values for the combined errors.}
\label{fig:cum.errors}
\end{center}
\end{figure}


\begin{figure}[t]
\begin{center}
\includegraphics[width=3.0in,angle=0]{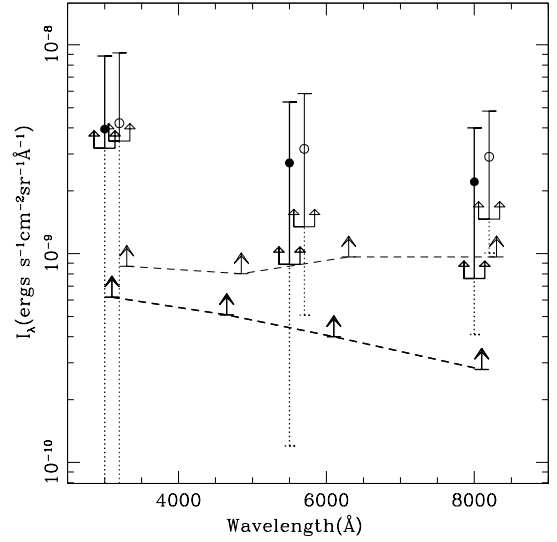}
\caption{\footnotesize 
The detected background light from galaxies
fainter than $V_{555}=23$\,AB\,mag is marked by large filled circles
with 2$\sigma$ error bars.  
The error bars are dotted where they extend below the flux recovered
from galaxies in the range $23<V_{555}<28$\,AB\,mag by the aperture
photometry method (see \S\ref{ebl.minim}) shown by the u-shaped lower
limit symbols.
The integrated flux from galaxies in the HDF, from 23\,AB\,mag to the
detection limit at each wavelength (29--30\,AB\,mag), is marked by
lower limit arrows connected by the thick, dashed line.  
The total flux in galaxy counts, HDF counts plus ground--based counts
brighter than $V_{555}=23$\,AB\,mag is shown by lower limit arrows
connected by a thin dashed line.  
The open circles (displaced by 200\AA\ for display purposes)
show the EBL we detect, plus the integrated ground--based counts
brighter than $V_{555}=23$\,AB\,mag. We have not included
uncertainties in the ground-based counts in the error bars shown.}
\label{fig:ebl.hdf.grnd}
\end{center}
\end{figure}

The detected EBL in each of the three WFPC2 bands is the difference
between the total sky flux in the WFPC2 images, $I_{\rm Total}$, and
the foreground flux from ZL and DGL.  As described in
\S\ref{dgl.struc},  the DGL contribution has no spectral features,
is relatively flat in $I_\lambda$, and  is easily
subtracted.  The spectrum of the zodiacal light is uniquely defined by
our measurement of the mean flux of the ZL at 4650\AA, the solar
spectrum, and the color of the ZL as follows:
\begin{equation}
I_{\rm ZL}({\lambda}) = I_{\odot}(\lambda)\;C(\lambda,4650)\;
		\frac{I_{\rm ZL}(4650)}{I_\odot(4650)}.
\label{eq:izl}
\end{equation}
(See the Appendix A for a discussion of the solar reference spectrum,
$I_{\odot}$, used and the convolution of $I_{\rm ZL}(\lambda)$ with the
WFPC2 bandpasses to obtain the ZL flux through each filter.)

Both the mean flux and color of the zodiacal light in the range
3900--5100\AA\ are uniquely determined by our LCO data.  However, the
color over the wavelength range of the WFPC2 data is less certain.
From our LCO data, we find $C(5100,3900)=1.05\pm0.01$ from absolute
surface spectrophotometry of the ZL.  As discussed in
\S\ref{fos.resul}, we find $C(7000,4000)=1.044\pm0.01$ from our FOS
spectra. Although this is in excellent agreement with the LCO results,
extragalactic contributions are included in the FOS spectrum.  We have
therefore devised a method to further constrain the ZL color using the
WFPC2 results.

Between 2000\AA\ and 1{\micron}, the color of the zodiacal light
relative to the solar spectrum as a function of wavelength is
empirically known to be roughly linear with wavelength; the mean flux
with respect to the Sun becomes redder by roughly 5\%/1000\AA\ over
the range from 2500\AA--1.4\micron.  No deviations from a linear color
are apparent in our FOS spectra, which are the first observations to
have the required broad--band calibration sensitivity to address this
question.  To constrain the ZL and EBL surface brightness in all three
bands, we therefore begin by assuming that the color of the ZL is a
linear function of the solar spectrum with wavelength and that any
small deviations from linearity average out over the 1000\AA\
bandwidths of the WFPC2 filters. 

Only if the EBL is flat in $I_\lambda$ would the color of the EBL plus
ZL be the appropriate color to adopt for the ZL.  If, for example, the
EBL is blue in $I_\lambda$, then the ZL is slightly redder.  At 3000,
5500, and 8000\AA, we have already identified minima for the EBL. If
we adopt these minima for the EBL at 3000 and 8000\AA, we can
obviously infer that the maximum possible flux for the ZL is what
remains when we subtract the DGL and the minimum EBL from the total
surface brightness detected in each band.  By considering the maximum
possible $I_{\rm ZL}$ at 3000 and 8000\AA, and comparing those maxima
to the absolute flux which we have measured at 4650\AA, we constrain
the color of the ZL to be 3.0\%/1000\AA\ $< C(\lambda,\lambda_o) <
$5.0\%/1000\AA. This estimate is in excellent agreement with the color
we find for the total background as measured by FOS, which was
4.4\%/1000\AA\ redder than solar.  Using these extrema for the color
of the ZL, we can constrain the EBL at all three WFPC2 wavelengths
(see Figure \ref{fig:ebl.min.cnts}).  We adopt this range as the
$3\sigma$ uncertainty in the ZL color.

We calculate final statistical and systematic uncertainties
from the WFPC2 and LCO measurements assuming a Gaussian probability
distribution for each source of random error listed in Tables
\ref{tab:wfpc2.errors} and \ref{tab:wfpc2.results} 
and in Table 1 of Paper II.  The final systematic
uncertainties are calculated assuming a flat probability distribution
for the systematic uncertainties listed in the same tables.
Cumulative random and systematic errors are quoted and plotted
separately to give a sense of the measurement limitations.  In
general, random errors reflect limitations in instrument sensitivity
and stability, while systematic errors are dominated by stability of
the flux calibration of each instrument and the accuracy of the flux
calibration for the instruments relative to each other.  Uncertainty
in the DGL subtraction is not explicitly included in these combined
errors, as the total DGL flux is relatively small and the error would
not contribute significantly to the total error.  The DGL estimate was
intended to be conservatively large, so that the EBL is, if anything,
underestimated due to errors in the DGL.

We have also calculated combined errors (systematic and statistical)
to obtain a final confidence interval. For this we have again assuming
a Gaussian probability distribution for all of the random sources of
error and a flat probability distribution for the systematic
uncertainties. The combined error probability distributions are shown
in
Figure \ref{fig:cum.errors}.  In general, non-Gaussian systematic
errors and Gaussian random errors cannot meaningfully be combined in
quadrature.  Nonetheless, given the large number (16) of individual
sources of uncertainty contributing to our final errors, the combined
uncertainty has a nearly Gaussian distribution, as expected.  
We therefore equate the $1\sigma$ combined
errors for each bandpass with the 68\% confidence intervals for the
mean EBL23 detections and the lower limits, minEBL23.  The final
errors are summarized in see Table \ref{tab:cum.errors}.

\section{\uppercase{Summary}}\label{summa}

We summarize our detection of the surface brightness of  EBL23
from resolved and unresolved galaxies fainter than
$V_{555}=23$\,AB\,mag as follows (in {\escsa}):
\begin{tabbing}
$I_\lambda$(F300W)= 4.0 $(\pm2.5)$ \tto{-9} \\
$I_\lambda$(F555W)= 2.7 $(\pm1.4)$ \tto{-9} \\
$I_\lambda$(F814W)= 2.2 $(\pm1.0)$ \tto{-9}. \\
\end{tabbing}
The quoted errors are $1\sigma$ combined uncertainties (statistical
and systematic) corresonding to 68\% confidence intervals, as
described in \S\ref{ebl.detec}.  We can also define strict lower
limits to the EBL from detected galaxies fainter than
$V_{555}=23$\,AB\,mag by computing the flux from all detected objects
using the ``aperture photometry'' method described in
\S\ref{ebl.minim}. The lower limits for EBL23 are (in {\escsa}):
\begin{tabbing}
 $I_{23<V<27.5}$(F300W) $\ge 3.2$  $(\pm 0.22)$ \tto{-9} \\
 $I_{23<V<27.5}$(F555W) $\ge 0.89$ $(\pm 0.01)$ \tto{-9} \\
 $I_{23<V<27.5}$(F814W) $\ge 0.76$ $(\pm 0.01)$ \tto{-9}.
\end{tabbing}

For comparison with predictions of the EBL based on the local metal
mass density and total star formation history of the universe, the
flux from galaxies brighter than $V_{555}=23$\,AB\,mag should be added
to these results, as shown in Figure \ref{fig:ebl.hdf.grnd}.  We
discuss these comparisons in Paper III.

\acknowledgments

It is a pleasure to thank A.\ Fruchter, R.\ Lyons, C.\ Keyes, A.\
Koratkar, L.\ Petro and D.\ Van Orsow for help in planning and
scheduling the HST observations, and J.\ Christensen, H.\ Ferguson,
and J.\ Keyes for help in understanding and improving the standard
pipeline reduction for both the WFPC2 and FOS data.  S.\ Baggett, S.\
Cassertano , R.\ Bohlin, and E.\ Smith also provided help with the
WFPC2 and FOS calibration. We have also benefited greatly from
discussions with J.\ Dalcanton, S.\ Shectman, T.\ Small, I.\ Smail,
J.\ Trauger, and B.\ Weiner.  R.\ Kurucz kindly provided an electronic
version of the Solar Atlas. Finally, we would like to thank Carnegie
Observatories and specifically L.\ Searle, A.\ Oemler, and I.\
Thompson for generous allocation of observing time at Las Campanas
Observatory; the referee, R. Windhorst, for helpful comments; R.\
Blandford, A.\ Readhead, and W.\ Sargent for financial support to RAB
during the first year of this work; and especially R. Williams for his
support of this project.  This work was supported by NASA through
grants NAG LTSA 5-3254 and GO-05968.01-94A to WLF.

\appendix
\setcounter{figure}{0}

\section{\uppercase{Mean Zodiacal Light Flux Through HST Filters}}

The fiducial solar spectrum we have used for the purposes of creating
a low--resolution spectrum of the zodiacal light from
2500\AA--1\micron\ is a composite of the UV solar spectrum of Woods
\etal (1997), the optical spectrum from NL84, and the infrared
spectrum produced by Arvesen \etal (1969), as recommended by Colina,
Bohlin \& Castelli (1996).  The accuracy of the absolute flux of this
solar spectrum is irrelevant to the accuracy of the zodiacal light
spectrum, $I_{\rm ZL}({\lambda})$; the absolute flux of $I_{\rm
ZL}({\lambda})$ is defined by the measured flux of the ZL in our own
LCO spectra at 4650\AA, and by a combination of that measurement plus
the color term, $C(\lambda,4650\AA)$, at all other wavelengths. The
color term itself is simply an empirical description of color of
whatever fiducial solar spectrum we adopted relative to the observed
color of the zodiacal light as measured in our FOS data. Thus, the
absolute flux of $I_{\rm ZL}({\lambda})$ is defined by the accuracy of
the broad--band flux calibration of the FOS and LCO spectra, and the
accuracy of the ZL measurement in the LCO spectra, as described in
Paper II.  The accuracy of the ZL measurement is, or course, dependent
on the Solar Flux Atlas as discussed in Paper II.

The spectrum $I_{\rm ZL}(\lambda)$ as expressed in equation
\ref{eq:izl} is then an absolute flux-calibrated spectrum of the
ZL, which we can convolve with the SYNPHOT throughput tables (using
the version released in May 1997) in the usual way to obtain the
absolute flux of the ZL through each of the filters.  The flux through
the WFPC2 band is given by
\begin{equation}
 I_{\rm WF}(\lambda) = \frac{\int{
T(\lambda)\;I_{\lambda}(\lambda)\;\lambda\;d\lambda}} {\int
T(\lambda)\;\lambda\; d\lambda},
\label{eq:ebl.flux.in.band}
\end{equation}
in which $T(\lambda)$ is the effective throughput of a WFPC2 filter
(including telescope and detector efficiencies),
$I_{\lambda}(\lambda)$ is the spectrum of the zodiacal light, and all
spectra are in units of {\escsa}. No additional error results
from convolving the flux calibrated ZL with the bandpasses that
define the WFPC2 system, as any error in the SYNPHOT synthetic
photometry is incorporated in our estimate of the WFPC2 systematic
uncertainty.  See Paper II for a discussion of the LCO measurement of
$I_{ZL}(4650\AA)$.

\section{\uppercase{
Flux from the low surface brightness wings of detected galaxies}}
\setcounter{figure}{0}

In order to estimate the contribution from the wings of detected
galaxies to the mean sky flux beyond $4r_{\rm iso}$ in a particular
image, we have constructed a Monte Carlo simulation.  In this
simulation, we sum the cumulative flux contributed by randomly placed,
detectable galaxies to a given point on the sky, such that the total galaxy
population simulated reproduces the appropriate surface number density as a
function of magnitude.  If a randomly placed galaxy is close enough to
the ``sky pixel'' that the pixel would fall within the detection
aperture of the galaxy, then the trial is rejected and another begins.
The extent of the galaxy apertures (how close galaxies can be to the
sky pixel in question before that pixel falls within the galaxy's
detection aperture) are determined based on the mean ``foreground''
sky level and noise characteristics assumed for the image being
simulated.  The simulation continues until we obtain 10,000 ``sky
pixels.''

\begin{figure}[t]
\begin{center}
\includegraphics[width=3.0in,angle=0]{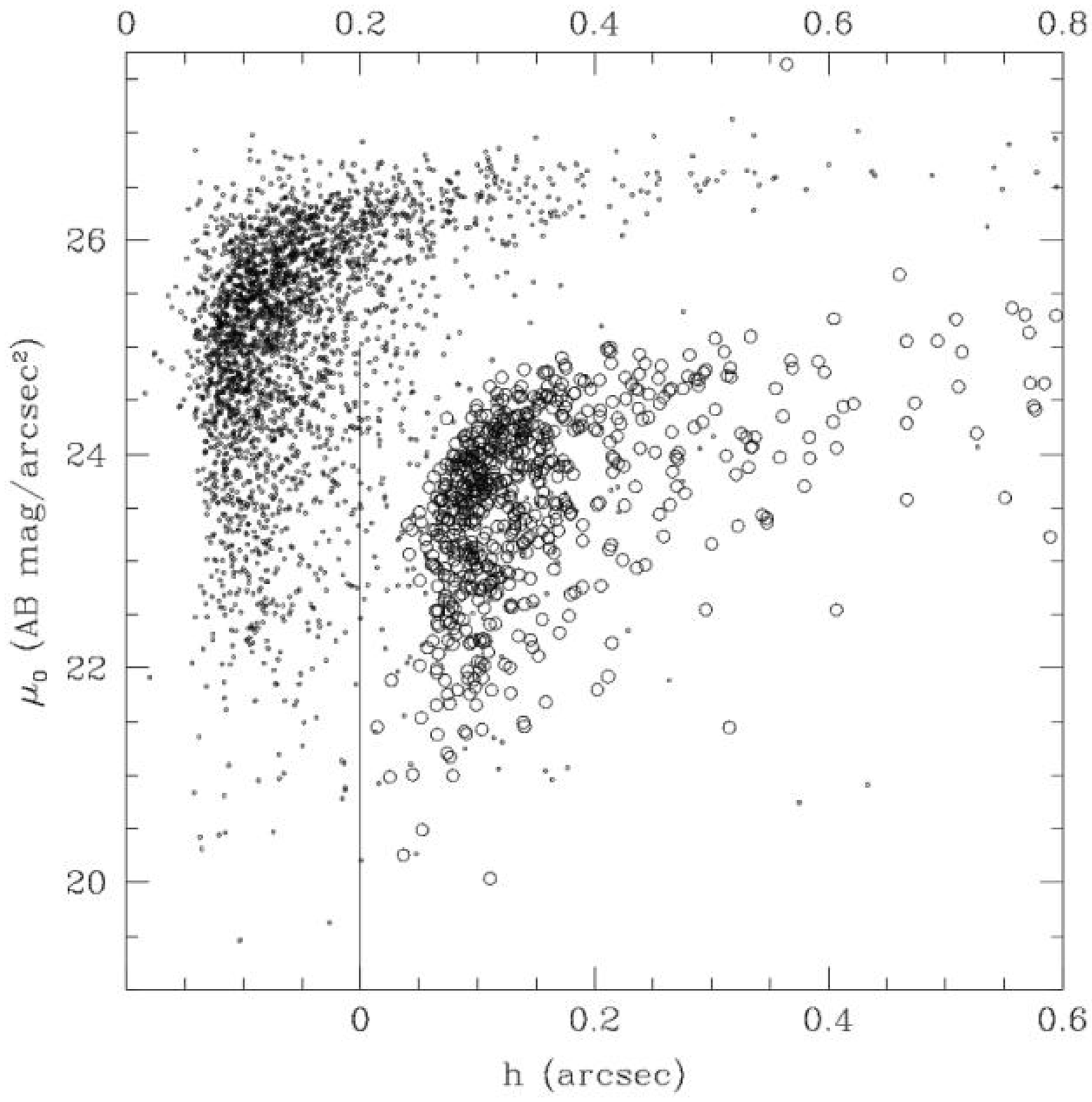}
\caption{\footnotesize
 Fitted values of scale length, $h$, and central surface
brightness, $\mu_0$, for galaxies from the Figures
\ref{fig:detec.rabmumumag} (EBL data, open circles) and
\ref{fig:detec.hdfmumumag} (HDF data, points) assuming exponential
light profiles.  The HDF galaxies are plotted relative to the X-axis
at the top of the plot; EBL galaxies are relative to the lower X-axis
for clarity. The feature in both data sets which trails off to high
$h$ at constant $\mu_0$ corresponds to galaxies at the detection limit
for which $\mu_{\rm iso}-\mu_0\lta 1$.}
\label{fig:detec.mu0a}
\end{center}
\end{figure}

As faint galaxies are weakly clustered on small projected scales
(Colley \etal 1996, 1997; Roche \etal 1993), we have simply placed
galaxies randomly in this simulation. For the surface density of
galaxies as a function of apparent magnitude, we have adopted the HDF
galaxy counts at $V>23$AB mag, which go 1.5 mag fainter than our own
data (see Figure \ref{fig:wfpc2.glxycnts}). We have assumed that the
observed light profiles of faint galaxies are adequately described by
an exponential profile, $\mu_r = \mu_0 \exp{(-r/h)}$, where $\mu_0$ is
the central surface brightness and $h$ is the scale length.  Using the
measured values of the core surface brightness in the central 4
(undrizzled) pixels, $\mu_{\rm core}$, the isophotal surface brightness,
$\mu_{\rm iso}$, and the isophotal radius, $r_{\rm iso}$ for the HDF
galaxies, we can determine $\mu_0$ and $h$ (see Figures
\ref{fig:detec.rabmumumag} and \ref{fig:detec.hdfmumumag}).  As can be
seen in Figures \ref{fig:detec.mu0a} and \ref{fig:detec.mu0b}, $h$ and
$\mu_0$ as a function of magnitude are consistent for our data and the
HDF images, given the relative surface brightness limits of each.

\subsection{ EBL images}

\begin{figure}[t]
\begin{center}
\includegraphics[width=3.0in,angle=0]{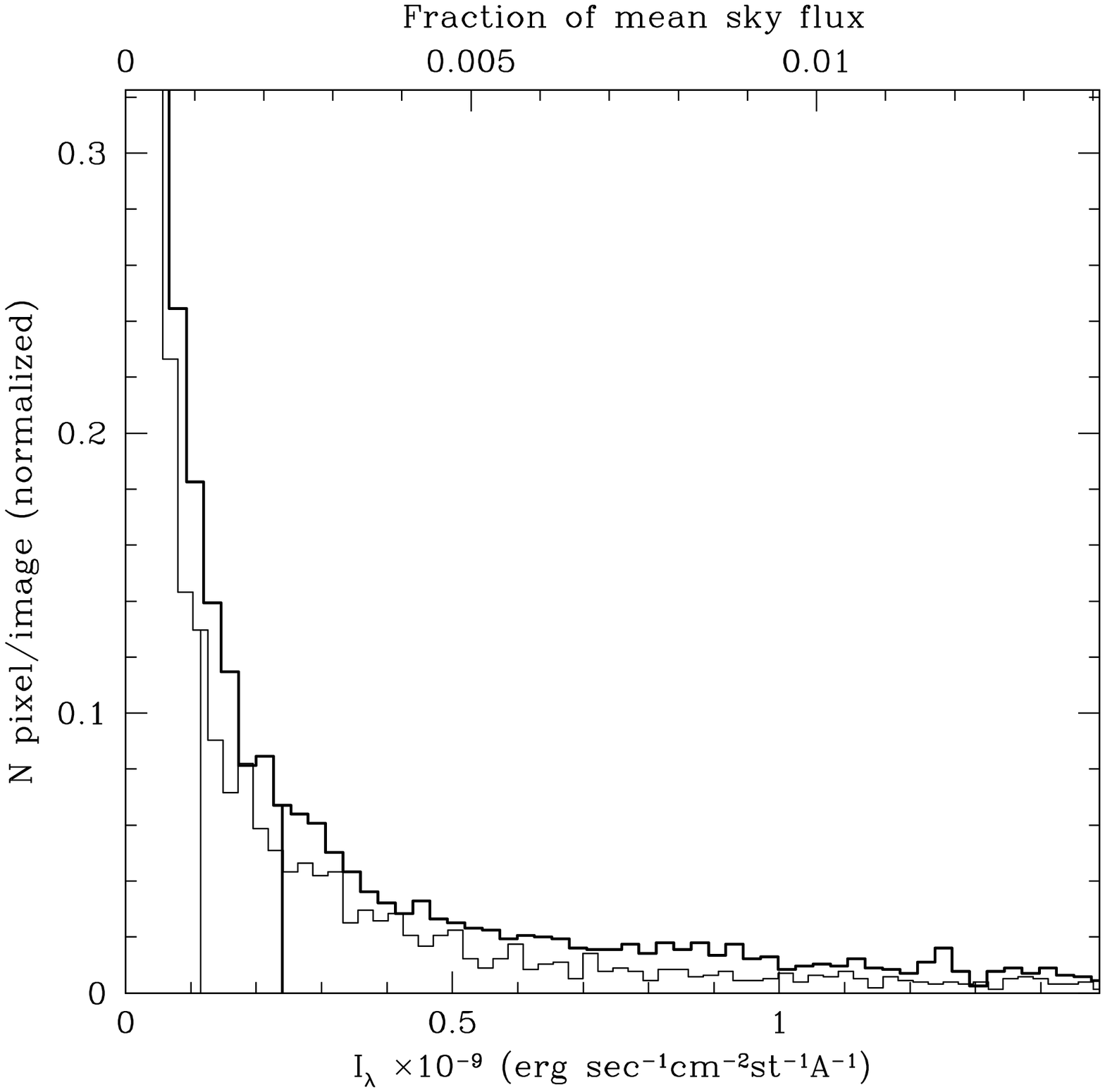}
\caption{\footnotesize
 Histogram of flux per ``sky pixel'' found in
two Monte Carlo simulations described in \S\ref{ebl.minim}.  Each
trial models the flux contributed to a random point on the sky by
nearby galaxies.  The lower X-axis shows flux in cgs units, while the
top X-axis shows the flux as a fraction of the the mean foreground sky
flux (dominated by ZL) at the levels observed in the EBL WFPC2 field.
Surface brightness limits and sky noise levels used in this
simulation to define the detection region around the simulated sources
reflect the parameters of the $V_{555}$ EBL images.  The thick line
shows the sky pixel histogram which results from extending the
``detection'' aperture around simulated source to the $1.4r_{\rm
iso}$. The thin line shows the histogram associated with detection
apertures extending to $4r_{\rm iso}$.  Vertical lines mark the
mean ``sky'' flux identified in each simulation.}
\label{fig:sky.hist.ebl}
\end{center}
\end{figure}


In Figure \ref{fig:sky.hist.ebl}, we plot histograms of the absolute
flux contributed to 10,000 sky pixels at $\approx 5500$\AA\ in two
simulations corresponding to the surface brightness limits of our
F555W EBL images, $\mu_{\rm iso}= 1\sigma_{\rm sky}=25.6 V_{555}$ AB
mag arcsec$^{-2}$.  In the first simulation, we define the regions
associated with galaxies by the standard aperture size used for
``total'' magnitudes, $\sqrt{2}r_{\rm iso}$. In the second simulation,
we extend the galaxy apertures to $4r_{\rm iso}$.  In the first
simulation, we find that the flux from wings of galaxies contributes
2.3\tto{-10}\escsa\ to the mean ``sky'' flux.  As more galaxy light is
excluded from the sky mean in the second simulation, the mean level of
simulated ``sky'' pixels drops by just over 50\% to
1.1\tto{-10}\escsa.  The difference in flux in the ``sky''pixels is
within 10\% of the flux we measured in the region $1.4 - 4 r_{\rm
iso}$ by ensemble aperture photometry, giving us confidence in our
estimate of the flux coming from beyond $4r_{\rm iso}$.  Note that the
total flux from beyond the standard galaxy photometry apertures,
2.3\tto{-10} \escsa, is roughly $<0.5$\% of the mean sky level from
diffuse zodiacal and galactic foregrounds, as indicated on the top
x-axis of Figure \ref{fig:sky.hist.ebl}.  Note further that while the
extragalactic pedestal from galaxies outside of  the $4 r_{\rm iso}$
detection apertures is only 0.1\% of the ZL plus DGL foreground flux
and 5\% of the detected EBL in \S\ref{ebl.detec}, it is 20\% of the
{\it recovered} flux from galaxies with $V>23$ \ABmag\ by standard methods of
galaxy photometry.

\subsection{Comparison with the detection limits of the HDF}
\begin{figure}[t]
\begin{center}
\includegraphics[width=3.0in,angle=0]{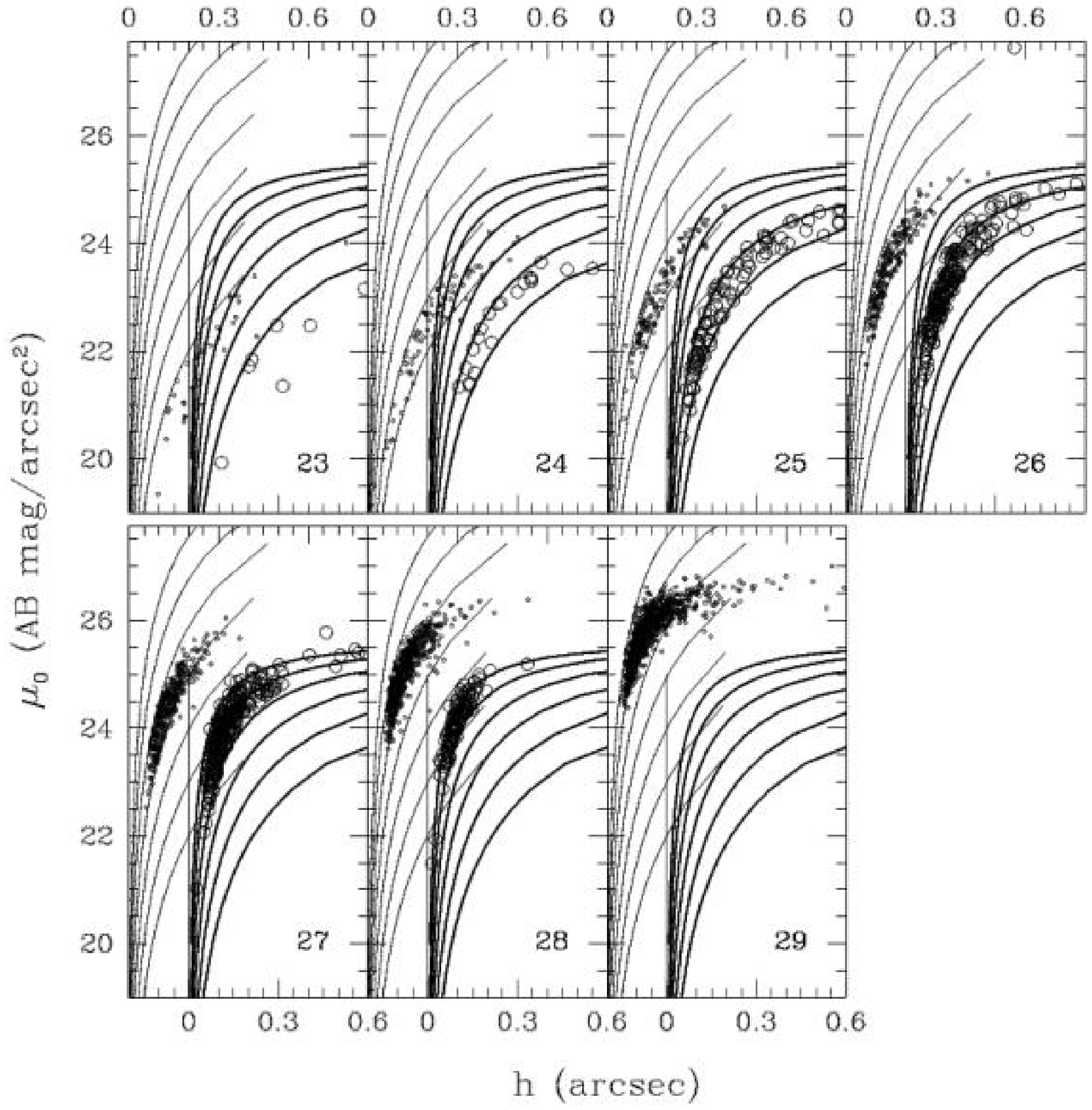}
\caption{\footnotesize
Each panel shows fitted $h$ and $\mu_0$ for
galaxies in the HDF and EBL fields in unit magnitude bins. Axes and
symbol types are as in Figure \ref{fig:detec.mu0a}.  Lines show the
$\mu_0$ and $h$ limits for galaxies with $m=23, 24, ...29$. Galaxies
in each magnitude bin can be seen to lie within the $\mu_0,h$ relation
for that magnitude range with the exception of galaxies near the 
surface brightness limits
of the data, for which profile solutions become ill-defined.}
\label{fig:detec.mu0b}
\end{center}
\end{figure}

\begin{figure}[t]
\begin{center}
\includegraphics[width=3.0in,angle=0]{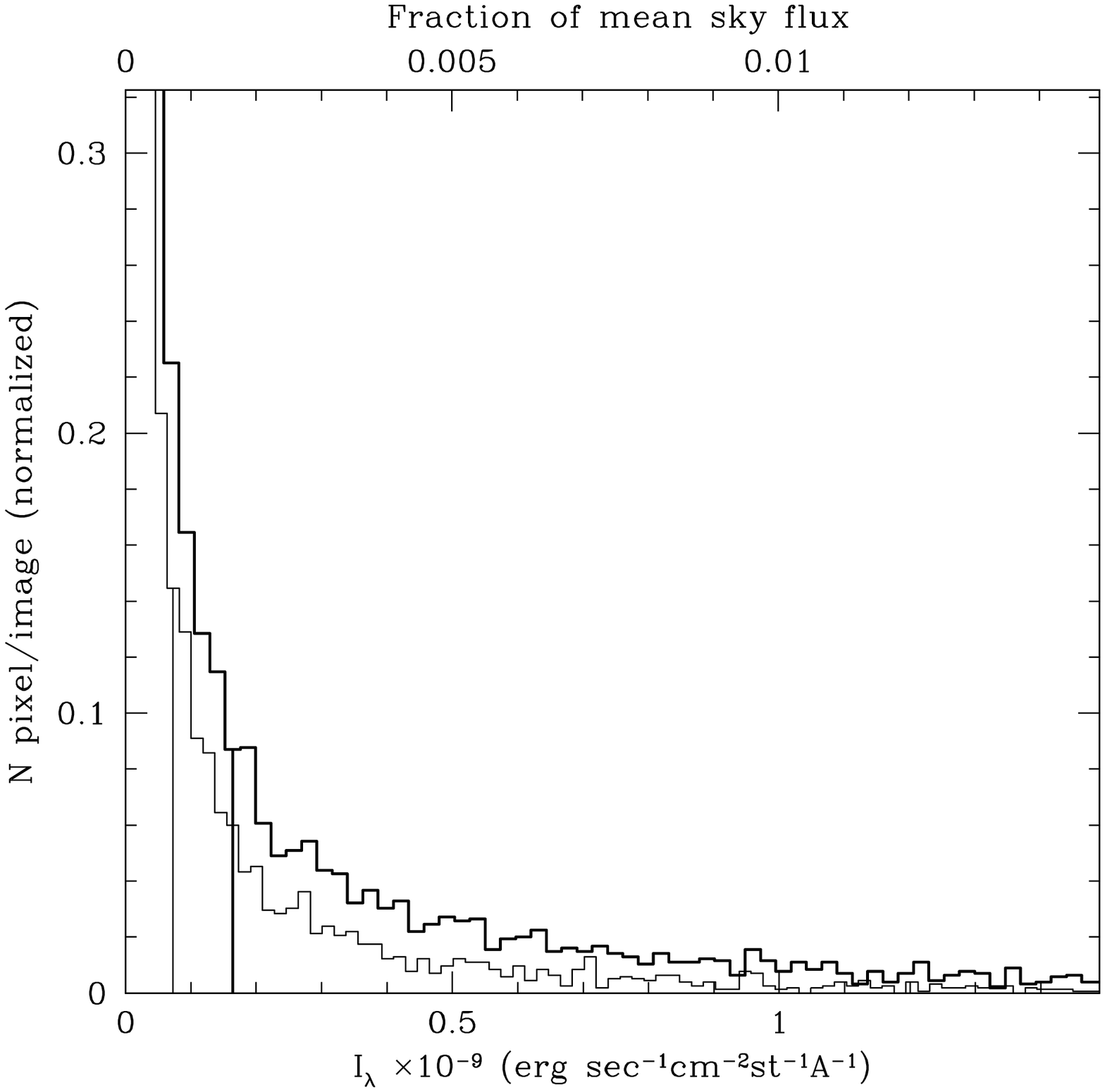}
\caption{\footnotesize
 Same as Figure \ref{fig:sky.hist.ebl}, but for
surface bright detection limits,  noise characteristics, and
mean sky flux of the HDF $V_{606}$ images.}
\label{fig:sky.hist.hdf}
\end{center}
\end{figure}

As discussed in \S\ref{ebl.minim}, the crucial parameter for
predicting the flux which will be recovered from a detected galaxy is
the difference between the sky (limiting isophotal) surface brightness
and the galaxy's core surface brightness, $\Delta\mu=\mu_{\rm
sky}-\mu_{\rm core}$.  The surface brightness detection limits of our
data and HDF data are demonstrated in the Figures
\ref{fig:detec.rabmumumag} and \ref{fig:detec.hdfmumumag}. 
Figure \ref{fig:detec.hdfmumumag} demonstrates that galaxies in the
HDF with $V_{\rm 606}=29$ AB mag have $\Delta\mu \lta 1.0$mag
arcsec$^{-2}$ on average, with some galaxies in that bin nearly
reaching the limit $\mu_{\rm sky} \approx \mu_{\rm core}$.  If these
galaxies have exponential profiles, then only $\sim 20$\% of their
total flux can be recovered by standard photometry methods.  By
analogy with our data, galaxy apertures which extend to
$\sqrt{2}r_{\rm iso}$ in the HDF will miss at least 20\% of the flux
from galaxies in the range $25<V_{606}<29.5$ AB mag.  This light will
be included in the estimate of the local sky and thus subtracted from
the flux within the galaxy aperture, doubling the total error.

We have run the same Monte Carlo simulations described in \S B1 using
parameters that describe the noise and sky statistics of the
HDF. Using $\mu_{\rm iso}= 1\sigma_{\rm sky}=27.1 V_{606}$ AB mag
arcsec$^{-2}$, we find an extragalactic contribution to the mean sky
flux of 1.6\tto{-10} and 0.8\tto{-10}\escsa\ for galaxy apertures
extending to $1.4r_{\rm iso}$ and $4r_{\rm iso}$, respectively, as
shown by the histograms in Figure \ref{fig:sky.hist.hdf}.  As for our
EBL images, this is in good agreement with estimates of the fractional
flux that should be recovered from galaxies as a function of magnitude
for the corresponding values of $\Delta\mu$.  In order to calculate
the total flux from galaxies detected in the HDF, we can sum the
flux in the individual sources within standard, $1.4r_{\rm iso}$
apertures, and add to the resulting surface brightness {\it twice} the
flux which lies outside of those apertures as identified from the
simulated sky histogram for HDF detection parameters.  Doing so, we
find that the total corrected flux from detected galaxies in the HDF
$23<V_{606}<31$ is 7.5\tto{-10}\escsa\ at 5850\AA. Converting to the
central wavelength of the $V_{555}$ filter, this is roughly
9.1\tto{-10}\escsa\ at 5250\AA, in good agreement with the total flux
from sources in our EBL field, estimated in \S\ref{minebldiscussion}.

\end{document}